\documentclass[aip,pop,reprint]{revtex4-1}

\usepackage{graphicx}                       % Graphics package
\usepackage[outdir=./]{epstopdf}
\usepackage{amsmath}
\usepackage{amssymb}
\usepackage{amsthm}
\usepackage{dsfont}       % Required for dsfont - blackboard bold numbers
\usepackage[usenames,dvipsnames,table]{xcolor}
\usepackage[format=plain, font=small, labelfont=bf, justification = raggedright]{caption}
\usepackage{array}  % Required to make custom columns
\usepackage{overpic}
\usepackage{placeins} % Required to use floatbarrier
\usepackage{fancyhdr}
\usepackage{colortbl}
\usepackage{multirow} % Required to use multirow/multicolumn
\usepackage{hyperref} % Required to have doi links in bibliography
\hypersetup{colorlinks = true,linkcolor = blue, breaklinks = true}

\newcommand{\eq}[1]{(\ref{#1})}
\newcommand{\Eq}[1]{Eq.~\eq{#1}}
\newcommand{\Eqs}[1]{Eqs.~\eq{#1}}
\newcommand{\Fig}[1]{Fig.~\ref{#1}}
\newcommand{\Sec}[1]{Sec.~\ref{#1}}

\renewcommand{\Ref}[1]{Ref.~\onlinecite{#1}}
\newcommand{\Refs}[1]{Refs.~\onlinecite{#1}}
\newcommand{\App}[1]{Appendix~\ref{#1}}

\newcommand{\eg}{{e.g., }}
\newcommand{\ie}{{i.e., }}

\newcommand{\mc}[1]{\mathcal{#1}}

\newcommand{\msf}[1]{\mathsf{#1}}

\newcommand{\mbb}[1]{\mathbb{#1}}

\newcommand{\bra}[1]{\langle#1 |}
\newcommand{\ket}[1]{|#1 \rangle}
\newcommand{\braket}[2]{\langle#1 |  #2 \rangle}

\newcommand{\oper}[1]{\smash{\hat{#1}}}
\newcommand{\unit}[1]{\smash{\check{#1}}}
\newcommand{\fourier}[1]{\smash{\widetilde{#1}}}

\newcommand{\pd}[1]{\partial_{#1}}

\newcommand{\dd}{\mathrm{d}}

\newcommand{\Vect}[1]{{\boldsymbol{\rm #1}}}
\newcommand{\VectOp}[1]{\oper{\Vect{#1}}}

\newcommand{\Mat}[1]{\msf{#1}}
\newcommand{\IMat}[1]{\Mat{I}_{#1}}
\newcommand{\OMat}[1]{\Mat{0}_{#1}}
\newcommand{\JMat}[1]{\Mat{J}_{#1}}

\newcommand{\Tr}{\text{tr}}

\newcommand{\Symb}[1]{\mc{#1}}
\newcommand{\Weyl}{\mbb{W}}
\newcommand{\WeylInv}{\mbb{W}^{-1}}

\newcommand{\NIMT}[1]{\mbb{N}^{}_{#1}}
\newcommand{\IdentOp}{\oper{\mathds{1}}}

\newcommand{\cont}[1]{\mc{C}_{#1}}
\newcommand{\curv}{\mc{K}}

\newcommand{\MTnorm}{\mc{N}_\Vect{t}}
\newcommand{\env}{\phi}

\renewcommand{\Re}{\textrm{Re}}

\newcommand{\nullFrac}{\vphantom{\frac{}{}}}

\newcommand{\Stroke}[1]{\text{\ooalign{ $#1$\cr \hidewidth\raise.225ex \hbox{$-\mkern.5mu$}\cr}}}

\begin{document}
\setlength{\parskip}{0pt}
\setlength{\belowcaptionskip}{0pt}
%\setlength{\abovedisplayskip}{7pt}
%\setlength{\belowdisplayskip}{7pt}

%%%%%%%%%%%%%%%

\title{Metaplectic geometrical optics for ray-based modeling of caustics: Theory and algorithms}
\author{N. A. Lopez}
\affiliation{Department of Astrophysical Sciences, Princeton University, Princeton, New Jersey 08544, USA}
\author{I. Y. Dodin}
\affiliation{Department of Astrophysical Sciences, Princeton University, Princeton, New Jersey 08544, USA}
\affiliation{Princeton Plasma Physics Laboratory, Princeton, New Jersey 08543, USA}

\begin{abstract}
The optimization of radiofrequency-wave (RF) systems for fusion experiments is often performed using ray-tracing codes, which rely on the geometrical-optics (GO) approximation. However, GO fails at caustics such as cutoffs and focal points, erroneously predicting the wave intensity to be infinite. This is a critical shortcoming of GO, since the caustic wave intensity is often the quantity of interest, \eg RF heating. Full-wave modeling can be used instead, but the computational cost limits the speed at which such optimizations can be performed. We have developed a less expensive alternative called metaplectic geometrical optics (MGO). Instead of evolving waves in the usual $\Vect{x}$ (coordinate) or $\Vect{k}$ (spectral) representation, MGO uses a mixed $\Vect{X} \equiv \Mat{A}\Vect{x} + \Mat{B}\Vect{k}$ representation. By continuously adjusting the matrix coefficients $\Mat{A}$ and $\Mat{B}$ along the rays, one can ensure that GO remains valid in the $\Vect{X}$ coordinates without caustic singularities. The caustic-free result is then mapped back onto the original $\Vect{x}$ space using metaplectic transforms. Here, we overview the MGO theory and review algorithms that will aid the development of an MGO-based ray-tracing code. We show how using orthosymplectic transformations leads to considerable simplifications compared to previously published MGO formulas. We also prove explicitly that MGO exactly reproduces standard GO when evaluated far from caustics (an important property which until now has only been inferred from numerical simulations), and we relate MGO to other semiclassical caustic-removal schemes published in the literature. This discussion is then augmented by an explicit comparison of the computed spectrum for a wave bounded between two cutoffs.
\end{abstract}

\maketitle

\pagestyle{fancy}
\lhead{Lopez \& Dodin}
\rhead{Metaplectic geometrical optics}
\thispagestyle{empty}

% ==================== %
% --- INTRODUCTION --- %
% ==================== %
\section{Introduction}

The use of ray-tracing codes to quickly model wave propagation in complex media is ubiquitous in nuclear fusion research, whether that be the radiofrequency (RF) waves used to heat and drive current in a tokamak plasma~\cite{Stix92,Fisch87,Prater08,Farina14,Poli15,Lopez18a} or the high-power lasers used to compress a fuel pellet~\cite{Marinak01,Lindl04,Hohenberger15,Robey18,Kritcher21}. Indeed, the speed of ray-tracing codes makes them invaluable in multiphysics simulation packages that are increasingly used to design and optimize future experimental campaigns. Unfortunately, the ray-tracing formalism, based on the geometrical-optics (GO) approximation~\cite{Kravtsov90,Tracy14}, breaks down at caustics and mode-conversion regions. This is a particular impediment to fusion research since the wave behavior at such regions is often precisely the quantity being optimized, for example, when attempting to mode-convert externally launched RF waves to electrostatic waves that can drive current in an overdense plasma~\cite{Igami06,Laqua07,Urban11}. There has been much recent progress to rigorously incorporate mode conversion into the GO framework, notably the normal-form approach~\cite{Tracy93,Tracy01,Tracy03a,Tracy07,Jaun07,Richardson08} and the extended GO (XGO) framework~\cite{Ruiz15a,Ruiz15b,Ruiz17a} with its beam-tracing generalization~\cite{Dodin19,Yanagihara19a,Yanagihara19b,Yanagihara21a,Yanagihara21b}. Comparatively less work has been dedicated to caustics.

The most promising approach to modeling caustics with rays are phase-space GO methods inspired by Maslov~\cite{Maslov81,Ziolkowski84,Thomson85}. In Maslov's method, caustics are avoided by occasionally Fourier-transforming the wavefield, since the Fourier transform (FT) of a caustic wavefield is locally nonsingular. However, the exact moment for performing the FT is only loosely specified. This makes Maslov's method cumbersome to implement in codes, since it would require supervision. More recent methods~\cite{Littlejohn85,Littlejohn86b,Kay94a,Zor96,Madhusoodanan98,Alonso97b,Alonso99} remedy this issue by replacing the occasional FT of Maslov's theory with a metaplectic transformation (MT) applied continually along a ray. However, these works either introduced additional free parameters or made overly restrictive assumptions on the class of solutions sought (\eg wavepackets). They also tended to simultaneously under- and over-emphasize the use of rays by expressing the wavefield as an integral taken over all points along every ray (rather than only the rays that actually arrive at a given observation point), whose integrand is determined entirely by the phase-space ray geometry (only true for scalar diffraction-free waves). See \Sec{sec:MGOcompare} for more details of this comparison.

As an attempt to resolve these remaining shortcomings, we have developed a new ray-tracing framework called metaplectic geometrical optics (MGO)~\cite{Lopez19,Lopez20,Lopez21a,Lopez21b,Donnelly21}. In essence, MGO is a formalism that \textbf{(i)} avoids caustics by construction, \textbf{(ii)} can be applied to any linear wave equation, including integro-differential wave equations that arise in kinetic treatments of plasma waves, \textbf{(iii)} includes an envelope equation along each ray that can readily incorporate diffraction and eventually mode conversion too, and \textbf{(iv)} is practical for implementation in a ray-tracing code, as exemplified by the several efficient algorithms for MGO that have already been developed. Although development is still ongoing, MGO has already been demonstrated as a robust scalar-wave theory via several examples, namely, fold and cusp caustics in various different types of wave equations~\cite{Lopez20,Lopez21a}.

This paper is organized as follows. In \Sec{sec:background} we summarize how caustic singularities arise in GO (\Sec{sec:GO}) and examine the five stable caustics that can occur in three dimensions ($3$-D) within the context of a paraxial wave propagating in uniform medium, (\Sec{sec:paraxial}). In \Sec{sec:MGOtheory} we provide a theoretical overview of the MGO approach to modeling caustics, which is actually `caustic-agnostic' in that the different caustic types discussed in \Sec{sec:paraxial} do not explicitly show up in the theory. We first review the metaplectic transform, with particular emphasis on orthosymplectic transformations (\Sec{sec:MT}). We then review the MGO formalism, and present a novel derivation using orthosymplectic transformations that leads to a considerably simplified final result (\Sec{sec:MGOrev}). We then show for the first time how MGO reduces analytically to GO when evaluated far from caustics (\Sec{sec:MGOtoGO}), and we illustrate how MGO can be understood in the context of other published semiclassical methods (\Sec{sec:MGOcompare}).

In \Sec{sec:MGOalgor} we discuss four recently developed algorithms for an MGO-based ray-tracing code: an adaptive discretization for MGO rays (\Sec{sec:ALGadapt}), an explicit construction of the desired phase-space rotation along the MGO rays (\Sec{sec:ALGgs}), a fast linear-time algorithm for computing near-identity MTs (\Sec{sec:ALGnimt}), and a numerical steepest descent quadrature rule for evaluating MT integrals (\Sec{sec:ALGgf}). We then use this last algorithm to numerically compute the MGO solution for a wave bounded in a parabolic cavity well and compare with the exact solution and other semiclassical models. Finally, in \Sec{sec:summary} we summarize the MGO procedure in a step-by-step list that might also serve as outline for an MGO-based ray-tracing code, and in \Sec{sec:concl} we conclude. Additional discussions are provided in appendices.

% ==================== %
% -- BACKGROUND -- %
% ==================== %

\section{Background}
\label{sec:background}

% ==================== %
% -- GO BACKGROUND -- %
% ==================== %

\subsection{General theory of caustics in geometrical optics}
\label{sec:GO}

The GO model aims to describe the propagation of waves in inhomogeneous media when the wavelength is the shortest relevant lengthscale (including those characterizing the media and the wavefield itself). To develop this idea more quantitatively, suppose a stationary wavefield $\psi$ is governed by a linear wave equation
\begin{equation}
    \oper{D}(\Vect{x}, -i \pd{\Vect{x}}) \psi(\Vect{x}) = 0
    .
    \label{eq:waveEQ}
\end{equation}

\noindent (We assume $\psi$ is a scalar wavefield for simplicity.) Suppose further that $\psi$ can be partitioned into a rapidly varying phase $\theta$ and a slowly varying envelope $\env$:
\begin{equation}
    \psi(\Vect{x})
    =\env(\Vect{x}) \exp[i \theta(\Vect{x})]
    .
    \label{eq:wave}
\end{equation}

\noindent Then, it is well-known~\cite{Tracy14,Dodin19} (and will be shown explicitly in \Sec{sec:MGOtheory}) that $\theta$ and $\env$ asymptotically satisfy the following two relations: (i) the local dispersion relation
\begin{equation}
    \Symb{D}[\Vect{x}, \pd{\Vect{x}} \theta(\Vect{x})] = 0
    ,
    \label{eq:goDISP}
\end{equation}

\noindent and the envelope transport equation
\begin{equation}
    2 \Vect{v}(\Vect{x})^\intercal \pd{\Vect{x}} \env(\Vect{x}) + \left[\nabla \cdot \Vect{v}(\Vect{x})\right] \env(\Vect{x}) = 0
    .
    \label{eq:goENV}
\end{equation}

\noindent Here, $\Symb{D}(\Vect{x}, \Vect{k})$ is the Weyl symbol of $\oper{D}(\Vect{x}, -i \pd{\Vect{x}})$ (\App{sec:APPwwt}) and 
\begin{equation}
    \Vect{v}(\Vect{x}) \doteq \left.\pd{\Vect{k}} \Symb{D}(\Vect{x}, \Vect{k})\right|_{\Vect{k} = \pd{\Vect{x}} \theta(\Vect{x})}
\end{equation}

\noindent is proportional to the local group velocity. (Note, all vectors are column vectors unless explicitly transposed via $^\intercal$.) For simplicity, we shall neglect dissipation in \Eq{eq:wave}. Specifically, we shall assume that $\oper{D}$ is Hermitian and consequently, both $\Symb{D}$ and $\Vect{v}$ are real. Then, \Eq{eq:goENV} can also be cast as a conservation relation:
\begin{equation}
    \nabla \cdot \left[|\env(\Vect{x})|^2 \Vect{v}(\Vect{x})  \right]
    = 0,
    \label{eq:actionCONS}
\end{equation}

\noindent where the quantity within square brackets is recognized as the wave action flux (or the wave energy flux, which for stationary waves is the same up to a constant factor)~\cite{Whitham74}.

The local dispersion relation naturally resides within the $2N$-D phase space with coordinates $(\Vect{x}, \Vect{k})$; \Eq{eq:goDISP} implicitly defines a $(2N-1)$-D volume within this phase space that describes the local momentum of the wavefield at a given point $\Vect{x}$ in configuration space. For coherent wavefields that have a single wavevector $\Vect{k}(\Vect{x})$ (or a finite superposition of such wavevectors), then one can identify
\begin{equation}
    \Vect{k}(\Vect{x}) = \pd{\Vect{x}} \theta(\Vect{x})
    \label{eq:goK}
\end{equation}

\noindent such that $\Vect{k}$ is actually restricted to an $N$-D surface contained within the $(2N-1)$-D volume defined by \Eq{eq:goDISP}. (The specific $N$-D surface is dictated by initial conditions.) This $N$-D surface is called the ray manifold, and by resulting from a gradient lift \eq{eq:goK} it is a Lagrangian manifold~\cite{Tracy14,Arnold89}. In particular, this means that all vectors $\{\Vect{T}_j \}$ tangent to it satisfy
\begin{equation}
    \Vect{T}_j^\intercal \JMat{2N} \Vect{T}_{j'} = 0
    ,
    \label{eq:tangentLAGRANG}
\end{equation}

\noindent where we have introduced the $2N \times 2N$ matrix
\begin{equation}
    \JMat{2N} = 
    \begin{pmatrix}
        \OMat{N} & \IMat{N} \\
        - \IMat{N} & \OMat{N}
    \end{pmatrix}
    ,
    \label{eq:Jmat}
\end{equation}

\noindent with $\IMat{N}$ and $\OMat{N}$ being the $N \times N$ identity and zero matrices, respectively. We shall make use of \Eq{eq:tangentLAGRANG} in \Sec{sec:MGOtheory}.

The ray manifold is a central object in GO and MGO. It is therefore useful for practical purposes to have an explicit construction of it, rather than relying on the formal construction described in the preceding paragraph. This explicit construction is provided by the ray (Hamilton's) equations~\cite{Tracy14}
\begin{equation}
    \pd{\xi} \Vect{x} = 
    \pd{\Vect{k}} \Symb{D}(\Vect{x}, \Vect{k})
    , \quad
    \pd{\xi} \Vect{k} =
    - \pd{\Vect{x}} \Symb{D}(\Vect{x}, \Vect{k})
    .
    \label{eq:goRAYS}
\end{equation}

\noindent The family of solution trajectories $(\Vect{x}(\xi), \Vect{k}(\xi))$ for a corresponding family of initial conditions $(\Vect{x}(0), \Vect{k}(0))$ then trace out the ray manifold.

Since the ray manifold is $N$-D, let us introduce a set of $N$-D coordinates $\Vect{\tau}$ such that it can be parameterized as $(\Vect{x}(\Vect{\tau}),\Vect{k}(\Vect{\tau}))$. We shall choose $\tau_1 = \xi$ as a `longitudinal' coordinate along each ray and the remaining $\Vect{\tau}_\perp \doteq (\tau_2,\ldots \tau_N)$ as `transverse' coordinates that describe the different initial conditions of each ray. Along this family of rays, the envelope equation \eq{eq:goENV} takes a simple form~\cite{Kravtsov90,Lopez20}:
\begin{equation}
    2 j(\Vect{\tau}) \pd{\tau_1} \env(\Vect{\tau})
    + \env(\Vect{\tau}) \pd{\tau_1} j(\Vect{\tau})
    = 0
    ,
    \label{eq:goENVray}
\end{equation}

\noindent where we have introduced the Jacobian determinant of the ray trajectories
\begin{equation}
    j(\Vect{\tau}) \doteq 
    \det \pd{\Vect{\tau}} \Vect{x}(\Vect{\tau})
    .
    \label{eq:jacDEF}
\end{equation}

Equation \eq{eq:goENVray} can be formally solved to yield the envelope evolution along a ray:
\begin{equation}
    \env(\Vect{\tau}) = \env_0(\Vect{\tau}_\perp) 
    \sqrt{
        \frac{j_0(\Vect{\tau}_\perp)}{j(\Vect{\tau})}
    }
    ,
    \label{eq:goENVsol}
\end{equation}

\noindent where $\env_0$ and $j_0(\Vect{\tau}_\perp)$ are determined by initial conditions. Intuitively, since the matrix determinant equals the (signed) volume spanned by the constituent column (or row) vectors, \Eq{eq:goENVsol} states that $|\env|^2 |\Vect{v}| \dd A$ is constant along a ray, where $\dd A$ is an infinitesimal cross-sectional area of a ray family. This is consistent with action conservation \eq{eq:actionCONS} for an infinitesimal `ray tube' volume centered on a specific ray. 

Having determined $\env$ from \Eq{eq:goENVsol} and $\theta$ from integrating the rays [\Eqs{eq:goRAYS}, then \eq{eq:goK}], the full field $\psi$ is constructed by summing over all rays that arrive at a given $\Vect{x}$, that is,
\begin{align}
    \psi(\Vect{x})
    &= \sum_{\Vect{t} \in \Vect{\tau}(\Vect{x})}
    \env(\Vect{t}) \exp[i\theta(\Vect{t})]
    \nonumber\\
    &\equiv
    \sum_{\Vect{t} \in \Vect{\tau}(\Vect{x})}
    \env_0(\Vect{t}_\perp) 
    \sqrt{
        \frac{j_0(\Vect{t}_\perp)}{j(\Vect{t})}
    } \exp\left(i \int \Vect{k}^\intercal \dd \Vect{x} \right)
    ,
    \label{eq:GO}
\end{align}

\noindent where $\Vect{\tau}(\Vect{x})$ is the formal function inverse of $\Vect{x}(\Vect{\tau})$ and is generally multi-valued (corresponding to the multiple rays whose interference pattern determines $\psi$). Clearly though, the GO field \eq{eq:GO} diverges where
\begin{subequations}
    \begin{equation}
       j(\Vect{t}) = 0
        ,
    \end{equation}

    \noindent or equivalently, where 
    \begin{equation}
        \det \pd{\Vect{x}}\Vect{k} 
        \equiv \det \pd{\Vect{x} \Vect{x}} \theta
        \to \infty
        .
        \label{eq:causticSING}
    \end{equation}
    \label{eq:causticDIV}
\end{subequations}

\noindent Such locations are called `caustics'. The accurate modeling of $\psi$ in the neighborhood of caustics is the primary goal of this work.

% ==================== %
% -- PARAXIAL CAUSTICS -- %
% ==================== %

\subsection{Case study: caustics in paraxial propagation}
\label{sec:paraxial}

Before discussing how to accurately model caustics, it is instructive to review the different types of caustics that can occur. Being singularities of a gradient map [\Eq{eq:causticSING}], caustics are optical catastrophes~\cite{Berry80b}, and as such, they can be systematically classified using catastrophe theory~\cite{Poston96}. For $3$-D systems, it turns out that only five unique caustics can occur: the fold ($A_2$), the cusp ($A_3$), the swallowtail ($A_4$), the hyperbolic umbilic ($D_4^+$), and the elliptic umbilic ($D_4^-$) catastrophe functions. The labels within parenthesis correspond to the commonly adopted Arnold classification~\cite{Arnold75}. As the labels suggests, the first three caustics in the list are members of a larger family of caustics called the `cuspoids' (represented by the label $A_{m+1}$ for $m \ge 1$), while the final two are members of the `umbilic' family (represented by the label $D_{m+1}^\pm$ for $m\ge 3$). Intuitively, $m$ is the minimum number of dimensions required to view the corresponding caustic in its entirety, and $m+1$ is the number of rays involved in creating the caustic pattern. For example, the fold caustic (which corresponds to a cutoff) can occur in $1$-D and involves two interfering rays (the incoming and reflected rays).

To see how these caustics can occur in practice, let us consider a wavefield propagating paraxially in an ($N + 1$)-D uniform medium according to the wave equation
\begin{equation}
    4 \pi i \pd{z} \psi(\Vect{x}, z)
	+ \lambda \pd{\Vect{x}}^2 \psi(\Vect{x}, z) 
	= 0
	,
	\label{eq:paraxial}
\end{equation}

\noindent where $z$ is the direction of propagation, $\Vect{x}$ are the $N$-D coordinates transverse to $z$ (\ie the optical axis corresponds to $\Vect{x} = \Vect{0}$), and $\lambda$ is the wavelength. The formal solution to \Eq{eq:paraxial} is readily obtained:
\begin{subequations}
    \begin{equation}
        \psi(\Vect{x}, z) = 
        \exp\left(
            i \frac{\lambda z}{4 \pi} 
            \pd{\Vect{x}}^2
        \right) \psi(\Vect{x},0)
        ,
        \label{eq:freePMT}
    \end{equation}
    
    \noindent or equivalently,
    \begin{equation}
        \psi(\Vect{x}, z) =
        \int \dd \Vect{y} \,
        \frac{
            \psi(\Vect{y}, 0) 
        }{ 
            \left( i \lambda z \right)^{N/2}
        }
        \exp\left( 
            i \pi\frac{
                \|\Vect{y} - \Vect{x}\|^2
            }{\lambda z} 
        \right)
        .
        \label{eq:freeMT}
    \end{equation}
\end{subequations}

\noindent One recognizes \Eq{eq:freeMT} as the well-known Fresnel diffraction integral~\cite{Born99}, but it is also an example of a metaplectic transform (\Sec{sec:MT}), which feature prominently in MGO.

The GO rays for \Eq{eq:paraxial} solve the local dispersion relation
\begin{equation}
    \Symb{D}(\Vect{x},z, \Vect{k}, k_z)
    = \frac{4\pi k_z}{\lambda} + \Vect{k}^\intercal \Vect{k}
    = 0
    .
\end{equation}

\noindent Hence, they are given explicitly as
\begin{align}
    \Vect{x}(z, \Vect{x}_0) &= \Vect{x}_0 + \frac{\lambda z}{2\pi} \Vect{k}_0(\Vect{x}_0)
    , \quad
    \Vect{k}(z, \Vect{x}_0) = \Vect{k}_0(\Vect{x}_0)
    ,
    \label{eq:paraxRAYS}
\end{align}

\noindent where in the notation of the previous section we have chosen to set $\Vect{\tau} = (z, \Vect{x}_0)$.

\subsubsection{Cuspoid caustics}

Consider first the case $N = 1$. An $A_{m+1}$-type cuspoid caustic can be generated by choosing the following initial conditions for $\psi$ (up to an arbitrary constant factor):
\begin{equation}
    \psi(x, 0)
    = \exp\left( i \, \frac{x^{m+2}}{\ell^{m+2}} + i \sum_{j = 1}^m a_j \frac{x^j}{\ell^j} \right)
    ,
    \label{eq:cuspoidINIT}
\end{equation}

\noindent which corresponds to
\begin{equation}
    k_{x,0}(x_0) = (m+2)\frac{x_0^{m+1}}{\ell^{m+2}} + \sum_{j = 1}^m j a_j \frac{x_0^{j-1}}{\ell^j}
    .
    \label{eq:cuspoidINITk}
\end{equation}

\noindent (Here $\{a_j\}$ are constant parameters, \eg lens aberrations, and $\ell$ determines the characteristic length.) Then, for $m = 1$, \Eq{eq:freeMT} leads to
\begin{align}
	&\psi(x, z) =
	\frac{\ell}{\sqrt{ i \lambda z} } \,
	\exp \left(
		i \frac{\pi x^2}{\lambda z} 
		+ i \frac{2 \pi^2 \ell^3 x}{3 \lambda^2 z^2}
		+ i \frac{2 \pi^3 \ell^6}{27 \lambda^3 z^3}
	\right)
	\nonumber\\
	&\hspace{5mm}\times
	\exp\left(
	    - i a_1 \frac{\pi \ell^2}{3 \lambda z}
	\right)
	A_2\left(
		a_1 
		- \frac{2 \pi \ell x}{\lambda z} 
		- \frac{\pi^2 \ell^4}{3 \lambda^2 z^2}
	\right)
	,
	\label{eq:cuspoidSOLm1}
\end{align}

\noindent and for $m > 1$, \Eq{eq:freeMT} leads to
\begin{align}
	&\psi(x, z) =
	\frac{\ell}{\sqrt{ i \lambda z} } \,
	\exp \left(
		i \frac{\pi x^2}{\lambda z} 
	\right)
	\nonumber\\
	&\hspace{7mm}\times
	A_{m+1}\left(
		a_1 - \frac{2\pi \ell x}{\lambda z}
		,
		a_2 + \frac{\pi \ell^2}{\lambda z}
		,
		a_3
		, \ldots
		,
		a_m
	\right)
	,
	\label{eq:cuspoidSOL}
\end{align}

\noindent where the $A_{m+1}$ `catastrophe integral' is defined as
\begin{align}
	&A_{m+1}\left(
		a_1
		, \ldots
		,
		a_m
	\right)
	\nonumber\\
	&\hspace{20mm}\doteq
	\int \dd y \,
	\exp\left(
		i y^{m+2} + i \sum_{j = 1}^m a_j y^j 
	\right)
	.
\end{align}

\begin{figure}
	\includegraphics[width=\linewidth,trim={3mm 22mm 13mm 21mm},clip]{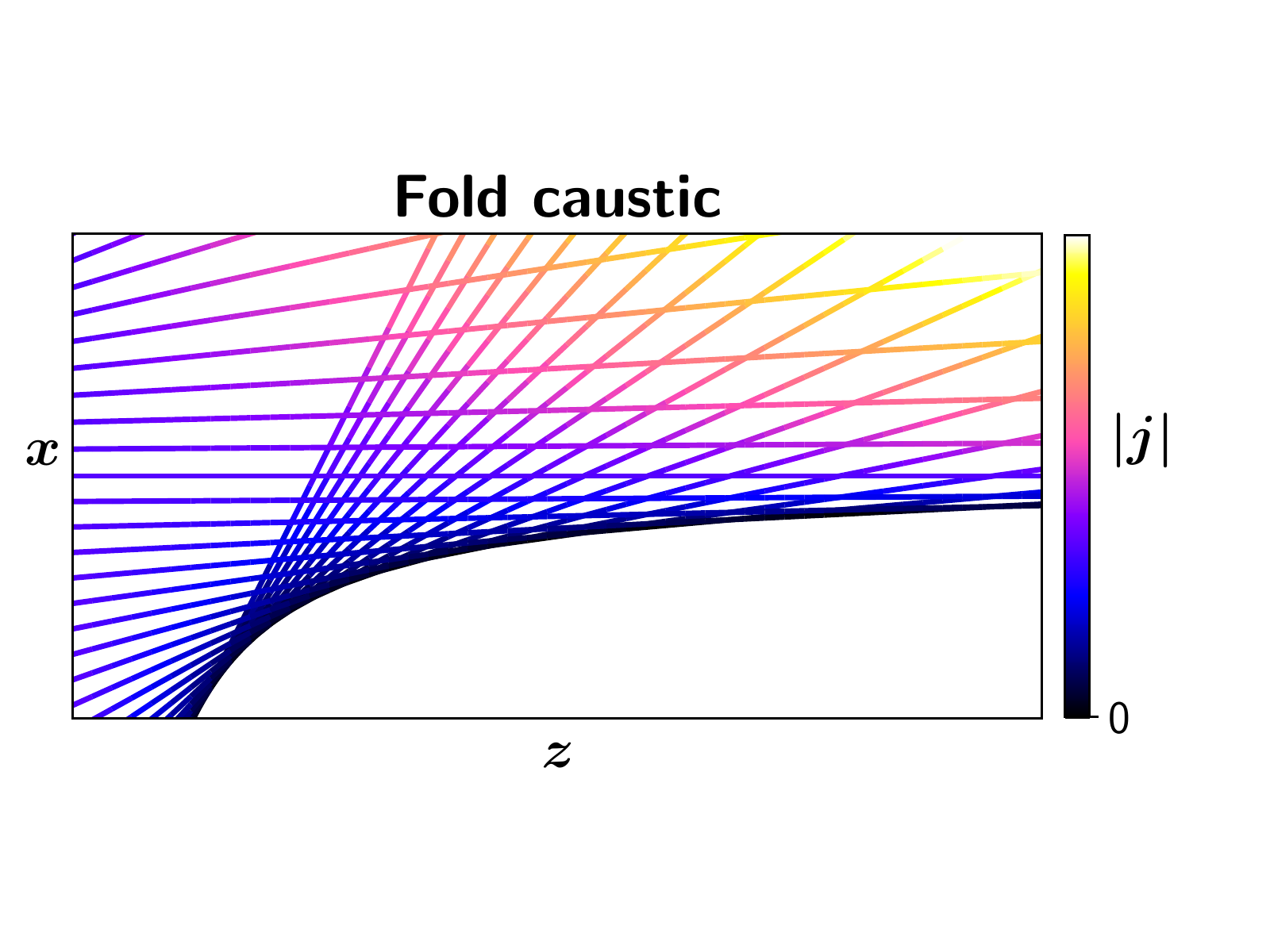}
	\caption{Ray trajectories for the fold caustic obtained via \Eq{eq:rayFOLD} with $\lambda = 2\pi$, $\ell = \sqrt[3]{3}$, and $a_1 = 0$. The color shows the magnitude of the Jacobian $j(\Vect{\tau})$ defined by \Eq{eq:jacDEF}. The caustic occurs where $j = 0$ (black curve).}
	\label{fig:fold}
\end{figure}

The simplest caustic of the cuspoid family is the $A_2$ fold caustic. The field near a fold caustic is given by \Eq{eq:cuspoidSOLm1}, and the underlying ray trajectories are given by \Eqs{eq:paraxRAYS} and \eq{eq:cuspoidINITk}; namely,
\begin{equation}
    x(z, x_0)
    = x_0
    + \frac{\lambda z}{2\pi \ell }
    \left(
        \frac{3 x_0^2}{\ell^2} + a_1
    \right)
    .
    \label{eq:rayFOLD}
\end{equation}

\noindent The fold caustic occurs where \Eq{eq:causticDIV} is satisfied, or equivalently, where $\pd{x_0}x(z,x_0) = 0$. This ultimately yields the caustic curve
\begin{equation}
    x_c(z)
    =
    a_1\frac{\lambda z}{2 \pi \ell}
    - \frac{\pi \ell^3}{6 \lambda z}
    .
    \label{eq:causticFOLD}
\end{equation}

\noindent The ray pattern \eq{eq:rayFOLD} is shown in \Fig{fig:fold}, in which the caustic curve \eq{eq:causticFOLD} is clearly visible.

Let us next consider the $A_3$ cusp caustic. The field near a cusp caustic is given by \Eq{eq:cuspoidSOL} with $m = 2$, and the underlying ray trajectories are given as
\begin{equation}
    x(z, x_0)
    = x_0
    + \frac{\lambda z}{2\pi \ell}
    \left(
        \frac{4 x_0^3}{\ell^3} 
        + a_1
        + a_2 \frac{2 x_0}{\ell}
    \right)
    .
    \label{eq:rayCUSP}
\end{equation}

\noindent The cusp caustic can be shown to occur along the curve
\begin{equation}
    x_c(z) = 
    a_1\frac{ \lambda z}{2 \pi \ell} 
	\pm
	\frac{
	    \sqrt{- 6 \lambda z (\pi \ell^2 + a_2 \lambda z )^3}
	}{
		9\pi \ell \lambda  z
	}
	.
	\label{eq:causticCUSP}
\end{equation}

\begin{figure}
	\includegraphics[width=\linewidth,trim={3mm 22mm 13mm 21mm},clip]{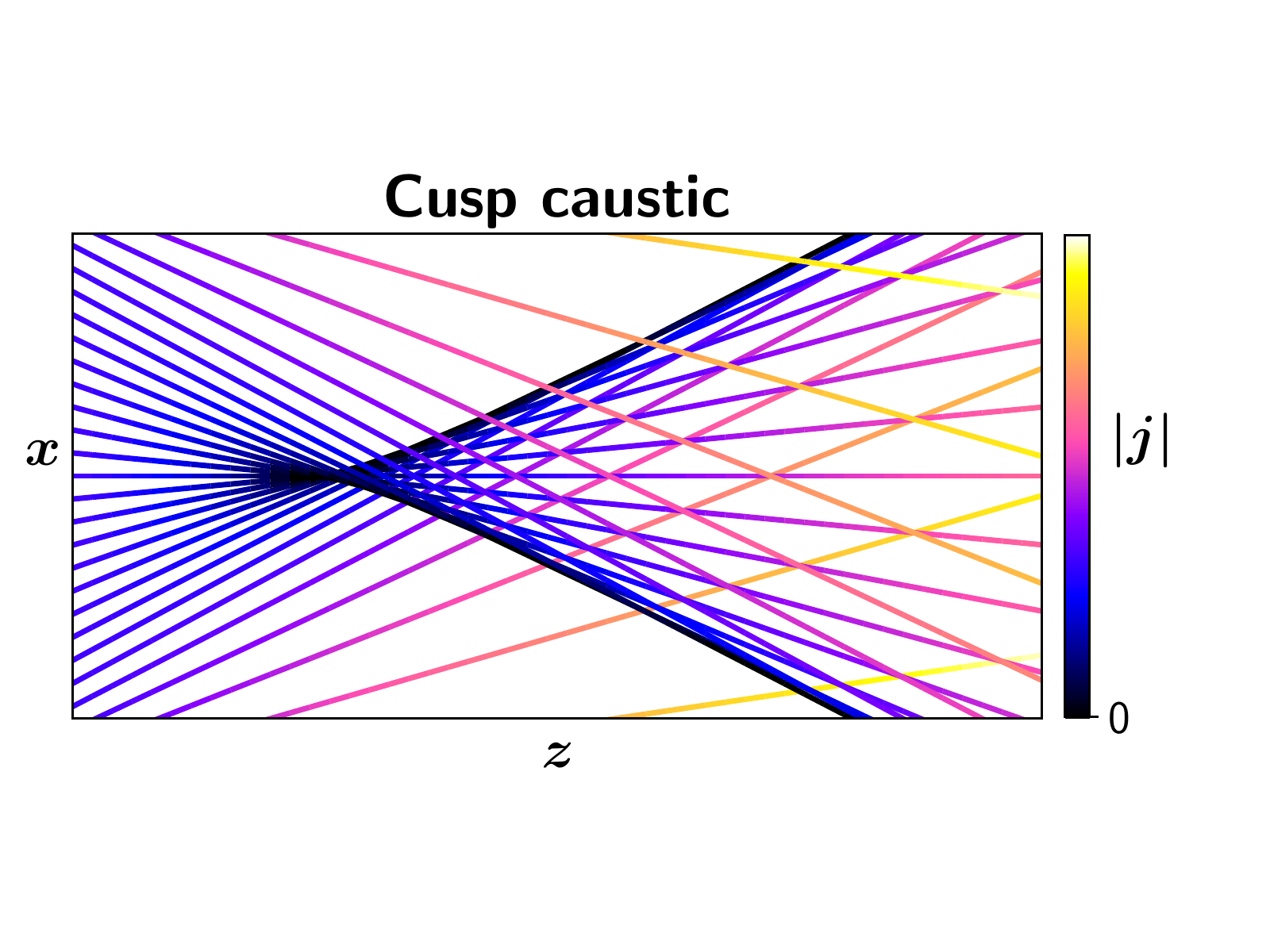}
	\caption{Same as \Fig{fig:fold} but for the cusp caustic \eq{eq:rayCUSP} with $\lambda = 2\pi$, $\ell = \sqrt{2}$, $a_1 = 0$, and $a_2 = -2$.}
	\label{fig:cusp}
\end{figure}

\noindent The ray pattern \eq{eq:rayCUSP} is shown in \Fig{fig:cusp}, in which the caustic curve \eq{eq:causticCUSP} is clearly visible.

\begin{figure}
	\includegraphics[width=\linewidth,trim={3mm 22mm 13mm 21mm},clip]{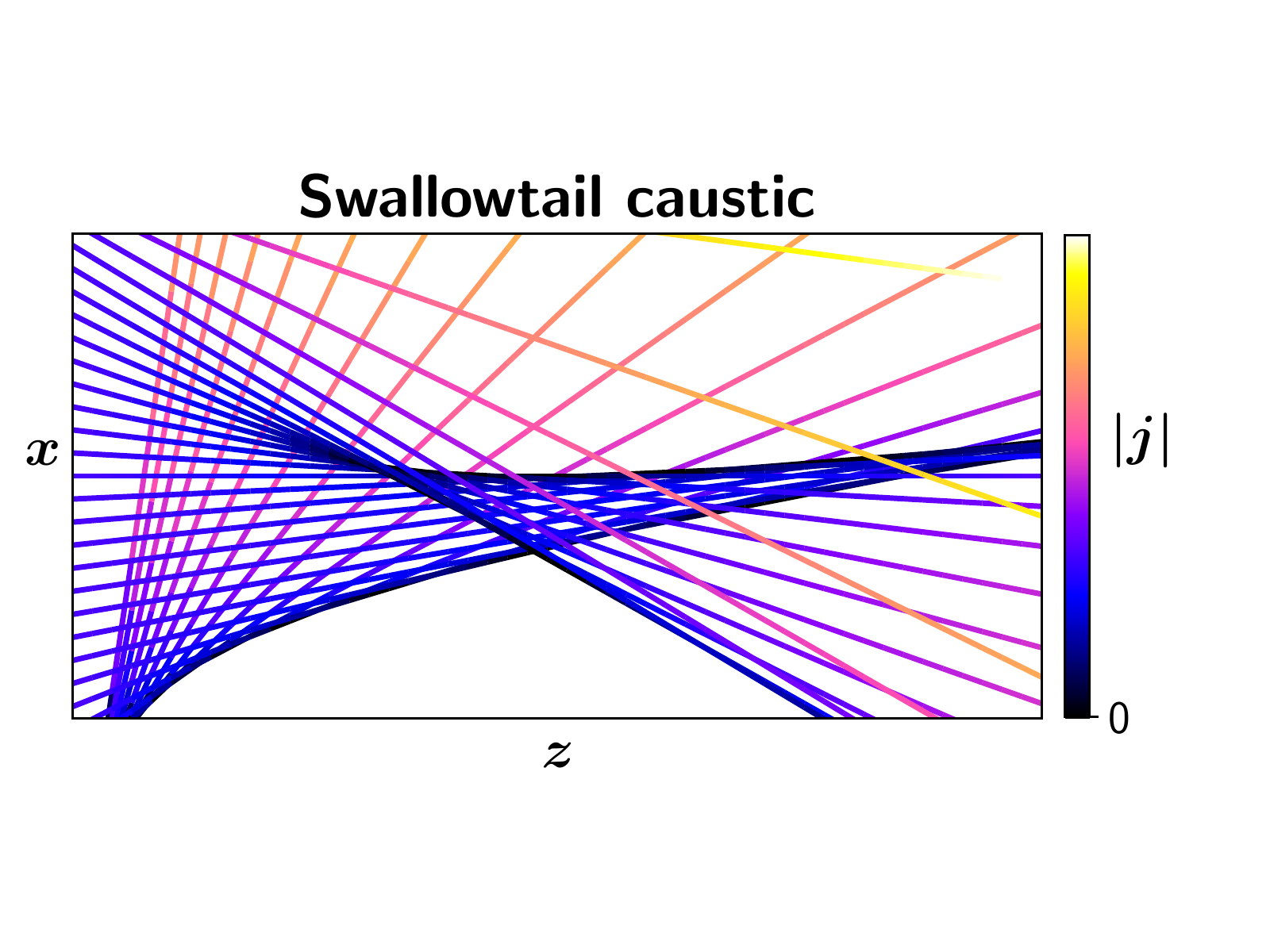}
	\caption{Same as \Fig{fig:fold} but for the swallowtail caustic \eq{eq:raySWTAIL} with $\lambda = 2\pi$, $\ell = \sqrt[5]{5}$, $a_1 = 0$, $a_2 = -1$, and $a_3 = -1$.}
	\label{fig:swtail}
\end{figure}

Let us next consider the final stable cuspoid in $3$-D, the $A_4$ swallowtail caustic. Here, we shall choose to generate the swallowtail via the mathematically simpler approach of having the a high-order $1$-D aberration, rather than with a low-order $2$-D aberration as would more commonly occur in practice. Correspondingly, the field near a swallowtail caustic is given by \Eq{eq:cuspoidSOL} with $m = 3$, and the underlying ray trajectories are given as
\begin{equation}
    \hspace{-2mm}x(z, x_0)
    = x_0
    + \frac{\lambda z}{2\pi \ell}
    \left(
        \frac{5 x_0^4}{\ell^4} 
        + a_1
        + a_2 \frac{2 x_0}{\ell}
        + a_3 \frac{3 x_0^2}{\ell^2}
    \right)
    .
    \label{eq:raySWTAIL}
\end{equation}

\noindent The cusp caustic can be shown to occur along the parametric curve
\begin{subequations}
    \label{eq:causticSWTAIL}
    \begin{align}
    	x_c(\zeta)
	    &=
    	\frac{a_1 \lambda z_c(\zeta)}{2 \pi \ell}
	    - \frac{3 \lambda z_c(\zeta)}{2 \pi \ell} (a_3 + 5 \zeta^2) \zeta^2
    	, \\
	    z_c(\zeta)
    	&= - \frac{\pi \ell^2}
	    {
		    \lambda \left[
			    a_2 + \zeta (3 a_3 + 10 \zeta^2)
    		\right]
	    }
	    ,
    \end{align}
\end{subequations}

\noindent where $\zeta \in (-\infty, \infty)$ is a $1$-D parameterization of the caustic curve in the ($x$,$z$) plane. The ray pattern \eq{eq:raySWTAIL} is shown in \Fig{fig:swtail} for parameters specifically chosen to have the eponymous swallowtail section of the caustic curve \eq{eq:causticSWTAIL} appear in the longitudinal plane.

\subsubsection{Umbilic caustics}

Consider now $N = 2$ (so the total number of spatial dimensions is three). In addition to the cuspoids, a new class of caustics, the umbilics, are now possible. A wavefield containing any caustic from the $D_{m+1}^\pm$ umbilic series can be generated by choosing
\begin{widetext}
    \begin{equation}
	    \psi(\Vect{x}, 0)
    	=
	    \exp\left(
		    i \frac{y^m}{\ell^m}
    		\pm i \frac{x^2 y}{\ell^3}
	    	+ i a_1 \frac{x}{\ell}
		    + i a_2 \frac{y}{\ell}
    		+ i a_3 \frac{x^2}{\ell^2}
	    	+ i \sum_{j = 4}^m a_j \frac{y^{j - 2}}{\ell^{j-2}}
    	\right)
	    .
	    \label{eq:umbilicINIT}
    \end{equation}

    \noindent The corresponding solution \eq{eq:freeMT} for $m \ge 4$ is
    \begin{align}
        \psi(\Vect{x}, z) =
        \frac{\ell^2 }{i \lambda z}
        \exp \left(
            i \frac{\pi \Vect{x}^\intercal \Vect{x}}{\lambda z} 
        \right)
        D_{m+1}^\pm
        \left(
            a_1 - \frac{2\pi \ell x}{\lambda z}
            ,
            a_2 - \frac{2\pi \ell y}{\lambda z}
            ,
            a_3 + \frac{\pi \ell^2 }{\lambda z}
            , 
            a_4 + \frac{\pi \ell^2 }{\lambda z}
            ,
            a_5
            , \ldots
            ,
            a_m
        \right)
        ,
        \label{eq:umbilicSOL}
    \end{align}

    \noindent while for $m = 3$ is given as
    \begin{align}
        \psi(\Vect{x}, z) =
        \frac{\ell^2 }{i \lambda z}
        \exp \left(
            i \frac{\pi \Vect{x}^\intercal \Vect{x}}{\lambda z} 
            -
            i a_2 \frac{\pi \ell^2}{3 \lambda z}
            + i \frac{2 \pi^2 \ell^3}{3 \lambda^2 z^2} y
            + i \frac{2 \pi^3 \ell^6}{27 \lambda^3 z^3}
        \right)
        D_{4}^\pm
    	\left(
            a_1 - \frac{2\pi \ell x}{\lambda z}
            ,
            a_2 - \frac{2\pi \ell y}{\lambda z} - \frac{\pi^2 \ell^4 }{3 \lambda^2 z^2}
            ,
            a_3 + \frac{3 \mp 1 }{3 \lambda z} \pi \ell^2
        \right)
	    ,
	    \label{eq:umbilicSOLm3}
    \end{align}
    
    \noindent where the $D_{m+1}^\pm$ catastrophe integral is defined as
    \begin{align}
	    D_{m+1}^\pm
    	\left(
	    	a_1
		    , \ldots
    		,
	    	a_m
    	\right)
	    \doteq 
    	\int \dd U \, \dd V
	    \exp\left[
		    i V^m
    		\pm i U^2 V
	    	+ i a_1 U
		    + i a_2 V
    		+ i a_3 U^2 
	    	+ i \sum_{j = 4}^m a_j V^{j - 2}
    	\right]
	    .
    \end{align} 

    \noindent Note that the initial condition \eq{eq:umbilicINIT} generates rays having
    \begin{equation}
        k_{x,0}(\Vect{x}_0) =
    		\pm 2 \frac{x_0 y_0}{\ell^3}
	    	+ \frac{a_1}{\ell}
    		+ 2 a_3 \frac{x_0}{\ell^2}
        , \quad
        k_{y,0}(\Vect{x}_0) = m\frac{y_0^{m-1}}{\ell^m}
    		\pm \frac{x_0^2}{\ell^3}
    		+ \frac{a_2}{\ell}
	    	+ \sum_{j = 4}^m (j - 2) a_j \frac{y_0^{j - 3}}{\ell^{j-2}}
	    .
        \label{eq:umbilicINITk}
    \end{equation}
\end{widetext}

The only umbilics that occur stably in $3$-D are the $D_4^+$ hyperbolic and the $D_4^-$ elliptic umbilic caustics. The field near these caustics are given by \Eq{eq:umbilicSOLm3}. The underlying ray trajectories for the hyperbolic umbilic are given as
\begin{subequations}
    \label{eq:rayHUMB}%
    \begin{align}
        x(z; \Vect{x}_0) &=
            x_0 +
            \frac{\lambda z}{2\pi \ell}
            \left(
                a_1
                + 2 a_3 \frac{x_0}{\ell}
                + 2 \frac{x_0 y_0}{\ell^2} 
            \right)
        , \\
        y(z; \Vect{x}_0) &= 
            y_0 +
            \frac{\lambda z}{2 \pi \ell}
            \left(
                a_2
                + \frac{3y_0^2 + x_0^2}{\ell^2}
            \right)
        ,
    \end{align}%
\end{subequations}%

\noindent and the caustic in the $(x,y)$ transverse plane at fixed propagation distance $z$ is given by the parametric curve 
\begin{subequations}
    \label{eq:causticHUmb}
    \begin{align}
        x_c(\zeta)
        &= 
            a_1 \frac{\lambda z}{2\pi \ell}
            - \frac{\sqrt{3} \lambda z}{4 \pi \ell}\left( 
                a_3 + \frac{2\pi \ell^2}{3\lambda z}
            \right)^2
            \nonumber\\
            &\hspace{29mm}\times
            \sinh(\zeta)
            \left[
                \cosh(\zeta) \pm 1
            \right]
        ,\\
        y_c(\zeta)
        &=
            a_2 \frac{\lambda z}{2\pi \ell}
            - \frac{\pi \ell^3}{6 \lambda z}
            + \frac{3 \lambda z}{4\pi \ell}
            \left( 
                a_3 + \frac{2\pi \ell^2}{3\lambda z}
            \right)^2
            \nonumber\\
            &\hspace{29mm}\times
            \cosh(\zeta)
            \left[
                \cosh(\zeta) \mp 1
            \right]
            .
    \end{align}
\end{subequations}

\begin{figure}
	\includegraphics[width=0.8\linewidth,trim={29mm 5mm 11mm 4mm}, clip]{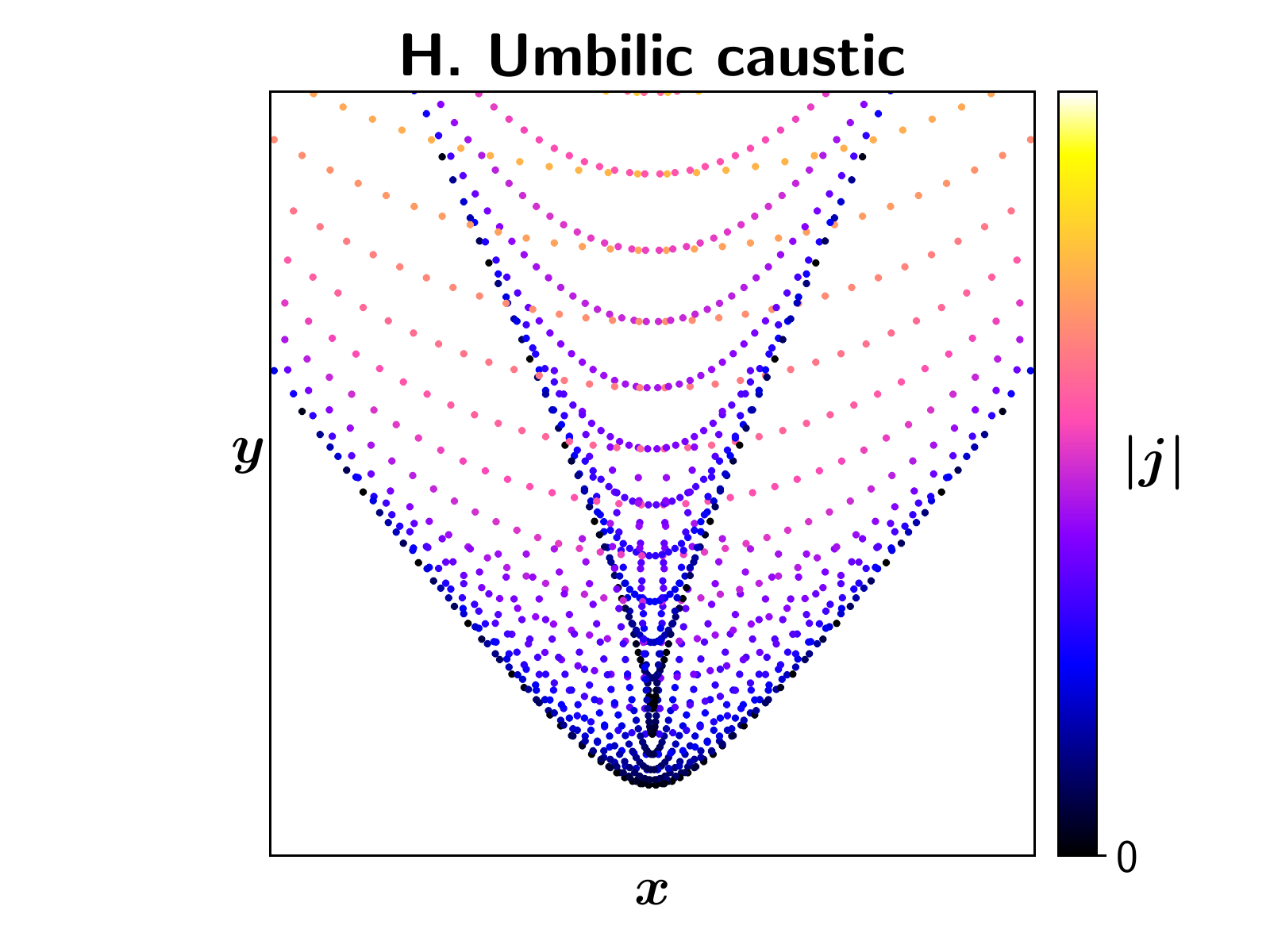}
	\caption{Intersection plot of the ray trajectories for the hyperbolic umbilic caustic obtained via \Eq{eq:rayHUMB} with $\lambda = 4\pi$, $\ell = 1$, $a_1 = 0$, $a_2 = -1.3$, and $a_3 = -1$ through the plane $z = 1$.}
	\label{fig:HUmb}
\end{figure}

\noindent Note that the hyperbolic umbilic actually contains two separate caustic curves (indicated by the $\pm$ terms above): a fold curve (top sign) and a cusp curve (bottom sign). These two curves lie in a squid-like orientation with the cusp residing within the bow of the fold. This is readily observed in \Fig{fig:HUmb}, which presents an intersection plot for the rays \eq{eq:rayHUMB} with respect to the plane $z = 1$. As each dot represents a single traversing ray, the $D_4^+$ caustic \eq{eq:causticHUmb} manifests as a visible increase in the ray density.

Similarly, the ray trajectories for the elliptic umbilic are given as
\begin{subequations}
    \label{eq:rayEUMB}
    \begin{align}
        x(z; \Vect{x}_0) &=
            x_0 +
            \frac{\lambda z}{2\pi \ell}
            \left(
                a_1
                + 2 a_3 \frac{x_0}{\ell}
                - 2 \frac{x_0 y_0}{\ell^2} 
            \right)
        , \\
        y(z; \Vect{x}_0) &=
            y_0 +
            \frac{\lambda z}{2\pi \ell}
            \left(
                a_2
                + \frac{3 y_0^2 - x_0^2}{\ell^2}
            \right)
        ,
    \end{align}
\end{subequations}

\noindent with the caustic at a fixed distance $z$ given by the parametric curve
\begin{subequations}
    \label{eq:causticEUmb}
    \begin{align}
        x_c(\zeta)
        &= 
            a_1 \frac{\lambda z}{2\pi \ell}
            + \frac{\sqrt{3} \lambda z}{2 \pi \ell}\left( 
                a_3 + \frac{4\pi \ell^2}{3\lambda z}
            \right)^2
            \nonumber\\
            &\hspace{36mm}\times
            \sin^2\left(\frac{\zeta}{2} \right)
            \sin(\zeta)
        , \\
        y_c(\zeta)
        &=
            a_2 \frac{\lambda z}{2\pi \ell}
            - \frac{\pi \ell^3}{6 \lambda z}
            + \frac{3 \lambda z}{2\pi \ell}
            \left( 
                a_3 + \frac{4\pi \ell^2}{3\lambda z}
            \right)^2
            \nonumber\\
            &\hspace{36mm}\times
            \cos^2\left(\frac{\zeta}{2} \right)
            \cos(\zeta)
            .
    \end{align}
\end{subequations}

\begin{figure}
	\includegraphics[width=0.8\linewidth,trim={29mm 5mm 11mm 4mm}, clip]{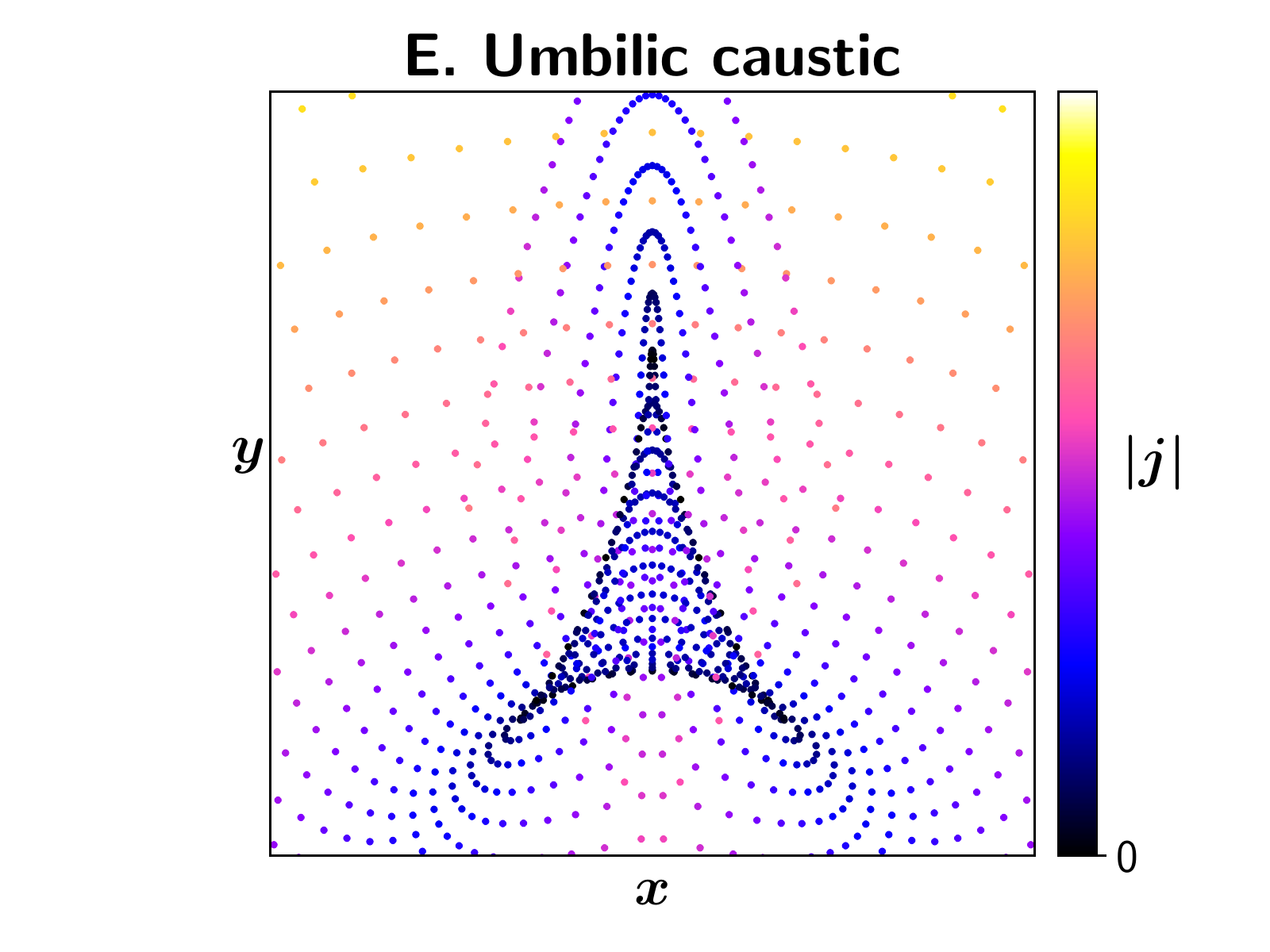}
	\caption{Same as \Fig{fig:HUmb} but for the elliptic umbilic caustic \eq{eq:rayEUMB} with $\lambda = 4\pi$, $\ell = 1$, $a_1 = 0$, $a_2 = -0.7$, and $a_3 = -2.1$.}
	\label{fig:EUmb}
\end{figure}

\noindent Figure \ref{fig:EUmb} presents an intersection plot of the ray trajectories \eq{eq:rayEUMB} that generate the $D_4^-$ caustic with respect to the plane $z = 1$. The characteristic tricorn shape of the caustic curve \eq{eq:causticEUmb} is readily observed by the visible increase in the ray density. 

As mentioned, these five caustics constitute the complete set of caustics that can occur in $3$-D systems. Hence, a popular means of modeling caustics along rays is the method of `uniform approximation'~\cite{Berry80b,Kravtsov93,Olver10a}, which consists of fitting a given wavefield to one of these normal forms. Although this method has been used effectively by previous authors, notably by \Ref{Colaitis19a} for modeling laser-plasma interactions near caustics, it suffers from the obvious disadvantage that the caustic type must be known beforehand or somehow guessed reliably. In the following section, we shall describe a different approach to dealing with caustics within ray optics. This alternate method is more general in that it does not assume any specific caustic type in advance, so it may be more convenient for practical calculations.

% ==================== %
% -- MGO THEORY -- %
% ==================== %

\section{Metaplectic geometrical optics: theory}
\label{sec:MGOtheory}

% ==================== %
% --- MT OPERATORS --- %
% ==================== %

\subsection{Metaplectic operators to `rotate' equations}
\label{sec:MT}

By definition, \Eq{eq:causticSING} is satisfied (and a caustic therefore occurs) where the ray manifold has a singular projection onto $\Vect{x}$-space. It stands to reason that removing caustics should be possible by continually rotating the ray manifold during the ray-tracing step \eq{eq:goRAYS} to maintain a good projection onto $\Vect{x}$-space at all points along a ray. This geometrical idea is the foundation for the MGO method of caustic-free ray tracing~\cite{Lopez20,Lopez21a}.

To develop this mathematically, it is necessary to introduce some new machinery. Rather than describing waves as propagating in some configuration space with coordinates $\Vect{x}$ according to a pseudo-differential equation of the form \eq{eq:waveEQ}, it is more natural to describe waves as state vectors $\ket{\psi}$ in a Hilbert space being acted upon by operators. Then, partial differential equations that govern wavefields can be understood as projections of the invariant wave equations
\begin{equation}
    \oper{D}(\VectOp{x}, \VectOp{k})\ket{\psi} = \ket{0}
    \label{eq:operGO}
\end{equation}

\noindent on a particular basis. In particular, \Eq{eq:waveEQ} is the projection of \Eq{eq:operGO} on the basis $\{\ket{\Vect{x}(\Vect{y})}\}$ formed by the orthonormal eigenvectors of the coordinate operator $\VectOp{x}$:
\begin{equation}
    \VectOp{x}\ket{\Vect{x}(\Vect{y})} = \Vect{y}\ket{\Vect{x}(\Vect{y})}
    ,\quad
    \braket{\Vect{x}(\Vect{y})}{\Vect{x}(\Vect{y}')} = 
    \delta(\Vect{y} - \Vect{y}'),
\end{equation}

\noindent and accordingly,
\begin{gather}
    \psi(\Vect{y}) = \braket{\Vect{x}(\Vect{y})}{\psi}.
\end{gather}

\noindent Likewise, the Fourier transform of \Eq{eq:waveEQ} can be understood as the projection of \Eq{eq:operGO} on the basis $\{\ket{\Vect{k}(\Vect{\kappa})}\}$ formed by the orthonormal eigenvectors of the wavevector operator $\VectOp{k}$:
\begin{equation}
    \VectOp{k}\ket{\Vect{k}(\Vect{\kappa})} = \Vect{\kappa}\ket{\Vect{k}(\Vect{\kappa})}
    ,\quad
    \braket{\Vect{k}(\Vect{\kappa})}{\Vect{k}(\Vect{\kappa}')} = 
    \delta(\Vect{\kappa} - \Vect{\kappa}')
    .
\end{equation}

\noindent As usual~\cite{Shankar94},
\begin{equation}
    \int \dd \Vect{y} \, \ket{\Vect{x}(\Vect{y})} \bra{\Vect{x}(\Vect{y})}
    =
    \int \dd \Vect{\kappa} \, \ket{\Vect{k}(\Vect{\kappa})} \bra{\Vect{k}(\Vect{\kappa})}
    = \IdentOp
    ,
    \label{eq:norm}
\end{equation}

\noindent where $\IdentOp$ is the identity operator, and also
\begin{subequations}
    \label{eq:matrixELEM}
    \begin{align}
        \bra{\Vect{x}(\Vect{y})} \VectOp{x} \ket{\Vect{x}(\Vect{y}')}
        &=
        \Vect{y}\delta(\Vect{y} - \Vect{y}')
        , \\
        \bra{\Vect{x}(\Vect{y})} \VectOp{k} \ket{\Vect{x}(\Vect{y}')}
        &= - i \pd{\Vect{y}} \delta(\Vect{y} - \Vect{y}')
        , \\
        \bra{\Vect{k}(\Vect{\kappa})} \VectOp{x} \ket{\Vect{k}(\Vect{\kappa}')}
        &=
        i \pd{\Vect{\kappa}} \delta(\Vect{\kappa} - \Vect{\kappa}')
        , \\
        \bra{\Vect{k}(\Vect{\kappa})} \VectOp{k} \ket{\Vect{k}(\Vect{\kappa}')}
        &= \Vect{\kappa}\delta(\Vect{\kappa} - \Vect{\kappa}')
        .
    \end{align}
\end{subequations}

Equations \eq{eq:matrixELEM} show that in the $\Vect{x}$ representation [\ie projection on the $\{\ket{\Vect{x}(\Vect{y})}\}$ basis], one has
\begin{equation}
    \VectOp{x} \mapsto \Vect{x}
    , \quad
    \VectOp{k} \mapsto - i\pd{\Vect{x}}
    ,
\end{equation}

\noindent while in the $\Vect{k}$ representation [\ie projection on the $\{\ket{\Vect{k}(\Vect{\kappa})}\}$ basis], one has
\begin{gather}
    \VectOp{x} \mapsto i\pd{\Vect{k}}
    , \quad
    \VectOp{k} \mapsto \Vect{k}
    .
\end{gather}

\noindent However, the $\Vect{x}$ representation and the $\Vect{k}$ representation are not the only ones possible. Instead, one can consider a different $\Vect{X}$ representation in which caustics of the wavefield vanish, at least locally. This is done as follows. 

Suppose $\Vect{X}$ is related to the original representation $\Vect{x}$ by a linear canonical transformation of the form
\begin{equation}
	\Stroke{\Vect{Z}}
	= \Mat{S} \Vect{z}
	,
	\label{eq:canonTRANS}
\end{equation}

\noindent where $\Stroke{\Vect{Z}}$ and $\Vect{z}$ are transformed and original phase-space coordinates, respectively, \ie
\begin{equation}
	\Stroke{\Vect{Z}} \doteq 
	\begin{pmatrix}
		\Vect{X} \\
		\Vect{K}
	\end{pmatrix}
	, \quad
	\Vect{z} \doteq
	\begin{pmatrix}
		\Vect{x} \\
		\Vect{k}
	\end{pmatrix}
\end{equation}

\noindent (where $\Vect{K}$ is the phase-space dual coordinate to $\Vect{X}$), and the $2N \times 2N$ transformation matrix $\Mat{S}$ is symplectic, satisfying
\begin{equation}
    \Mat{S}^\intercal \JMat{2N} \Mat{S} = \JMat{2N}
    .
    \label{eq:symplecDEF}
\end{equation}

\noindent As we shall show in \Sec{sec:ALGgs}, for MGO we can further impose that $\Mat{S}$ be orthogonal in addition to being symplectic (orthosymplectic). This means that $\Mat{S}$ also satisfies
\begin{equation}
    \Mat{S}^\intercal = \Mat{S}^{-1}
    .
    \label{eq:orthoDEF}
\end{equation}

\noindent Together, \Eqs{eq:symplecDEF} and \eq{eq:orthoDEF} imply that $\Mat{S}$ can be represented by block decomposition~\cite{Littlejohn86a}
\begin{equation}
    \Mat{S}
    =
    \begin{pmatrix}
		\Mat{A} & \Mat{B} \\
		- \Mat{B} & \Mat{A}
	\end{pmatrix}
	,
\end{equation}

\noindent where the $N \times N$ submatrices $\Mat{A}$ and $\Mat{B}$ satisfy
\begin{subequations}%
    \begin{align}%
       \Mat{A} \Mat{A}^\intercal + \Mat{B} \Mat{B}^\intercal &= \IMat{N} 
       , \\
       \Mat{A}^\intercal \Mat{A} + \Mat{B}^\intercal \Mat{B} &= \IMat{N}
       , \\
       \Mat{A} \Mat{B}^\intercal - \Mat{B} \Mat{A}^\intercal &= \OMat{N}
       , \\
       \Mat{B}^\intercal \Mat{A} - \Mat{A}^\intercal \Mat{B} &= \OMat{N}
       .
    \end{align}%
\end{subequations}%

Linear symplectic phase-space transformations are special because they have an exact operator analogue in a Hilbert space; these are the unitary metaplectic operators~\cite{Moshinsky71,Littlejohn86a}, denoted as $\oper{M}(\Mat{S})$. Specifically, the corresponding operators $\VectOp{z}$ and $\VectOp{Z} \doteq (\VectOp{X}, \VectOp{K})^\intercal$ are related as 
\begin{equation}
	\Stroke{\VectOp{Z}} = 
	\oper{M}^\dagger(\Mat{S}) \VectOp{z} \oper{M}(\Mat{S})
	\equiv
	\Mat{S} \VectOp{z}
	,
	\label{eq:operTRANS}
\end{equation}

\noindent or equivalently in terms of the constituent operators
\footnote{Here and in the following, we assume that $\Mat{S}$ is orthosymplectic; however, we note that metaplectic operators only require $\Mat{S}$ be symplectic, regardless of orthogonality.},%
\begin{subequations}
    \begin{align}
        \VectOp{X} &= \oper{M}^\dagger(\Mat{S}) \VectOp{x} \oper{M}(\Mat{S})
	    \equiv
    	\Mat{A} \VectOp{x} + \Mat{B} \VectOp{k}
	    , \\
    	\VectOp{K} &= \oper{M}^\dagger(\Mat{S}) \VectOp{k} \oper{M}(\Mat{S})
	    \equiv
    	- \Mat{B} \VectOp{x} + \Mat{A} \VectOp{k}
	    ,
    \end{align}
\end{subequations}

\noindent Also, $\oper{M}$ directly generates the basis transformation as
\begin{equation}
    \ket{\Vect{X}(\Vect{y})} = \oper{M}^\dagger(\Mat{S}) \ket{\Vect{x}(\Vect{y})}
    .
\end{equation}

The matrix elements of $\oper{M}$ in the $\Vect{x}$ representation, which govern the overlap between the $\Vect{x}$ and $\Vect{X}$ representations, are given for invertible $\Mat{B}$ as~\cite{Moshinsky71,Littlejohn86a}
\begin{align}
    \braket{\Vect{X}(\Vect{Y})}{\Vect{x}(\Vect{y})}
    &= \bra{\Vect{x}(\Vect{Y})} \oper{M}(\Mat{S}) \ket{\Vect{x}(\Vect{y})}
    \nonumber\\
    &= 
    \frac{ 
        \sigma 
        \exp[i G(\Vect{y}, \Vect{Y}) ] 
    }{ 
        (2 \pi i)^{N/2} \sqrt{\det \Mat{B}} 
    }
    ,
    \label{eq:MTkernel}
\end{align}

\noindent where $G(\Vect{y}, \Vect{Y})$ is the quadratic generator function
\begin{align}
    \hspace{-1mm}G(\Vect{y}, \Vect{Y})
    &=
    \frac{1}{2} \Vect{y}^\intercal \Mat{B}^{-1} \Mat{A} \Vect{y}
    - \Vect{y}^\intercal \Mat{B}^{-1} \Vect{Y}
    + \frac{1}{2} \Vect{Y}^\intercal \Mat{A} \Mat{B}^{-1} \Vect{Y}
    ,
    \label{eq:gener1}
\end{align}

\noindent and $\sigma \doteq \pm 1$. The sign ambiguity in $\sigma$ is of fundamental importance, as it contributes to the well-known phase shifts that a wavefield acquires upon touching a caustic~\cite{Littlejohn86a,Lopez19,Berry72,Littlejohn87,Ciobanu21}, the $\pi/2$ phase shift following reflection being a famous example. Note that when $\det{\Mat{B}} = 0$, the right-hand side of \Eq{eq:MTkernel} becomes a delta function over the singular subspace of $\Mat{B}$~\cite{Moshinsky71,Littlejohn86a}.

Accordingly, wavefunctions (\ie \textit{projections} of state vectors $\ket{\psi}$, not the invariant state vectors themselves) are transformed via \Eq{eq:norm} as
\begin{align}
    \hspace{-1mm}\fourier{\psi}(\Vect{Y}) \doteq
    \braket{\Vect{X}(\Vect{Y})}{\psi}
    &=
    \int \dd \Vect{y} \braket{\Vect{X}(\Vect{Y})}{\Vect{x}(\Vect{y})} \braket{\Vect{x}(\Vect{y})}{\psi}
    \nonumber\\
    &=
    \int \dd \Vect{y} \,
    \frac{
        \sigma 
        \exp[i G(\Vect{y}, \Vect{Y}) ]
    }{
        (2 \pi i)^{N/2} \sqrt{\det \Mat{B}}
    } \, \psi(\Vect{y})
    .
    \label{eq:MT}
\end{align}

\noindent Equation \eq{eq:MT} is called the metaplectic transform (MT). Special cases include the $N$-D Fourier transform ($\Mat{A} = \OMat{N}$, $\Mat{B} = \IMat{N}$) and the Fresnel transform \eq{eq:freeMT} (for which $\Mat{S}$ is not orthogonal, but is instead given by the ray transfer matrix for propagation in uniform media~\cite{Kogelnik66}). The inverse MT is given similarly as
\begin{equation}
    \psi(\Vect{y}) =
    \int \dd \Vect{Y} \, 
    \frac{
        \sigma 
        \exp[- i G(\Vect{y}, \Vect{Y}) ]
    }{
        (-2 \pi i)^{N/2} \sqrt{\det \Mat{B}}
    } \, \fourier{\psi}(\Vect{Y})
    .
\end{equation}

\noindent Let us now discuss how these tools enable the development of a caustic-free GO formalism called metaplectic geometrical optics.

% ==================== %
% --- MGO FORMULAS --- %
% ==================== %

\subsection{MGO formulas for scalar wavefields}
\label{sec:MGOrev}

Suppose that $\ket{\psi}$ exhibits a caustic in the $\Vect{x}$ representation but has an eikonal form in the $\Vect{X}$ representation:
\begin{equation}
    \fourier{\psi}(\Vect{Y})
    =
    \fourier{\env}(\Vect{Y}) \exp[i \Theta(\Vect{Y})]
    ,
    \label{eq:eikX}
\end{equation}

\noindent where $\Theta$ is a rapidly varying and $\fourier{\env}$ is a slowly varying function of $\Vect{Y}$, analogous to \Eq{eq:wave}. In the invariant form,  \Eq{eq:eikX} can be written as
\begin{equation}
    \ket{\psi} = \oper{U}(\VectOp{X})\ket{\phi}
    , \quad
    \oper{U}(\VectOp{X})
    \doteq \exp[i \Theta(\VectOp{X})]
    .
\end{equation}

\noindent The unitary operator $\oper{U}$ can be considered as a transformation connecting the complete wavefield $\fourier{\psi}$ with its envelope $\fourier{\env}$, since
\begin{align}
    \bra{\Vect{X}(\Vect{Y})} \oper{U}^\dagger(\VectOp{X}) \ket{\psi}
    &=
    \exp[- i \Theta(\Vect{Y})]
    \fourier{\env}(\Vect{Y}) \exp[i \Theta(\Vect{Y})]
    \nonumber\\
    &\equiv \fourier{\env}(\Vect{Y})
    .
\end{align}

\noindent Then, the wave equation \eq{eq:waveEQ} can be written as an envelope equation
\begin{equation}
    \oper{U}^\dagger(\VectOp{X}) \oper{D}(\Mat{S}^{-1} \Stroke{\VectOp{Z}}) \oper{U}(\VectOp{X}) \ket{\phi} = \ket{0}
    ,
    \label{eq:operMGO}
\end{equation}

\noindent where we used \Eq{eq:operTRANS} for $\VectOp{z}$.

Following \Refs{Dodin19,Lopez20}, \Eq{eq:operMGO} can be approximated in the GO limit by use of the Wigner--Weyl transform and the Moyal product (\App{sec:APPwwt}), much like they are used for deriving traditional GO. Indeed, the GO-approximated envelope operator is obtained by first calculating the appropriate symbol, then approximating the symbol in the GO limit using familiar Taylor expansions, and finally mapping the symbol back to obtain the correspondingly approximated operator. Mathematically, this reads
\begin{align}
    &\oper{U}^\dagger(\VectOp{X}) \oper{D}(\Mat{S}^{-1} \Stroke{\VectOp{Z}}) \oper{U}(\VectOp{X})
    \nonumber\\
    &\hspace{0mm}= \WeylInv\left\{ 
        \Weyl\left[ 
            \oper{U}^\dagger(\VectOp{X}) \oper{D}(\Mat{S}^{-1} \Stroke{\VectOp{Z}}) \oper{U}(\VectOp{X})
        \right]
    \right\}
    \nonumber\\
    &\hspace{0mm}=
    \WeylInv\left[
            \Symb{U}^*(\Vect{X}) \star \Symb{D}(\Mat{S}^{-1} \Stroke{\Vect{Z}}) \star \Symb{U}(\Vect{X})
    \right]
    \nonumber\\
    &\hspace{0mm}\approx
    \WeylInv\left\{
            \Symb{D}[\Mat{S}^{-1} \Stroke{\Vect{Z}}(\Vect{X})] 
            + \Vect{K}^\intercal \Vect{V}(\Vect{X})
            + \ldots
    \right\}
    \nonumber\\
    &\hspace{0mm}= \Symb{D}[\Mat{S}^{-1} \Stroke{\Vect{Z}}(\VectOp{X})] 
        + \Vect{V}(\VectOp{X})^\intercal \VectOp{K}
        - \frac{i}{2} \nabla \cdot \Vect{V}(\VectOp{X})
        + \ldots
    ,
    \label{eq:approxOPER}
\end{align}

\noindent where we have introduced
\begin{subequations}
    \begin{align}
        \Stroke{\Vect{Z}}(\Vect{X}) &\doteq 
        \begin{pmatrix}
            \Vect{X} &
            \pd{\Vect{X}} \Theta(\Vect{X})
        \end{pmatrix}^\intercal
        , \\
        \Vect{V}(\Vect{X}) &\doteq
        \pd{\Vect{K}} \Symb{D}\left[ \Mat{S}^{-1} \Stroke{\Vect{Z}}(\Vect{X}) \right]
        .
    \end{align}
\end{subequations}

\noindent (Only first-order terms are included in the expansion \eq{eq:approxOPER} for simplicity; higher-order terms to enable beam-tracing of arbitrarily structured light~\cite{Yanagihara19a,Yanagihara19b,Yanagihara21a,Yanagihara21b} can be readily included within this framework~\cite{Lopez20}.) Hence, the GO-approximated equation that governs $\fourier{\env}$ is obtained by projecting \Eq{eq:approxOPER} onto the $\ket{\Vect{X}(\Vect{Y})}$ basis: 
\begin{align}
    \left\{
        \Symb{D}[\Mat{S}^{-1} \Stroke{\Vect{Z}}(\Vect{Y})] 
        - i \Vect{V}(\Vect{Y})^\intercal \pd{\Vect{Y}}
        - \frac{i}{2} \left[\nabla \cdot \Vect{V}(\Vect{Y})\right]
    \right\}
    \fourier{\env}(\Vect{Y})
    = 0
    .
    \label{eq:operMGOapprox}
\end{align}

Analogous to \Eqs{eq:goDISP} and \eq{eq:goENV}, \Eq{eq:operMGOapprox} is solved in two parts. First, the rotated dispersion relation,
\begin{equation}
    \Symb{D}[\Mat{S}^{-1} \Stroke{\Vect{Z}}(\Vect{X})]
    = 0
    ,
    \label{eq:mgoDISP}
\end{equation}

\noindent is solved (via ray-tracing) to obtain $\Theta$. Then, the rotated envelope equation,
\begin{equation}
    2 \Vect{V}(\Vect{X})^\intercal \pd{\Vect{X}} \fourier{\env}(\Vect{X}) + \left[\nabla \cdot \Vect{V}(\Vect{X})\right] \fourier{\env}(\Vect{X}) = 0
    ,
    \label{eq:mgoENV}
\end{equation}

\noindent is solved to obtain $\fourier{\env}$. Having the same formal structure as \Eqs{eq:goDISP} and \eq{eq:goENV}, the solutions to \Eqs{eq:mgoDISP} and \eq{eq:mgoENV} can be inferred from the known solutions $\Vect{z}(\Vect{\tau})$ and $\env(\Vect{\tau})$ for the unrotated equations: \Eq{eq:mgoDISP} is solved by the rotated rays (equivalently, by the rotated ray manifold)
\begin{equation}
    \Stroke{\Vect{Z}}(\Vect{\tau})
    = \Mat{S} \Vect{z}(\Vect{\tau})
    ,
    \label{eq:mgoRAYsol}
\end{equation}

\noindent and \Eq{eq:mgoENV} is solved along the rotated rays as
\begin{equation}
    \fourier{\env}(\Vect{\tau})
    =
    \fourier{\env}_0(\Vect{\tau}_\perp) 
    \sqrt{
        \frac{J_0(\Vect{\tau}_\perp)}{J(\Vect{\tau})}
    }
    ,
    \label{eq:mgoENVsol}
\end{equation}

\noindent where the Jacobian
\begin{equation}
    J(\Vect{\tau}) \doteq 
    \det \pd{\Vect{\tau}} \Vect{X}(\Vect{\tau})
    \label{eq:rotJAC}
\end{equation}

\noindent governs the projection properties of the ray manifold onto the rotated $\Vect{X}$-space rather than $\Vect{x}$-space. 

Equation \eq{eq:mgoENVsol} implies that a caustic at some position $\Vect{\tau} = \Vect{t}$ on the ray manifold can be avoided by choosing $\Vect{X}$ to be the tangent plane at $\Vect{t}$, since $J(\Vect{t}) \neq 0$ is then guaranteed by definition. This is the logic of the MGO framework, which is summarized in three steps. First, the rotated GO equations \eq{eq:mgoDISP} and \eq{eq:mgoENV} are solved in the tangent plane at a given ray position $\Vect{t}$. Next, the obtained solution $\fourier{\psi}(\Vect{X})$ is rotated into the following tangent plane at $\Vect{t} + \delta \Vect{t}$ using an infinitesimal near-identity MT (NIMT) to provide initial conditions for the corresponding next set of rotated GO equations to be solved. This step improves the continuity of the global solution. In parallel, $\fourier{\psi}(\Vect{X})$ is mapped back to the original $\Vect{x}$ coordinates using an inverse MT. This is repeated for all points on the ray manifold, resulting in an approximate solution of the form
\begin{equation}
    \psi(\Vect{x})
    =
    \sum_{\Vect{t} \in \Vect{\tau}(\Vect{x})} 
    \MTnorm(\Vect{x}) \,
    \Upsilon_\Vect{t}(\Vect{x})
    ,
    \label{eq:MGO}
\end{equation}

\noindent where the sum is taken over all ray contributions that arrive at a given point $\Vect{x}$, similar to \Eq{eq:GO}.

The analysis to obtain specific expressions for $\MTnorm(\Vect{x})$ and $\Upsilon_\Vect{t}(\Vect{x})$ is quite involved; let us therefore simply quote the final results when $\det \Mat{B} \neq 0$~\cite{Lopez20}:
\begin{align}%
    \label{eq:MTpre}
    \MTnorm(\Vect{x})
    &=
    \frac{
        \sigma_\Vect{t} \, \alpha_\Vect{t} 
        \exp
        \left\{ 
            - i G_\Vect{t}[\Vect{x}, \Vect{X}_\Vect{t}(\Vect{t})]
            \nullFrac
        \right\}
    }
    {
        (-2\pi i)^{N/2} \sqrt{ \det{\Mat{B}_\Vect{t}}}
    }
    , \\
    \label{eq:upsilon}
    \Upsilon_\Vect{t}(\Vect{x})
    &=
        \int_{\cont{0}} \dd \Vect{\epsilon} \,
        \fourier{\psi}\left[ 
            \Vect{\epsilon} + \Vect{X}_\Vect{t}(\Vect{t})
        \right] 
        \exp\left[ 
            - i \gamma_\Vect{t}(\Vect{\epsilon}, \Vect{x})
        \right]
    ,
\end{align}

\noindent where we have introduced $\Mat{S}_\Vect{t}$ as the matrix that rotates $\Vect{x}$ to the tangent plane at $\Vect{t}$, $\sigma_\Vect{t}$ as an overall sign such that $\sigma_\Vect{t}/\sqrt{\det \Mat{B}_\Vect{t}}$ is continuous across any branch cuts, and $\cont{0}$ as the steepest-descent contour that passes through the origin $\Vect{\epsilon} = \Vect{0}$. The functions $G_\Vect{t}$ and $\Vect{X}_\Vect{t}$ are determined by \Eqs{eq:gener1} and \eq{eq:mgoRAYsol} for $\Mat{S}_\Vect{t}$, respectively. We have also defined
\begin{align}
    \gamma_\Vect{t}(\Vect{\epsilon}, \Vect{x}) 
    &\doteq 
        \frac{1}{2} \Vect{\epsilon}^\intercal \Mat{A}_\Vect{t} \Mat{B}_\Vect{t}^{-1} \Vect{\epsilon}
        +
        \Vect{\epsilon}^\intercal
        \Mat{B}_\Vect{t}^{-\intercal}
        \left[   
            \Mat{A}_\Vect{t}^\intercal \Vect{X}_\Vect{t}(\Vect{t})
            -
            \Vect{x}
        \right]
    ,
\end{align}

\noindent along with the continuity factor $\alpha_\Vect{t}$ that can be evolved along a ray via the NIMT as
\begin{equation}
    \alpha_{\Vect{t}+\Vect{h}} = 
    \alpha_\Vect{t}
    \left.\NIMT
    {
        \Mat{S}_{\Vect{t}+\Vect{h}} \Mat{S}^{-1}_\Vect{t}
    }\left[ 
        \fourier{\psi}(\Vect{X})
    \right]\right|_{\Vect{X} = \Vect{X}_{\Vect{t}+\Vect{h}}( \Vect{t}+\Vect{h} )}
    ,
    \label{eq:alphaEQ}
\end{equation}

\noindent where $\NIMT{\Mat{S}}$ denotes the NIMT corresponding to a given near-identity symplectic transformation $\Mat{S}$. Equivalently, as shown in \App{app:deriv}, $\alpha_\Vect{t}$ can be computed as
\begin{align}
    \label{eq:alphaINT}
    \alpha_\Vect{t} 
    &= 
    f(\Vect{t}_\perp)
    \exp
    \left[
        \frac{i}{2} \Vect{K}^\intercal_\Vect{t}(\Vect{t}) \Vect{X}_\Vect{t}(\Vect{t})
        + \int_0^{t_1} \dd \xi \, \eta_{(\xi, \Vect{t}_\perp)}
    \right]
    , \\
    \eta_\Vect{t}
    &\doteq
    - \frac{i}{2} \Vect{z}^\intercal(\Vect{t})
	\,
	\JMat{2N} 
	\,
	\dot{\Vect{z}}(\Vect{t})
	%%%
	\nonumber\\
	&\hspace{4mm}
	+ \left[
		\Mat{A}_\Vect{t} \dot{\Vect{x}}(\Vect{t})
		+ \Mat{B}_\Vect{t} \dot{\Vect{k}}(\Vect{t})
	\right]^\intercal
	\pd{\Vect{X}} \fourier{\phi}\left[
		\Vect{X}_\Vect{t}(\Vect{t})
	\right]
	,
	\label{eq:etaDEF}
\end{align}

\noindent where $f(\Vect{t}_\perp)$ is set by initial conditions and the dot $\cdot$ denotes $\pd{t_1}$, \ie the directional derivative along a ray. Accordingly, \Eq{eq:MTpre} can be shown to take the form
\begin{equation}
    \MTnorm(\Vect{x})
    =
    \frac{
        \sigma_\Vect{t} \, f(\Vect{t}_\perp) 
        \exp
        \left[ 
            \frac{i}{2} \Vect{x}^\intercal \Vect{k}(\Vect{t})
            + \int_0^{t_1} \dd \xi \, \eta_{(\xi, \Vect{t}_\perp)}
        \right]
    }
    {
        (-2\pi i)^{N/2} \sqrt{ \det{\Mat{B}_\Vect{t}}}
    }
    .
    \label{eq:MTpreSIMPLE}
\end{equation}

Equations \eq{eq:alphaINT}--\eq{eq:MTpreSIMPLE} are considerably simpler than the expressions obtained previously in \Refs{Lopez20, Lopez21a}; this is largely because the orthosymplecticity of $\Mat{S}$ implies that $\dot{\Mat{S}} \Mat{S}^{-1}$ is an \textit{antisymmetric} Hamiltonian matrix, rather than simply a Hamiltonian matrix when $\Mat{S}$ is merely symplectic.

Note that $\fourier{\psi}(\Vect{X})$ should be normalized identically for all tangent planes, since any discrepancy in the normalization is already accounted for by $\alpha_\Vect{t}$ [specifically by the final term in \Eq{eq:etaDEF}]. We also note that the final simplification \eq{eq:MTpreSIMPLE} holds even when $\Mat{B}$ is not invertible. For MGO formulas when $\det \Mat{B} = 0$ or an alternate MGO formulation that is insensitive to $\det \Mat{B}$, see \App{sec:APPsingular} and \App{sec:APPcoherent}, respectively.

The steepest-descent integration of \Eq{eq:upsilon} ensures that only points around $\Vect{t}$ on the ray manifold contribute to the MT integral, since only these ray contributions are guaranteed to be caustic-free by our construction. Such `saddlepoint filters' tend to differ between otherwise similar phase-space rotation schemes; for example, \Ref{Littlejohn85} uses phase-space translation operators and a delta-shaped envelope model (\Sec{sec:MGOcompare}). Our formalism is flexible enough to allow modeling a variety of caustics in a variety of wave systems; to date, MGO has been used to model a caustic-free plane wave, an isolated fold caustic, a simple fold-caustic network, and an isolated cusp caustic in unidirectional, Helmholtz-like, and paraxial wave equations~\cite{Lopez20,Lopez21a,Donnelly21}. However, modeling the higher-order caustics discussed in \Sec{sec:paraxial} requires a numerical implementation of MGO, which we shall discuss in \Sec{sec:MGOalgor}.

% ==================== %
% -- MGO TO GO -- %
% ==================== %
\subsection{Reducing MGO to standard GO away from caustics}
\label{sec:MGOtoGO}

In \Refs{Lopez20, Lopez21a}, it was shown numerically in a series of examples that the MGO formula \eq{eq:MGO} remains finite at caustics and agrees with the standard GO formula \eq{eq:GO} away from caustics. The fact that \eq{eq:MGO} remains finite at caustics follows immediately from the observation that all terms in \Eqs{eq:MGO}--\eq{eq:MTpreSIMPLE} remain finite at caustics by construction; however, the fact that MGO reproduces GO away from caustics is less immediately obvious. In this section, we shall prove this property explicitly.

Let us assume that $\Mat{B}_\Vect{t}$ is invertible for simplicity. Assuming that $\Vect{x}$ is located far from a caustic, $\Upsilon_\Vect{t}(\Vect{x})$ can be evaluated by the standard (quadratic) saddlepoint method as follows:
\begin{align}
    \Upsilon_\Vect{t}(\Vect{x})
	&\approx
	\int_{\cont{0}} \dd \Vect{\epsilon} \,
	\exp\left\{ 
		i \Vect{\epsilon}^\intercal \Vect{K}_\Vect{t}(\Vect{t})
		- i \gamma_\Vect{t}(\Vect{\epsilon}, \Vect{x})
	\right\}
	%%%%%%%%
	\nonumber\\
	&=
	\int_{\cont{0}} \dd \Vect{\epsilon} \,
	\exp\left\{ 
		- \frac{i}{2} \Vect{\epsilon}^\intercal \Mat{A}_\Vect{t} \Mat{B}_\Vect{t}^{-1} \Vect{\epsilon}
	\right\}
	%%%%%%%
	\nonumber\\
	&=
	\frac{(-2 \pi i)^{N/2} \sqrt{\det \Mat{B}_\Vect{t} }}{ \sqrt{\det{\Mat{A}_\Vect{t}}} }
	,
	\label{eq:UpsilonSADDLE}
\end{align}

\noindent where we have used the consistent normalization $\fourier{\psi}\left[  \Vect{X}_\Vect{t}(\Vect{t})\right] = 1$ [see discussion following \Eq{eq:MTpreSIMPLE}], along with the fact that the second derivatives of the phase vanish when evaluated at the ray position $\Vect{X}_\Vect{t}(\Vect{t})$, \ie 
\begin{equation}
    \pd{\Vect{X}} \Vect{K}_\Vect{t}\left[ \Vect{X}_\Vect{t}(\Vect{t}) \right] = \OMat{N}
    ,
    \label{eq:tangent}
\end{equation}

\noindent which is true by definition of the tangent plane to a Lagrangian manifold as shown in \App{app:tangent}. [Note that any overall sign in \Eq{eq:UpsilonSADDLE} that results from branch cuts can be absorbed into the overall sign $\sigma_\Vect{t}$.]

Next, we must integrate $\eta_\Vect{t}$ given by \Eq{eq:etaDEF}. To do so, note that \Eq{eq:mgoENV} implies
\begin{align}
    &\left[
		\Mat{A}_\Vect{t} \dot{\Vect{x}}(\Vect{t})
		+ \Mat{B}_\Vect{t} \dot{\Vect{k}}(\Vect{t})
	\right]^\intercal 
	\pd{\Vect{X}} \fourier{\phi}_\Vect{t}\left[
		\Vect{X}_\Vect{t}(\Vect{t})
	\right]
	\nonumber\\
	&\hspace{35mm}=
	\left.
		- \frac{1}{2} \pd{\tau_1} \left[ \log J_\Vect{t}(\Vect{\tau}) \right]
	\right|_{\Vect{\tau} = \Vect{t}}
	,
\end{align}

\noindent where $J_\Vect{t}(\Vect{\tau})$ is given by \Eq{eq:rotJAC} for $\Vect{X}_\Vect{t}$. To evaluate this expression further, recall that $\Mat{S}_\Vect{t}$ is the orthosymplectic matrix that maps $\Vect{x}$ to the tangent plane of the ray manifold at $\Vect{t}$. Since the tangent plane at $\Vect{t}$ is spanned by vectors of the form $\{ \pd{\tau_j} \Vect{z}(\Vect{t}) \}$ for $j = 1, \ldots , N$, an orthogonal basis can be obtained via the QR decomposition
\begin{equation}
    \pd{\Vect{\tau}} \Vect{z}(\Vect{t})
    \doteq 
    \begin{pmatrix}
		\uparrow  & & \uparrow\\
		\pd{\tau_1} \Vect{z}(\Vect{t}) & \ldots & \pd{\tau_N} \Vect{z}(\Vect{t}) \\
		\downarrow & & \downarrow \\
	\end{pmatrix}
	= \Mat{Q}_\Vect{t} \Mat{R}_\Vect{t}
	,
	\label{eq:zQR}
\end{equation}

\noindent where the arrows emphasize that the constituent vectors are column vectors%
%%%
\footnote{Equation \eq{eq:zQR} implies the convention $(\pd{\Vect{\tau}} \Vect{z})_{ij} \equiv \pd{\tau_j} z_i$; a similar convention will also be assumed for gradients of other matrices.}. %
%%%
Here, $\Mat{R}_\Vect{t}$ is an upper triangular matrix of size $N \times N$, while $\Mat{Q}_\Vect{t}$ is an orthogonal rectangular matrix of size $2N \times N$:
\begin{equation}
    \Mat{Q}_\Vect{t}^\intercal \Mat{Q}_\Vect{t}
    = \IMat{N}
    ,
\end{equation}

\noindent which we can then relate to $\Mat{S}_\Vect{t}$ as
\begin{equation}
    \Mat{Q}_\Vect{t}
    =
    \begin{pmatrix}
		\Mat{A}_\Vect{t}^\intercal \\ 
		\Mat{B}_\Vect{t}^\intercal
	\end{pmatrix}
	.
	\label{eq:QRab}
\end{equation}

\noindent Hence, the rotated ray Jacobian matrix can be compactly expressed as
\begin{equation}
    \pd{\Vect{\tau}} \Vect{X}_\Vect{t}(\Vect{\tau})
    \doteq 
    \Mat{A}_\Vect{t} \pd{\Vect{\tau}} \Vect{x}(\Vect{\tau})
    + \Mat{B}_\Vect{t} \pd{\Vect{\tau}} \Vect{k}(\Vect{\tau})
    = \Mat{Q}_\Vect{t}^\intercal
	\Mat{Q}_\Vect{\tau} \Mat{R}_\Vect{\tau}
	.
	\label{eq:pdX}
\end{equation}

\noindent Using Jacobi's formula for the derivative of the matrix determinant, we can therefore perform the simplification
\begin{align}
    \left.
		\pd{\tau_1} \left[ \log J_\Vect{t}(\Vect{\tau}) \right]
	\right|_{\Vect{\tau} = \Vect{t}}
	&=
	\left.
		\Tr\left\{
			\left[ \Mat{Q}_\Vect{t}^\intercal \Mat{Q}_\Vect{\tau} \Mat{R}_\Vect{\tau}\right]^{-1}
			\pd{\tau_1} \left[\Mat{Q}_\Vect{t}^\intercal \Mat{Q}_\Vect{\tau} \Mat{R}_\Vect{\tau} \right]
		\right\}
	\right|_{\Vect{\tau} = \Vect{t}}
	%%%
	\nonumber\\
	&=
	\Tr\left(
		\Mat{R}_\Vect{t}^{-1} \Mat{Q}_\Vect{t}^\intercal \dot{\Mat{Q}}_\Vect{t} \Mat{R}_\Vect{t}
	\right)
	+
	\Tr\left(
		\Mat{R}_\Vect{t}^{-1} \dot{\Mat{R}}_\Vect{t}
	\right)
	%%%
	\nonumber\\
	&=
	\pd{t_1} \left[ \log \det \Mat{R}_\Vect{t} \right]
	,
\end{align}

\noindent where we have used the fact that $\Mat{Q}_\Vect{t}^\intercal \dot{\Mat{Q}}_\Vect{t}$ is antisymmetric and thereby traceless, and we have invoked the Jacobi formula again in the final line. The final expression is a total derivative along a ray, and hence it can be trivially integrated. Indeed, following an integration by parts, we can now compute 
\begin{align}
    &f(\Vect{t}_\perp) 
    \exp
    \left[ 
        \frac{i}{2} \Vect{x}^\intercal \Vect{k}(\Vect{t})
        + \int_0^{t_1} \dd \xi \, \eta_{(\xi, \Vect{t}_\perp)}
    \right]
    \nonumber\\
    &\hspace{5mm}=
    \frac{
        g(\Vect{t}_\perp)
    }{\sqrt{ \det \Mat{R}_\Vect{t}} }
    \exp
	\left[ 
		i \int_0^{t_1} \dd \xi \, \Vect{k}^\intercal(\xi, \Vect{t}_\perp) \dot{\Vect{x}}(\xi, \Vect{t}_\perp)
	\right]
	,
\end{align}

\noindent where $g$ is determined by the initial conditions and incorporates the boundary term from integrating by parts, specifically, $g(\Vect{t}_\perp) \doteq f(\textbf{t}_\perp) \exp\left[ \frac{i}{2} \textbf{x}(0, \textbf{t}_\perp)^\intercal \textbf{k}(0, \textbf{t}_\perp) \right]$.

Lastly, note that \Eq{eq:zQR} implies the relation
\begin{equation}
    \pd{\Vect{\tau}} \Vect{x}(\Vect{t}) = \Mat{A}_\Vect{t}^\intercal \Mat{R}_\Vect{t}
    .
\end{equation}

\noindent Hence, upon noting that
\begin{equation}
    \det{\Mat{A}_\Vect{t}} \det \Mat{R}_\Vect{t}
	= \det{\Mat{A}_\Vect{t}^\intercal } \det \Mat{R}_\Vect{t}
	= \det{ \pd{\Vect{t}} \Vect{x}(\Vect{t}) }
	\doteq j(\Vect{t})
	,
\end{equation}

\noindent we can simplify the MGO formula \eq{eq:MGO} as
\begin{equation}
    \psi(\Vect{x})
	=
	\sum_{\Vect{t} \in \Vect{\tau}(\Vect{x})} 
	\frac{
		g(\Vect{t}_\perp) 
		\exp
		\left(
			i \int \, \Vect{k}^\intercal \dd \Vect{x}
		\right)
	}
	{
		\sqrt{ j(\Vect{t}) }
	}
	.
	\label{eq:MGOfinal}
\end{equation}

\noindent Note that the arbitrary function $g(\Vect{t}_\perp) $ encodes the initial conditions along a ray. This means that \Eq{eq:MGOfinal} is equivalent to the standard GO formula \eq{eq:GO}, thus concluding our proof that MGO reduces to GO away from caustics.

% ==================== %
% -- MGO COMPARISON -- %
% ==================== %

\subsection{Relation between MGO and other semiclassical integral expressions}
\label{sec:MGOcompare}

Let us now relate the MGO method we have just outlined [\Eqs{eq:MGO}-\eq{eq:etaDEF}] with the related caustic-removal scheme presented in \Ref{Littlejohn85}, which shares the same underlying idea with MGO of using continual phase-space rotations as the rays propagate, and has also been previously related to semiclassical methods based on wavepackets~\cite{Littlejohn86b, Kay94a, Zor96, Madhusoodanan98}. There, the following $1$-D expression for $\psi$ was obtained [Eqs.~(5) and (7) in \Ref{Littlejohn85}]:
\begin{widetext}
\begin{align}
    \psi(x)
    &= c 
    \int \dd \tau \,
    \exp\left[
		\frac{i}{2} K_\tau(\tau) X_\tau(\tau)
        - \frac{i}{2}
		\int_0^\tau \dd \xi \,
		\Vect{z}(\xi)^\intercal \, \JMat{2} \, \dot{\Vect{z}}(\xi)
	\right]
	\int \dd X \, 
	\frac{
		\delta\left[ X - X_\tau(\tau) \right]
	}{
		\sqrt{-2 \pi i B_\tau}
	}
	\exp\left[
		- i G_\tau(x, X)
	\right]
	\nonumber\\
	%%%%%
	&=
	c \int \frac{  \dd \tau }{\sqrt{-2 \pi i B_\tau}} \,
    \exp\left\{
	    \frac{i}{2} k(\tau) x
		+ \frac{i}{2} k(\tau) \left[x - x(\tau) \right]
		- \frac{iA_\tau}{2B_\tau} [x - x(\tau)]^2
		- \frac{i}{2}
		\int_0^\tau \dd \tau' \,
		\Vect{z}(\xi)^\intercal \JMat{2} \dot{\Vect{z}}(\xi)
	\right\}
	,
	\label{eq:LittlejohnEQ}
\end{align}
%\end{widetext}

\noindent where $c$ is a constant. (Note that we have replaced the original notation with our notation.) Using the well-known delta-function manipulations, we can similarly express the MGO solution \eq{eq:MGO} by the integral
\begin{align}
    \psi(x)
    =
    \int \dd t \,
    \delta \left[ t - \tau(x) \right]
    \exp
    \left[
        \frac{i}{2} K_t(t) X_t(t)
        - \frac{i}{2}
		\int_0^t \dd \xi \,
		\Vect{z}(\xi)^\intercal \JMat{2} \dot{\Vect{z}}(\xi)
    \right]
    \int_{\cont{0}} \dd \epsilon \,
    \frac{
        \fourier{\alpha}_t \fourier{\psi}\left[ 
            \epsilon + X_t(t)
        \right]
    }{
        \sqrt{ - 2 \pi i B_t}
    }
    \exp\left\{
        - i G_t[x, \epsilon + X_t(t)]
        \nullFrac
    \right\}
    ,
    \label{eq:MGOlittlejohn}
\end{align}
%\end{widetext}

\noindent where $\fourier{\alpha}_t \doteq \alpha_t \exp\left[\frac{i}{2}\int_0^t \dd \xi \, \Vect{z}(\xi)^\intercal \JMat{2} \dot{\Vect{z}}(\xi) - \frac{i}{2} K_t(t) X_t(t) \right]$ is the envelope continuity factor [see \Eqs{eq:alphaINT} and \eq{eq:etaDEF}]. By comparing \Eq{eq:MGOlittlejohn} with the top line of \Eq{eq:LittlejohnEQ}, one finds that choosing
\begin{equation}
    \fourier{\alpha}_t \fourier{\psi}\left[\epsilon + X_t(t)\right] = c \, \delta(\epsilon)
    \label{eq:LittlejohnPSI}
\end{equation}

\noindent in \Eq{eq:MGOlittlejohn} yields an expression similar to \Eq{eq:LittlejohnEQ}:
\begin{equation}
    \psi(x)
    =
    c \int \dd t \, \frac{ \delta \left[ t - \tau(x) \right]}{\sqrt{-2 \pi i B_t}} \,
    \exp\left\{
	    \frac{i}{2} k(t) x
		+ \frac{i}{2} k(t) \left[x - x(t) \right]
		- \frac{iA_t}{2B_t} [x - x(t)]^2
		- \frac{i}{2}
		\int_0^t \dd \xi \,
		\Vect{z}(\xi)^\intercal \JMat{2} \dot{\Vect{z}}(\xi)
	\right\}
	.
	\label{eq:MGOhybrid}
\end{equation}
\end{widetext}

This observation reveals the following difference between MGO and the method of \Ref{Littlejohn85}: while MGO uses the GO solution in the tangent plane and the contributions only from the saddle points, \Ref{Littlejohn85} constructs the solution from delta-shaped envelopes and includes contributions from all locations. (Similar statements also hold when comparing MGO and wavepacket methods, since \Ref{Littlejohn85} is a special case.) That said, we must caution against using \Eq{eq:MGOhybrid} to estimate $\psi$ in a ray-tracing code: it diverges at caustics and does not even reproduce the standard GO formula \eq{eq:GO} away from caustics. In this sense, the $\delta$-shaped envelope model assumed in \Ref{Littlejohn85} seems incompatible with MGO, so one should not attempt to mix the two theories in this manner. More broadly, it is yet to be understood how the various semiclassical methods that appear in the literature~\cite{Littlejohn85, Littlejohn86b, Kay94a, Zor96, Madhusoodanan98}, including MGO, can be interchanged with each other, \ie which saddlepoint (ray) filters can be used with which ray summation (integration) schemes. We shall continue our comparison of MGO and semiclassical integral methods in \Sec{sec:QHOex}.

% ==================== %
% -- MGO ALGORITHMS -- %
% ==================== %

\section{Metaplectic geometrical optics: algorithms}
\label{sec:MGOalgor}

Four algorithms have been designed thus far to aid in the development of an MGO-based ray-tracing code. These algorithms address the four main steps of the MGO framework: \textbf{(i)} tracing rays [\Eq{eq:goRAYS}], \textbf{(ii)} solving the GO equations in the tangent plane [\Eqs{eq:mgoRAYsol} and \eq{eq:mgoENVsol}], \textbf{(iii)} the small-angle rotations to ensure continuity [\Eq{eq:alphaEQ}], and \textbf{(iv)} the inverse MT to obtain the $\Vect{x}$-space wavefunction [\Eq{eq:upsilon}]. The corresponding algorithms are: \textbf{(i)} a curvature-dependent adaptive ray-tracing scheme~\cite{Lopez20}, \textbf{(ii)} a Gram--Schmidt orthogonalization scheme for constructing the rotation matrices $\Mat{S}_\Vect{t}$~\cite{Lopez20}, \textbf{(iii)} a fast linear-time algorithm for performing the small-angle NIMTs from one tangent plane to the next~\cite{Lopez19,Lopez21b}, and \textbf{(iv)} a numerical steepest-descent quadrature rule for the efficient computation of $\Upsilon_\Vect{t}$~\cite{Donnelly21}. We shall now briefly describe each algorithm separately.

% ==================== %
% -- ADAPTIVE DISCRETIZATION -- %
% ==================== %

\subsection{Curvature-dependent adaptive discretization of ray trajectories}
\label{sec:ALGadapt}

Since MGO relies on evolving the tangent plane of the ray manifold as the rays propagate, it is desirable to develop a discretization of the rays that naturally congregates in regions where the tangent plane changes quickly. This would ensure that the angle between neighboring tangent planes is always small even when discretized, as is necessary for the accuracy of MGO.

The procedure to develop adaptive discretizations for Hamiltonian systems is actually well known~\cite{Hairer97}: simply replace the Hamiltonian with a new Hamiltonian possessing the same root structure (so the dispersion relation $\Symb{D} = 0$ is the same) but different gradients (since ray velocities are set by $\pd{\Vect{z}} \Symb{D}$). In other plasma contexts, this method of adaptive discretization has been useful in developing ray equations that naturally slow down as they approach mode-conversion regions~\cite{Tracy07,Jaun07}, analogous to what we desire here for caustics. For our purpose, the new Hamiltonian has the form
\begin{equation}
    \bar{\Symb{D}}(\Vect{z}) = f[\curv(\Vect{z})] \Symb{D}(\Vect{z})
    ,
    \label{eq:modHAM}
\end{equation}

\noindent where $\curv$ is the local curvature of the ray manifold and $f$ is a monotonically decreasing, positive-definite function that also satisfies $f(0) = 1$. 

The rays that result from \Eq{eq:modHAM} are essentially identical to the original rays, but with a new longitudinal parameterization
\begin{equation}
    \fourier{\tau}_1 = \frac{\tau_1}{f[\curv(\Vect{z})]}
    .
\end{equation}

\noindent The conditions of $f$ thereby ensure that the rays indeed slow down (but never stop entirely) in regions where $\curv$ is large and $f$ is correspondingly small. In locally flat regions where $\curv = 0$, there is no difference between the two parameterizations. Said another way, a fixed time step $\Delta \fourier{\tau}_1$ corresponds to a variable time-step $\Delta \tau_1 = f \Delta \fourier{\tau}_1$ in the original coordinates that naturally shortens where $f$ is smaller (and $\curv$ larger) without any external input or monitoring. Performing the adaptive discretization in this manner maintains the Hamiltonian structure of the ray equations; consequently, their (adaptive) integration should be amenable to symplectic methods~\cite{Hairer97,Richardson12}.

% ==================== %
% -- GramSchmidt -- %
% ==================== %

\subsection{Symplectic Gram--Schmidt construction of ray tangent plane}
\label{sec:ALGgs}

It is likewise desirable to have an algorithm that computes the necessary symplectic rotation matrix $\Mat{S}_\Vect{t}$ which maps $\Vect{x}$-space to the tangent plane at $\Vect{t}$ on the ray manifold. Ideally, this construction should also be done using only local information from the ray trajectories. The following symplectic Gram-Schmidt orthogonalization scheme performs just this. 

Suppose that rays have already been traced to obtain the ray manifold $\Vect{z}(\Vect{\tau})$. (This step need only be done locally, not globally.) A basis for the tangent plane at $\Vect{\tau} = \Vect{t}$ is given by the set $\{ \pd{\tau_j} \Vect{z}(\Vect{t}) \}$, which can be orthogonalized via classical Gram--Schmidt~\cite{Trefethen97} to yield $N$ orthonormal tangent vectors at $\Vect{t}$, denoted $\{ \unit{\Vect{T}}_j(\Vect{t}) \}$. The symplectically dual normal vectors are obtained as
\begin{equation}
    \unit{\Vect{N}}_j(\Vect{t}) = - \JMat{2N} \unit{\Vect{T}}_j(\Vect{t}),
\end{equation}

\noindent where $\JMat{2N}$ is defined in \Eq{eq:Jmat}. Then, the desired symplectic matrix $\Mat{S}_\Vect{t}$ is
\begin{equation}
    \Mat{S}_\Vect{t} = 
    \begin{pmatrix}
        \uparrow & & \uparrow & \uparrow & & \uparrow \\
        \unit{\Vect{T}}_1(\Vect{t}) & \ldots & \unit{\Vect{T}}_N(\Vect{t}) & \unit{\Vect{N}}_1(\Vect{t}) & \ldots & \unit{\Vect{N}}_N(\Vect{t}) \\
        \downarrow & & \downarrow & \downarrow & & \downarrow
    \end{pmatrix}^\intercal
    .
\end{equation}
%row $m$, column $n$

\noindent Equivalently, the entries of $\Mat{A}_\Vect{t}$ and $\Mat{B}_\Vect{t}$ can be directly identified as
\begin{equation}
    \left[\Mat{A}_\Vect{t}\right]_{mn} = \left[\unit{\Vect{T}}_m(\Vect{t})\right]_n
    , \quad
    \left[\Mat{B}_\Vect{t}\right]_{mn} = \left[\unit{\Vect{T}}_m(\Vect{t})\right]_{n+N}
    ,
\end{equation}

\noindent for $m, n = 1, \ldots N$.

As noted in \Sec{sec:MGOtoGO}, $\Mat{S}_\Vect{t}$ can be generally constructed from the QR decomposition of the tangent vectors $\{ \pd{\tau_j} \Vect{z}(\Vect{t}) \}$, with $\Mat{A}_\Vect{t}$ and $\Mat{B}_\Vect{t}$ being determined by \Eq{eq:QRab}. If the upper-triangular matrix $\Mat{R}_\Vect{t}$ is restricted to having strictly positive diagonal elements, then the QR decomposition \eq{eq:zQR} is unique~\cite{Trefethen97}; the Gram--Schmidt algorithm presented here is then simply one means of computing this decomposition, but in principle any QR decomposition algorithm will suffice.

In practice, the tangent vectors $\{ \pd{\tau_j} \Vect{z}(\Vect{t}) \}$ can be computed via finite difference formulas, although maintaining low truncation errors might require launching a sufficiently large number of rays to ensure the ray bundle is dense at all desired locations of the ray manifold. We should emphasize that the symplectic property of $\Mat{S}_\Vect{t}$ is ensured precisely because the ray manifold is a Lagrangian manifold, a fact whose importance was alluded to in \Sec{sec:GO}.

% ==================== %
% -- NIMT -- %
% ==================== %

\subsection{Fast near-identity metaplectic transform}
\label{sec:ALGnimt}

Since MGO relies on performing frequent NIMTs to rotate from one tangent plane to the next, it is desirable to have an efficient NIMT algorithm. Although the MT is most commonly encountered as the integral transform \eq{eq:MT}, it can also be formulated explicitly in terms of the operators $\VectOp{x}$ and $\VectOp{k}$ as~\cite{Lopez21b}
\begin{align}
    \hspace{-2mm}\oper{M}(\Mat{S})
    =
    &\pm 
    \exp\left[ 
        i \frac{
            \VectOp{x}^\intercal \left(\log \Mat{A}^{-\intercal}\right) \VectOp{k} 
            + \VectOp{k}^\intercal \left(\log \Mat{A}^{-1}\right) \VectOp{x}
        }{2}
    \right] 
    \nonumber\\
    &\times\exp\left(
        -i \frac{\VectOp{x}^\intercal \Mat{A}^\intercal \Mat{B} \VectOp{x}}{2}
    \right)
    \, 
    \exp\left(
        - i \frac{\VectOp{k}^\intercal \Mat{A}^{-1} \Mat{B} \VectOp{k}}{2}
    \right)
    .
    \label{eq:MToper}
\end{align}

\noindent This representation is advantageous because the exponential operators can be expanded for small argument when $\Mat{S} \approx \IMat{2N}$. Hence, one can develop NIMT algorithms that require fewer floating-point operations (FLOPs) than either the direct integration or FFT-based computation of \Eq{eq:MT}, which require $O(M^2)$ and $O(M \log M)$ FLOPs for $M$ sample points of $\psi$, respectively.

One option is to use a Taylor expansion to approximate the rightmost exponential of \Eq{eq:MToper}, \ie let
\begin{equation}
    \exp\left(
        - \frac{i}{2} \VectOp{k}^\intercal \Mat{A}^{-1} \Mat{B} \VectOp{k}
    \right)
    \approx
    \IdentOp - \frac{i}{2} \VectOp{k}^\intercal \Mat{A}^{-1} \Mat{B} \VectOp{k} + \ldots
    \, .
\end{equation}

\noindent The spatial representation of the resulting operator is 
\begin{align}
    \fourier{\psi}(\Vect{X}) &= 
    \frac{
        \exp\left( - \frac{i}{2} \Vect{X}^\intercal \Mat{B}\Mat{A}^{-1} \Vect{X} \right)
    }{
        \sqrt{\det \Mat{A} }
    }
    \nonumber\\
    &\times
    \left.
        \left\{
            \psi(\Vect{x})
           + \frac{i}{2} \Tr\left[
                \Mat{A}^{-1} \Mat{B} \nabla \nabla \psi(\Vect{x})
            \right]
        \right\}
    \right|_{\Vect{x} = \Mat{A}^{-1} \Vect{X}}
    .
    \label{eq:NIMTtaylor}
\end{align}

\noindent This operator is local and can therefore be used to perform pointwise transformations along a ray without needing global solutions of the GO equations in each tangent plane. This aspect of the NIMT is not immediately obvious from the integral representation \eq{eq:MT}. The pointwise nature of \Eq{eq:NIMTtaylor} also means that it can be computed in $O(M)$ FLOPs, which is faster than the two alternate methods mentioned previously. Unfortunately, \Eq{eq:NIMTtaylor} is no longer unitary. A detailed analysis~\cite{Lopez19} showed that this loss of unitarity results in the unbounded growth of high-wavenumber oscillations, which can be partially mitigated by using low-pass smoothing methods.

To preserve the unitarity of the NIMT while also maintaining the fast $O(M)$ scaling, a diagonal Pad\'e approximation~\cite{Press07,Olver10a} for the exponential operators can alternatively be used. The lowest-order approximation has
\begin{align}
    &\exp\left(
        - \frac{i}{2} \VectOp{k}^\intercal \Mat{A}^{-1} \Mat{B} \VectOp{k}
    \right)
    \nonumber\\
    &\approx
    \left(\IdentOp + \frac{i}{4} \VectOp{k}^\intercal \Mat{A}^{-1} \Mat{B} \VectOp{k}\right)^{-1}
    \left(\IdentOp - \frac{i}{4} \VectOp{k}^\intercal \Mat{A}^{-1} \Mat{B} \VectOp{k}\right)
    ,
\end{align}

\noindent and likewise for the leftmost exponential in \Eq{eq:MToper}, \ie the dilation term. It is well-established that the diagonal Pad\'e approximation preserves the unitarity of exponential operators, a fact that has been used recently in plasma physics to develop unitary pitch-angle collision operators~\cite{Zhang20,Fu22}. However, the operator inversion means the Pad\'e-based NIMT is no longer local, but it can still be computed in $O(M)$ FLOPs in certain situations.

Consider $1$-D transformations on a discrete grid consisting of $M$ equally spaced points. The NIMT maps the length-$M$ vector $\Vect{\psi}$ to the new length-$M$ vector $\fourier{\Vect{\psi}}$ as
\begin{equation}
    \fourier{\Vect{\psi}}
    = \Mat{N}(\Mat{S}) 
    \Vect{\psi}
    ,
    \label{eq:dNIMTmap}
\end{equation}

\noindent where the Pad\'e NIMT matrix is given as
\begin{align}
    \Mat{N}(\Mat{S}) 
    =
    &\left(
        \IMat{M}
        + \log A
        \frac{
            \Mat{x} \, \delta_1
            + \delta_1 \Mat{x}
        }{4}
    \right)^{-1}
    \nonumber\\
    &\times
    \left(
        \IMat{M}
        - \log A
        \frac{
            \Mat{x} \, \delta_1
            + \delta_1 \Mat{x}
        }{4}
    \right)
    \exp
    \left(
        \frac{- i A B}{2} \Mat{x}^2
    \right)
    \nonumber\\
    &\times
    \left(
        \IMat{M}
        - \frac{i B}{4A} \Delta_2
    \right)^{-1}
    \left(
        \IMat{M}
        + \frac{i B}{4A} \Delta_2
    \right)
    .
    \label{eq:dNIMT}
\end{align}

\noindent Here we have introduced the diagonal coordinate matrix
\begin{equation}
    \Mat{x}
    \doteq
    \begin{pmatrix}
        x_1 & & \\
        & \ddots & \\
        & & x_M
    \end{pmatrix}
\end{equation}

\noindent and the matrices $\delta_1$ and $\Delta_2$ as finite-difference matrices for $\pd{x}$ and $\pd{x}^2$, respectively. One can use the same discretization for both by setting $\Delta_2 = \delta_1^2$ if desired, but this is not strictly necessary. What is necessary, however, is that $\delta_1$ be anti-Hermitian and $\Delta_2$ be Hermitian negative semi-definite to ensure that $\Mat{N}$ is unitary. This prohibits the use of forward- or backward-difference matrices, but central-difference matrices such as
\begin{equation}
    \hspace{-1mm}
    \delta_1 = \frac{1}{2h}
    \begin{pmatrix}
        0 & 1 & \\
        -1 & \ddots & 1 \\
        & -1 & 0
    \end{pmatrix}
    , \quad
    \Delta_2 = \frac{1}{h^2}
    \begin{pmatrix}
        -2 & 1 & \\
        1 & \ddots & 1 \\
        & 1 & -2
    \end{pmatrix}
    \label{eq:triMAT}
\end{equation}

\noindent are allowed, where $h$ is the constant step size.

By representing $\delta_1$ and $\Delta_2$ with banded matrices, \Eq{eq:dNIMTmap} can be computed in $O(M)$ FLOPs via banded matrix multiplication and banded linear system solves. For example, when $\Delta_2$ is given by \Eq{eq:triMAT}, then computing the first matrix-vector product of \Eq{eq:dNIMTmap},
\begin{equation}
    \Vect{\psi}_1 = 
    \left(
        \IMat{M}
        + \frac{i B}{4A} \Delta_2
    \right) \Vect{\psi}
    ,
\end{equation}

\noindent requires $O(3M)$ multiplications, being the direct matrix-vector multiplication for a tridiagonal matrix. Computing the second matrix vector product
\begin{equation}
    \Vect{\psi}_2
    =
    \left(
        \IMat{M}
        - \frac{i B}{4A} \Delta_2
    \right)^{-1}
    \Vect{\psi}_1
\end{equation}

\noindent is best done by solving the linear system
\begin{equation}
    \left(
        \IMat{M}
        - \frac{i B}{4A} \Delta_2
    \right)
    \Vect{\psi}_2
    = \Vect{\psi}_1
    .
\end{equation}

\noindent Since the matrix prefactor is tridiagonal, $\Vect{\psi}_2$ can be obtained in $O(M)$ FLOPs using standard tridiagonal Gaussian elimination algorithms~\cite{Press07}. The remaining matrix-vector products in \Eq{eq:dNIMTmap} are computed analogously, thus yielding an $O(M)$ scheme for obtaining $\fourier{\Vect{\psi}}$.

The NIMT can be used `as is' for performing a single small-angle rotation, but it can also be used iteratively to perform a single large-angle MT. This is because any symplectic matrix $\fourier{\Mat{S}}$ can be decomposed as
\begin{equation}
    \fourier{\Mat{S}} = \Mat{S}_K \ldots \Mat{S}_1
    ,
\end{equation}

\noindent where the near-identity iterates
\begin{equation}
    \Mat{S}_j \doteq 
    \Mat{S}\left( \frac{j}{K} \right)
    \Mat{S}^{-1} \left( \frac{j - 1}{K} \right)
\end{equation}

\noindent are obtained from the `trajectory' $\Mat{S}(t)$ that smoothly connects the identity $\Mat{S}(0) = \IMat{2N}$ with the desired transformation $\Mat{S}(1) = \fourier{\Mat{S}}$. The corresponding MT can then be approximated by the iterated NIMT as%
%%%
~\footnote{The winding number of $\Mat{S}(t)$ must also be consistent with the desired overall sign of the MT \eq{eq:MT}. An odd winding number results in a sign flip, while an even winding number does not; see \Refs{Lopez19,Littlejohn86a} for more discussion.}%
%%%
\begin{equation}
    \oper{M}(\Mat{S}) = \oper{M}(\Mat{S}_K) \ldots \oper{M}(\Mat{S}_1)
    \approx
    \oper{N}(\Mat{S}_K) \ldots \oper{N}(\Mat{S}_1)
    ,
    \label{eq:iterNIMT}
\end{equation}

\noindent where $\oper{N}(\Mat{S})$ denotes the NIMT operator, either given by the Taylor expansion \eq{eq:NIMTtaylor} or the Pad\'e expansion \eq{eq:dNIMTmap}. In particular, the exact unitarity of the Pad\'e NIMT means that \Eq{eq:iterNIMT} is a convergent scheme, with a single-step convergence rate of $3$ and a global convergence rate of $2$ for the lowest-order approximation \eq{eq:dNIMT}~\cite{Lopez21b}.

% ==================== %
% -- GaussFreud -- %
% ==================== %

\subsection{Gauss--Freud numerical steepest descent}
\label{sec:ALGgf}

\subsubsection{Derivation}

The computation of $\Upsilon_\Vect{t}(\Vect{x})$ [\Eq{eq:upsilon}] involves a highly oscillatory integral evaluated along a certain steepest-descent contour. Such an integral requires a specialized quadrature rule to compute efficiently. The quadrature rule we have developed~\cite{Donnelly21} proceeds in two steps: first, the correct steepest-descent contour $\cont{0}$ is identified, then the integral is efficiently computed along it using an appropriately designed Gaussian quadrature rule~\cite{Press07}.

For simplicity, let us consider a $1$-D system. At fixed $t = t_0$ and $x = x_0$, the phase and envelope of the integrand \eq{eq:upsilon} can be defined as
\begin{subequations}
    \begin{align}
        \vartheta(\epsilon; t_0, x_0) &\doteq
        \Theta[\epsilon + X_{t_0}(t_0)] - \gamma_{t_0}(\epsilon, x_0)
        , \\
        \varphi(\epsilon; t_0, x_0) &\doteq 
        \fourier{\env}[\epsilon + X_{t_0}(t_0)]
        .
    \end{align}
\end{subequations}

\noindent When $\vartheta$ is analytic, the steepest-descent contour that passes through $\epsilon = 0$ is given by the values of $\epsilon$ in the complex plane that satisfy
\begin{equation}
    \Re\left[\vartheta(\epsilon; t_0, x_0) \right] = \Re\left[\vartheta(0; t_0, x_0) \right]
    \equiv 0
    .
    \label{eq:steepCONT}
\end{equation}

\noindent Equation \eq{eq:steepCONT} defines $\cont{0}$, and can be readily computed using any standard contour-finding routine, \eg marching squares~\cite{Newman06}, near the known solution $\epsilon = 0$. As will be clear shortly, only a small portion of $\cont{0}$ is needed for the quadrature rule, which reduces the computational cost needed to complete this initial step.

Having obtained $\cont{0}$ in the neighborhood of the saddlepoint $\epsilon = 0$, the next step is to approximate it with a straight line. However, it is well-known that steepest-descent lines are not necessarily smooth across saddlepoints, exhibiting a finite-angle kink for odd-order saddles~\cite{Deano09}. (An order $\alpha$ saddle has local functional form $\epsilon^{\alpha+2}$, with $\alpha = 0$ being non-degenerate.) We must therefore use a bilinear approximation of $\cont{0}$ near $\epsilon = 0$. Let us denote $\cont{0}$ by $\epsilon(\ell)$, where $\ell$ is a $1$-D parameterization of $\cont{0}$. The bilinear approximation of $\cont{0}$ then reads
\begin{equation}
    \epsilon(\ell; t_0, x_0) \approx
    |\ell| \times
    \left\{
        \begin{array}{cc}
            \exp\left[i \phi_-(t_0, x_0) \right], & \ell \le 0 \\
            \exp\left[i \phi_+(t_0, x_0) \right], & \ell > 0
        \end{array}
    \right.
    ,
\end{equation}

\noindent where $\phi_\mp$ are the angles that the incoming and outgoing branches of $\cont{0}$ make with the $\Re(\epsilon)$ axis, respectively. Although these angles can be computed in many ways, we shall discuss a particularly convenient choice shortly.

Next, we must choose a quadrature rule to apply along the approximated $\cont{0}$. Along the exact steepest-descent contour, 
\begin{equation}
    i \vartheta[\epsilon(\ell); t_0, x_0] = 
    - F(\ell; t_0, x_0)
    ,
    \label{eq:Fdef}
\end{equation}

\noindent where $F$ is a strictly increasing function of $|\ell|$. Catastrophe theory predicts that locally $F$ will be a polynomial in $|\ell|$ (the various structurally stable caustics discussed in \Sec{sec:paraxial}). In particular, when $x$ is not at a caustic, $F$ will be a quadratic function of $|\ell|$. 

At this point, a compromise needs to be made: on one hand, the majority of points in $x$ are not caustics, so it is reasonable to optimize the quadrature rule for quadratic phase functions; on the other hand, the main usefulness of MGO depends heavily on the accuracy in computing $\Upsilon$ at these special points. The compromise proposed in \Ref{Donnelly21} is to use Gaussian quadrature~\cite{Press07} with respect to a fitted quadratic phase function
\begin{equation}
    F(\ell; t_0, x_0)
    = \ell^2 \times
    \left\{
        \begin{array}{cc}
            s_-(t_0, x_0), & \ell \le 0 \\
            s_+(t_0, x_0), & \ell > 0
        \end{array}
    \right.
    ,
    \label{eq:Ffit}
\end{equation}

\noindent where $s_\pm$ are determined by the quadratic fit
\begin{equation}
    s_\pm(t_0, x_0) = \frac{F(\ell_\pm; t_0, x_0) - F(0; t_0, x_0)}{\ell_\pm^2}
\end{equation}

\noindent and $\ell_\pm$ are determined by a threshold condition along the exact $\cont{0}$:
\begin{equation}
    F(\ell_\pm; t_0, x_0) - F(0; t_0, x_0) \ge 1
    .
\end{equation}

\noindent The threshold condition can then be used as a secant-line approximation to also obtain the rotation angles $\phi_\pm$ as
\begin{equation}
    \phi_\pm(t_0, x_0) = \arg\left[ \epsilon(\ell_\pm; t_0, x_0) \right]
    .
    \label{eq:PHIfit}
\end{equation}

\noindent Equations \eq{eq:Ffit}-\eq{eq:PHIfit} are exact when $F$ is locally quadratic and are still well-defined at caustics when $F$ is not locally quadratic, by virtue of the quadratic interpolation between $\epsilon(0)$ and $\epsilon(\ell_\pm)$.

An $n$-point Gaussian quadrature rule based on Freud polynomials~\cite{Steen69} is then readily developed for \Eq{eq:upsilon}:
\begin{subequations}
    \label{eq:GFquad}
    \begin{align}
        \Upsilon_{t_0}(x_0) &=
        \sum_{j = 1}^n w_j \exp(\ell_j^2)
        \left[
            h_+(\ell_j) - h_-(\ell_j)
            \nullFrac
        \right]
        , \\
        h_\pm(\ell) &\doteq
        I\left[
            \frac{\ell \exp(i \phi_\pm)}{\sqrt{s_\pm}}
        \right]
        \frac{\exp(i \phi_\pm)}{\sqrt{s_\pm}}
        ,
    \end{align}
\end{subequations}

\noindent where we have denoted the integrand by $I \doteq \varphi \exp(i \vartheta)$. The dependence of $I$, $\phi_\pm$, and $s_\pm$ on $t_0$ and $x_0$ have been suppressed for brevity. Also, $\{ w_j \}$ and $\{ \ell_j \}$ are respectively the quadrature weights and nodes for the Freud polynomials, which are the unique family of orthogonal polynomials for the inner product 
\begin{equation}
    \langle f , g \rangle \doteq \int_0^\infty \dd x \exp(-x^2) f(x) g(x)
    .
\end{equation}

\noindent Tables of $\{ w_j \}$ and $\{ \ell_j \}$ for $n \le 20$ are provided in \Refs{Donnelly21,Steen69}. 

Now let $t$ and $x$ vary. The steepest-descent contours are continuous functions of $t$ and $x$, so the calculation of $\cont{0}$ at some $(t_j, x_j)$ can inform the next calculation at $(t_{j+1}, x_{j+1})$ in a `memory feedback loop' that reduces the total computational cost. First, the MGO simulation is initialized far from a caustic such that $F(\ell; t, x)$ is locally quadratic in $\ell$ with initial angles $\sigma_\pm^{(0)}$ given as
\begin{equation}
    \sigma_\pm^{(0)} \approx
    - \frac{\pi}{4} 
    - \frac{\arg[\pd{\epsilon}^2 \vartheta(\epsilon; t_0, x_0)]}{2}
    \pm \frac{\pi}{2}
    .
\end{equation}

\noindent The search for the initial $\cont{0}^{(0)}$ can therefore be limited to some small angular window about $\sigma_\pm^{(0)}$. By continuity, the search for $\cont{0}^{(j)}$ can be similarly restricted to a small angular window about the previously calculated $\sigma_\pm^{(j-1)}$. Using this feedback system, the correct steepest-descent contours along which to evaluate $\Upsilon_t(x)$ can be unambiguously identified at caustics.

Thus far we have assumed $1$-D, but the generalization to $N$-D should be straightforward since multivariate analytic functions are analytic in each variable separately; the $N$-D steepest-descent surface is the union of (continuous families of) individual steepest-descent curves for each variable. Hence, it might be possible to evaluate $N$-D integrals as a nested series of $1$-D integrals~\cite{Press07} sequentially evaluated using our $1$-D algorithm along that variable's steepest-descent curve. This approach is best done when the nested integrals are sufficiently simple and the multidimensional weight function factors cleanly into a product of univariate weight functions, which is expected to be true for MGO. If not, though, the more complicated method of $N$-D Gaussian cubature~\cite{Cools97} must be used. We shall investigate this in future work.

\begin{figure*}
	\begin{overpic}[width=0.44\linewidth,trim={6mm 18mm 3mm 23mm},clip]{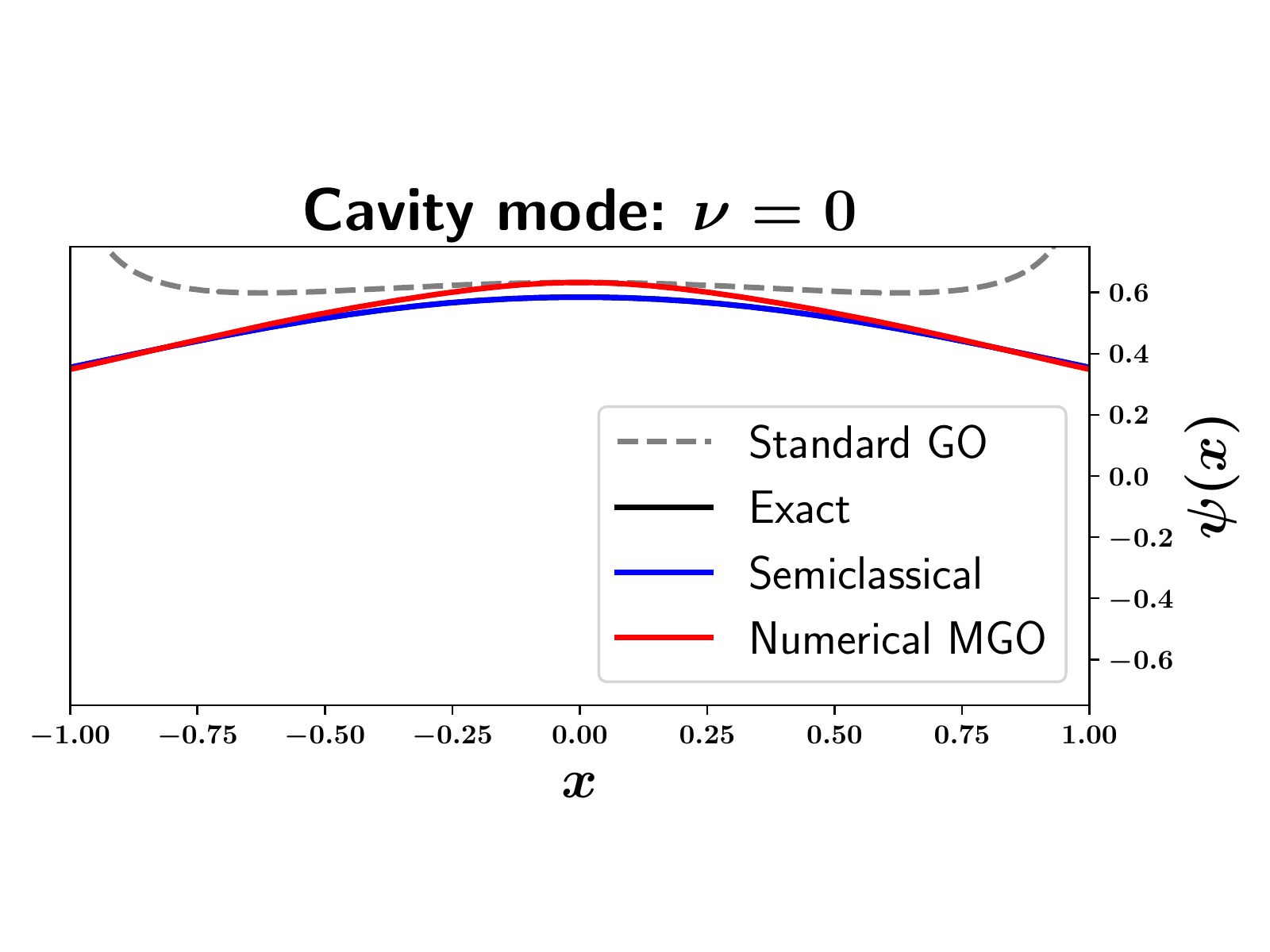}
        \put(5,10){\textbf{\small(a)}}
    \end{overpic}
	\hspace{4mm}
	\begin{overpic}[width=0.44\linewidth,trim={4mm 18mm 3mm 23mm},clip]{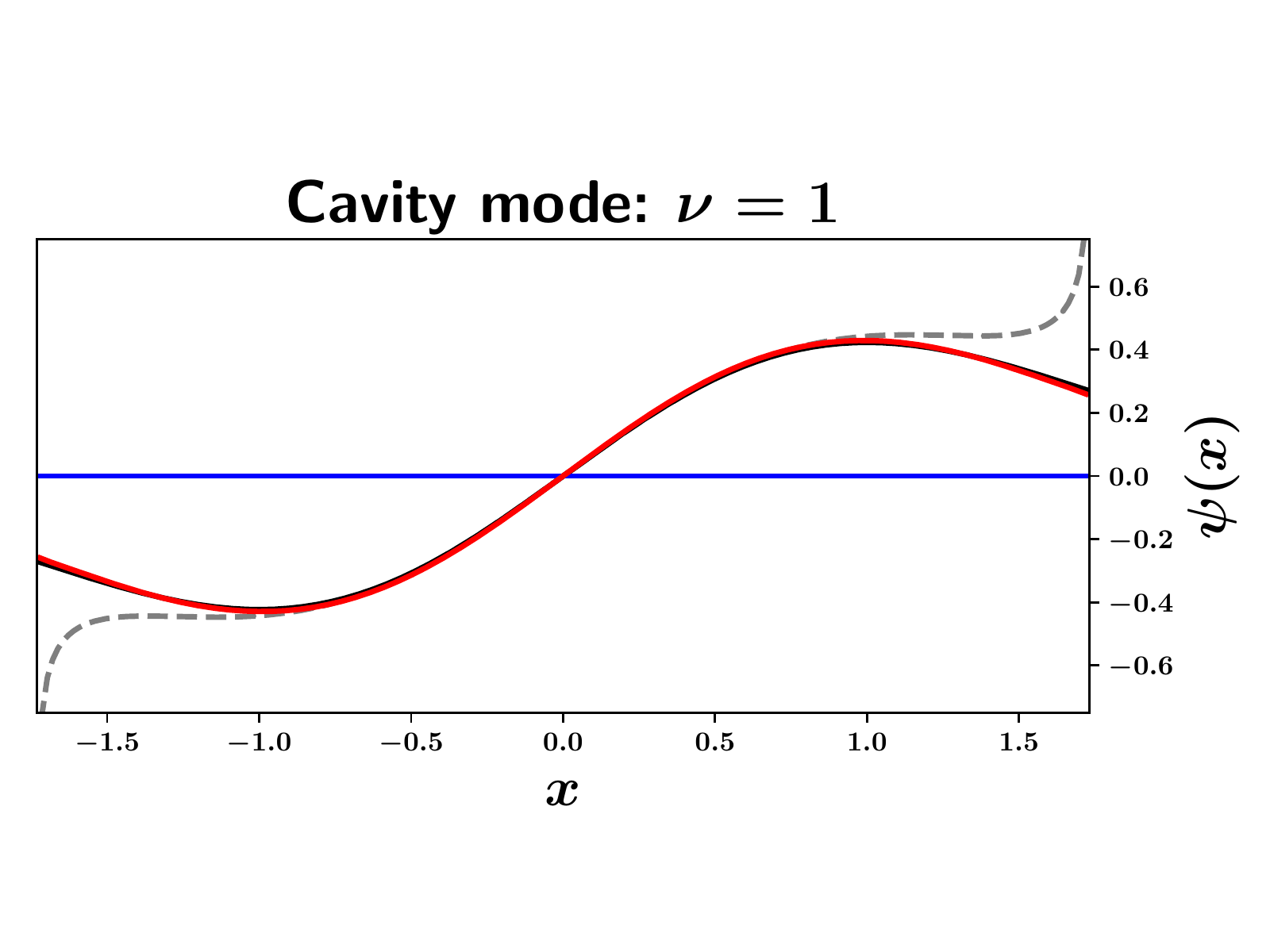}
        \put(5,10){\textbf{\small(b)}}
    \end{overpic}
	
	\vspace{2mm}
	\begin{overpic}[width=0.44\linewidth,trim={4mm 18mm 3mm 23mm},clip]{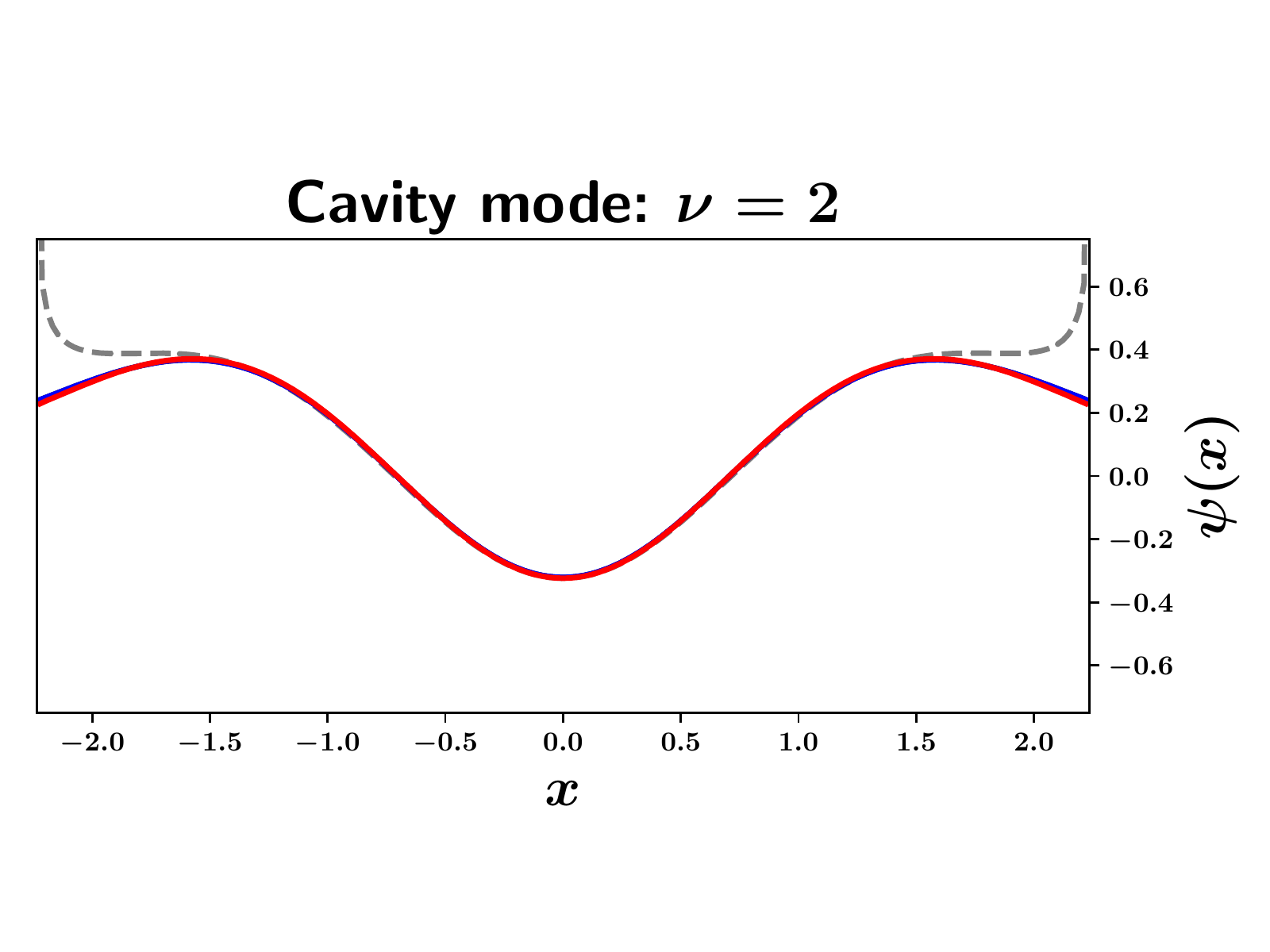}
        \put(5,10){\textbf{\small(c)}}
    \end{overpic}
	\hspace{4mm}
	\begin{overpic}[width=0.44\linewidth,trim={4mm 18mm 3mm 23mm},clip]{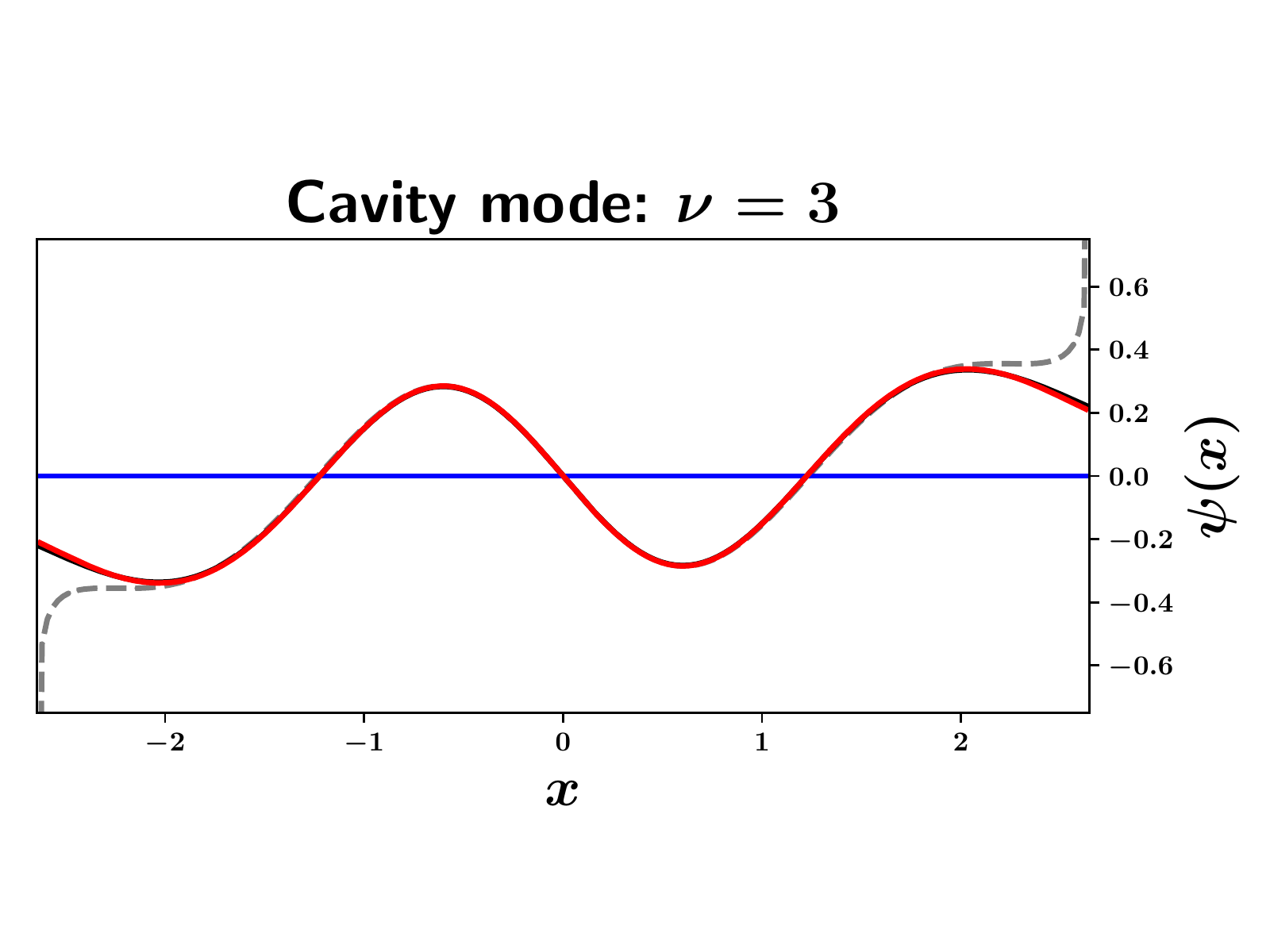}
        \put(5,10){\textbf{\small(d)}}
    \end{overpic}
	
	\vspace{2mm}
	\begin{overpic}[width=0.44\linewidth,trim={4mm 18mm 3mm 23mm},clip]{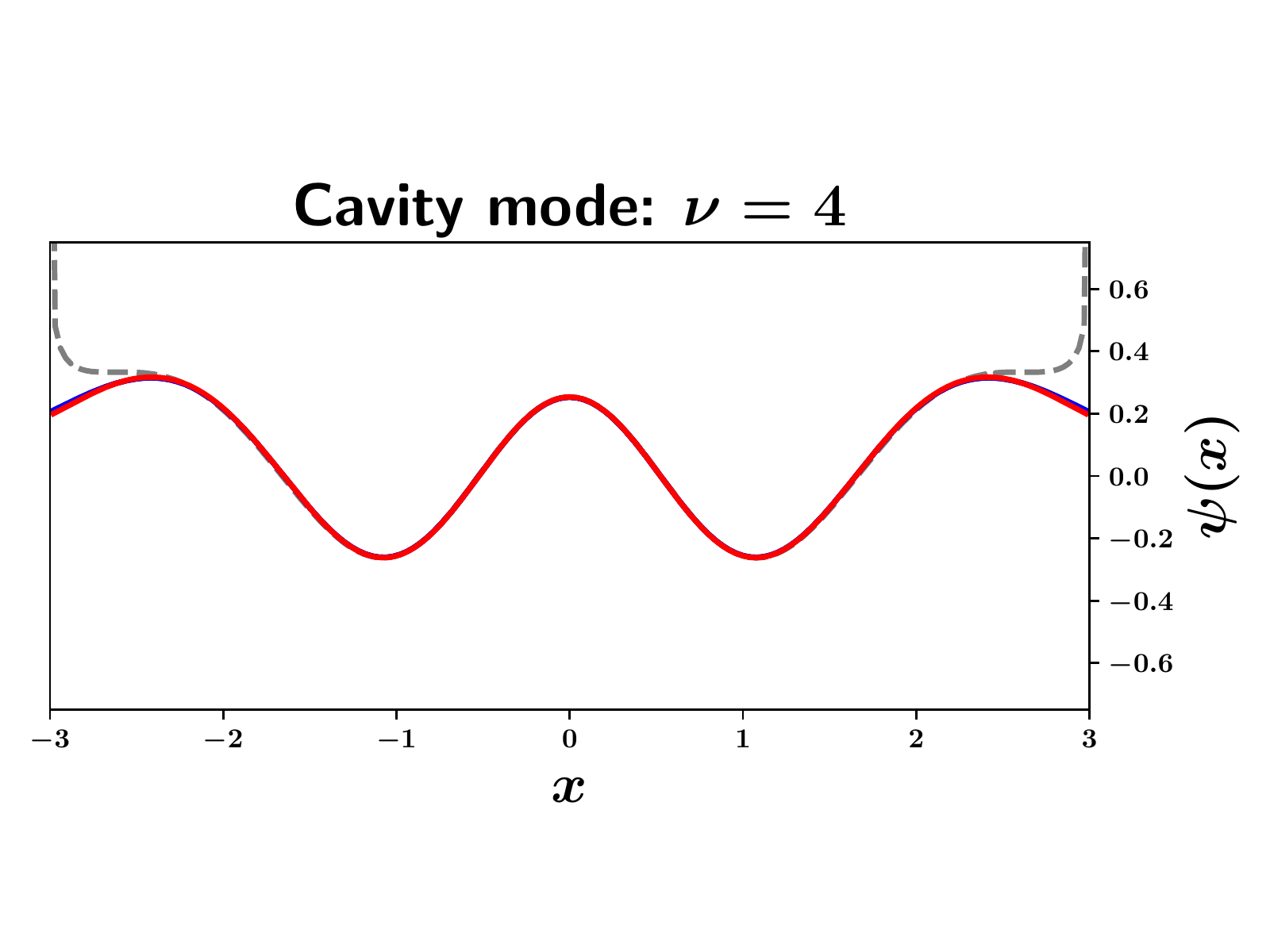}
        \put(5,10){\textbf{\small(e)}}
    \end{overpic}
	\hspace{4mm}
	\begin{overpic}[width=0.44\linewidth,trim={4mm 18mm 3mm 23mm},clip]{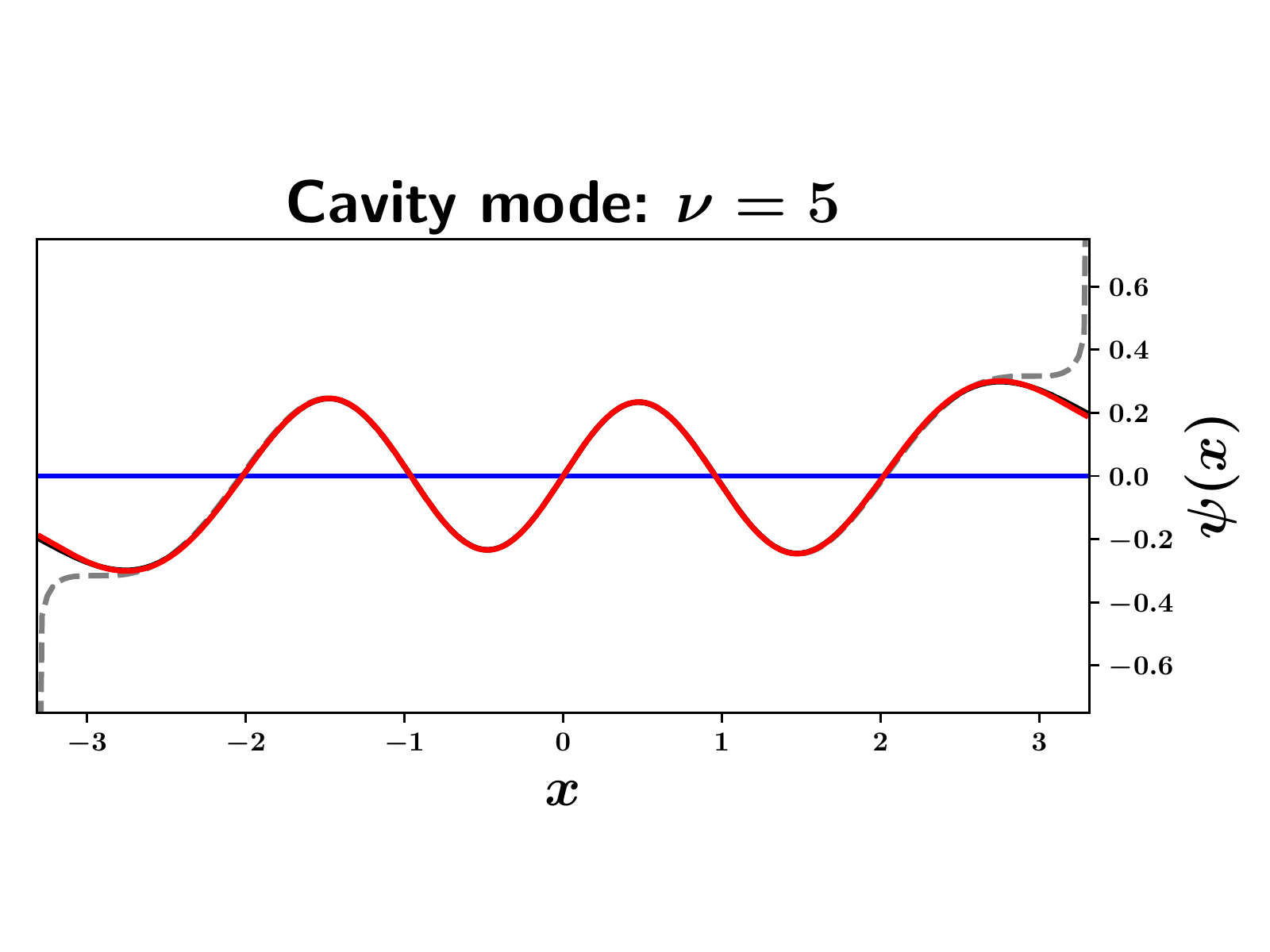}
        \put(5,10){\textbf{\small(f)}}
    \end{overpic}

	\caption{\textbf{(a)}--\textbf{(f)} Comparison between the standard GO solution \eq{eq:GOqho}, the exact solution \eq{eq:EXACTqho}, the semiclassical integral solution \eq{eq:SEMIqho}, and the MGO solution \eq{eq:MGOqho} computed via \Eq{eq:GFquad} with $n = 2$ for the eigenmodes of a wave bounded within a quadratic cavity potential (\ie the quantum harmonic oscillator problem). The first six modes are presented (with $\nu$ the mode number). The GO solution diverges at the caustics, but the MGO solution remains finite and agrees well with the exact solution, even though a low quadrature order was used. In contrast, the semiclassical integral method gives the exact answer for even mode numbers, but erroneously gives zero for odd mode numbers.}
	\label{fig:QHO}
\end{figure*}

\subsubsection{Example}
\label{sec:QHOex}

Lastly, let us illustrate \Eq{eq:GFquad} with a simple example. Consider an electromagnetic wave bounded within a 1-D quadratic density well. In suitable coordinates, such a system is described by the quantum harmonic oscillator (QHO) equation
\begin{equation}
    \pd{x}^2 \psi(x) + (2 \nu + 1 - x^2) \psi(x) = 0
    ,
    \label{eq:QHOeq}
\end{equation}

\noindent where the integer $\nu$ denotes the cavity mode number. In \Ref{Lopez20}, the MGO solution for \Eq{eq:QHOeq} was derived; the derived solution can be written as
\begin{equation}
	\psi_\text{MGO}(x) = \text{Im}\left\{  
		\frac{\Upsilon(x) \exp[i \beta(x)] }{\pi (2 R)^{1/3} \sqrt{|x|} } 
	\right\}
	,
	\label{eq:MGOqho}
\end{equation}

\noindent where $R \doteq \sqrt{2 \nu + 1}$, and
\begin{subequations}
    \begin{align}
    	\label{eq:UPSILONqho}
    	\Upsilon(x) 
	    &\doteq \int_{\cont{0}} \dd \epsilon \,
    	\frac{\exp\left[ i \vartheta(\epsilon, x) \right]}{\left[1 - (\epsilon/R)^2 \right]^{1/4}}
	    , \\
    	%%%
	    \beta(x) &\doteq
    	\frac{R^2}{2} \cos^{-1}\left( \frac{x}{R} \right) 
	    - \frac{x}{2} \sqrt{R^2 - x^2}
    	\nonumber\\
	    &\hspace{28mm}
    	+ \frac{\pi}{4} \left[ \text{sgn}(x) + 1 \right]
	    , \\
    	%%%
	    \vartheta(\epsilon, x) 
    	&\doteq
	    \frac{\epsilon}{2} \sqrt{R^2 - \epsilon^2}
    	+ \frac{R^2}{2} \tan^{-1} \left( \frac{\epsilon}{\sqrt{R^2 - \epsilon^2} } \right)
    	\nonumber\\
    	&\hspace{22mm}
	    - R \epsilon
    	- \frac{\epsilon^2}{2 x} \sqrt{R^2 - x^2}
	    .
    \end{align}
\end{subequations}

\noindent For comparison purposes, the exact solution to \Eq{eq:QHOeq} is given as
\begin{equation}
    \psi_\text{exact}(x) = \frac{\textrm{Ai}(0)}{\sqrt{R}} \, 
    \frac{
        \textrm{D}_{\nu} (\sqrt{2} \, x)
    }{
        \textrm{D}_\nu(\sqrt{2}\,R) 
    } ,
    \label{eq:EXACTqho}
\end{equation}

\noindent (where $\textrm{D}_\nu(x)$ is Whittaker's parabolic cylinder function~\cite{Olver10a}), the GO solution is given as
\begin{equation}
    \psi_\text{GO}(x) 
    = \frac{2^{1/6} \cos\left[ \frac{x}{2}\sqrt{R^2 - x^2} - \frac{R^2 }{2}\cos^{-1}\left(\frac{x}{R} \right)  + \frac{\pi}{4} \right]}{\sqrt{\pi} \, R^{1/3} (R^2 - x^2)^{1/4}} 
    ,
    \label{eq:GOqho}
\end{equation}

\noindent and as shown in \Ref{Littlejohn85}, the semiclassical integral method \eq{eq:LittlejohnEQ} yields
\begin{equation}
    \psi_\text{semi}(x)
    = 
    \left\{
        \begin{array}{ll}
            \frac{\textrm{Ai}(0)}{\sqrt{R}} \, 
            \frac{
                \textrm{D}_{\nu} (\sqrt{2} \, x)
            }{
                \textrm{D}_\nu(\sqrt{2}\,R) 
            },
            & \nu \text{ even} \\
            0,  & \nu \text{ odd}
        \end{array}
    \right.
    .
    \label{eq:SEMIqho}
\end{equation}

\noindent Note that we do not include the MGO-semiclassical hybrid method \eq{eq:MGOhybrid} in the comparison, since it will clearly have singularities at the $k$-space caustic $x = 0$.

To compute the MGO solution \eq{eq:MGOqho}, we use the Gauss--Freud quadrature rule \eq{eq:GFquad} with quadrature order $n = 2$ to calculate $\Upsilon(x)$ given by \Eq{eq:UPSILONqho}. The resulting solution is plotted in \Fig{fig:QHO} for the first six eigenmodes, \ie for $\nu = 0, \ldots, 5$. The agreement between the numerical MGO solution and the exact eigenmodes is remarkable, even for such a low quadrature order. Both MGO and the semiclassical integral method \eq{eq:LittlejohnEQ} remain finite at caustics. However, although \Eq{eq:LittlejohnEQ} obtains the exact result when $\nu$ is even, it fails when $\nu$ is odd, making it unreliable as a general method for removing caustics in ray-tracing codes. In contrast, MGO remains accurate for both even and odd mode numbers. This success is a direct result of the flexibility allowed in MGO for the shape of the tangent-plane wavefield; no artificial symmetry constraints are introduced~%
%%%
\footnote{ It was identified in \Ref{Littlejohn85} that the failure of \Eq{eq:LittlejohnEQ} to model odd-parity fields arose from the assumption that the tangent-plane wavefield was symmetric and delta-shaped [see \Eq{eq:LittlejohnPSI}.] }. %
%%%
The demonstrated robustness of MGO bodes well for the success of an MGO-based ray-tracing code.

%
% ==================== %
% ---- Summary ---- %
% ==================== %
%\vspace{3mm}
\section{Outline of the MGO procedure}
\label{sec:summary}

Let us now briefly outline how the MGO formalism can be used in a ray-tracing code. For simplicity, suppose an incident eikonal wavefield is prescribed on a plane $x_1 = 0$, that is, \begin{equation}
    \psi_\textrm{in}(\Vect{x}_\perp) 
    = \phi_0(\Vect{x}_\perp) \exp[i \theta_0(\Vect{x}_\perp)]
    ,
    \label{eq:psiINIT}
\end{equation}

\noindent where $\Vect{x}_\perp$ is a vector containing the remaining $N-1$ spatial coordinates besides $x_1$. [It is straightforward to generalize the following procedure for curvilinear initial surfaces, and even $\Vect{k}$-space surfaces in the event that $\psi(0,\Vect{x}_\perp)$ contains an $\Vect{x}$-space caustic.] Equation \eq{eq:psiINIT} provides the following initial conditions for the rays:
\begin{equation}
    \Vect{z}(0, \Vect{\tau}_\perp) = 
    \begin{pmatrix}
        0 \\
        \Vect{\tau}_\perp \\
        k_1(0, \Vect{\tau}_\perp) \\
        \nabla \theta_0(\Vect{\tau}_\perp)
    \end{pmatrix}
    ,
    \label{eq:zINIT}
\end{equation}

\noindent where $\Vect{\tau}_\perp \equiv \Vect{x}_\perp$ are coordinates on the initial plane and $k_1(0, \Vect{\tau}_\perp)$ solves the dispersion relation \eq{eq:goDISP} when $\tau_1 = 0$, \ie
\begin{equation}
    \Symb{D}[0, \Vect{\tau}_\perp, k_1(0, \Vect{\tau}_\perp), \nabla \theta_0(\Vect{x}_\perp)]
    = 0
    .
\end{equation}

\noindent The next step is to evolve the rays via \Eq{eq:goRAYS}, or more compactly,
\begin{equation}
    \dot{\Vect{z}}(\Vect{\tau})
    = \JMat{2N} \, \pd{\Vect{z}} \Symb{D}[\Vect{z}(\Vect{\tau})]
    ,
\end{equation}

\noindent subject to the initial conditions \eq{eq:zINIT}. (As a reminder, the dot $\cdot$ denotes $\pd{\tau_1}$.) This can be done using the adaptive time-stepping scheme presented in \Sec{sec:ALGadapt}. For each timestep $\Vect{\tau} = \Vect{t}$ along a ray, one computes the rectangular matrix $\pd{\Vect{\tau}}\Vect{z}(\Vect{t})$ (\eg via finite difference) and subsequently performs a QR decomposition to obtain $\Mat{Q}_\Vect{t}$ and $\Mat{R}_\Vect{t}$, as described in \Eq{eq:zQR}. This can be accomplished using the Gram--Schmidt procedure outlined in \Sec{sec:ALGgs}, but note that in practice reorthogonalization techniques may be required to ensure the norm of $\Mat{Q}_\Vect{t}$ is sufficiently close to unity. Having obtained $\Mat{Q}_\Vect{t}$, the symplectic submatrices $\Mat{A}_\Vect{t}$ and $\Mat{B}_\Vect{t}$ can be obtained via \Eq{eq:QRab}.

To allow for $\det \Mat{B}_\Vect{t} = 0$, as discussed in \App{sec:APPsingular} we perform an SVD of $\Mat{B}_\Vect{t}$ to obtain its rank $\rho$ along with the submatrices $\Mat{\Lambda}_{\rho \rho}$, $\Mat{a}_{\rho \rho}$, and $\Mat{a}_{\varsigma \varsigma}$ [\Eqs{eq:bSVD}--\eq{eq:aSVD}]. One then computes the MGO prefactor function
\begin{align}
    \MTnorm
    &= 
    \psi_\textrm{in}(\Vect{t}_\perp) \sqrt{\dot{x}_1(0, \Vect{t}_\perp)}
    \nonumber\\
    &\hspace{4mm}\times
    \frac{
        \sigma_\Vect{t} \,
        \exp\left[i \int_0^{t_1} \dd \xi \, \Vect{k}(\xi, \Vect{t}_\perp)^\intercal \dot{\Vect{x}}(\xi, \Vect{t}_\perp) \right]
    }{
        (-2\pi i)^{\rho/2}
        \sqrt{\det \Mat{\Lambda}_{\rho\rho} \det \Mat{a}_{\varsigma \varsigma} \det \Mat{R}_\Vect{t} }
    }
    .
\end{align}

\noindent Note that the phase integral has a clear incremental structure, \ie 
\begin{equation}
    \int_0^{t_1 + \Delta t} \dd \xi 
    = \int_0^{t_1} \dd \xi 
    +
    \int_{t_1}^{t_1 + \Delta t} \dd \xi  
    ,
\end{equation}

\noindent that can be leveraged for efficient evaluation along a ray. Also, the overall sign ambiguity $\sigma_\Vect{t}$ is chosen to maintain the continuity of $\MTnorm$ along a ray, and should be initialized such that $\MTnorm \Upsilon_\Vect{t} = \psi_\textrm{in}$ when $t_1 = 0$. 

Next, one computes the tangent-space wavefield $\fourier{\psi}(\Vect{X})$ by integrating $\Vect{K}_\Vect{t}(\Vect{X})$ to obtain the phase, and by using either \Eq{eq:mgoENV} or \Eq{eq:mgoENVsol} to compute the envelope [noting that $J(\Vect{\tau}) = \det \Mat{Q}_\Vect{t}^\intercal \Mat{Q}_\Vect{\tau} \Mat{R}_\Vect{\tau}$]. One can then take the integral
\begin{align}
    \hspace{-1mm}\Upsilon_\Vect{t}(\Vect{x})
    &=
    \int_{\cont{0}} \dd \Vect{\epsilon}_\rho \,
    \fourier{\psi}
    \left[
        \Mat{L}_\textrm{s}
        \begin{pmatrix}
            \Vect{X}_\Vect{t}^\rho(\Vect{t}) + \Vect{\epsilon}_\rho \\
            \Mat{a}_{\varsigma \varsigma} \Vect{x}_\varsigma
        \end{pmatrix}
    \right]
    \nonumber\\
    &\hspace{9mm}\times
    \exp
    \left[
        - \frac{i}{2} \Vect{\epsilon}_\rho^\intercal \, \Mat{a}_{\rho \rho} \Mat{\Lambda}_{\rho \rho}^{-1} \, \Vect{\epsilon}_\rho
        - i \Vect{\epsilon}_\rho^\intercal \Vect{K}_\Vect{t}^\rho(\Vect{t})
    \right]
\end{align}

\noindent over the invertible subspace of $\Mat{B}_\Vect{t}$ using the Gauss--Freud quadrature rule described in \Sec{sec:ALGgs} (where the notation is defined in \App{sec:APPsingular}). Finally, each branch of $\Vect{k}(\Vect{x})$ is summed over to obtain the MGO solution via \Eq{eq:MGO}. 

This final step may be difficult to perform numerically, because one must be able to identify the multivaluedness of the ray map when interpolating $\Vect{\tau}(\Vect{x})$. Essentially, ray discretization can obscure this feature through a type of aliasing, yielding a sampling of $\Vect{\tau}(\Vect{x})$ that appears single-valued but highly oscillatory as the discretization randomly samples from the different branches. Using an inverse ray-tracing framework in place of the standard forward ray-tracing framework can improve this situation by decoupling the field evaluation points from the ray discretization (sampling). Moreover, as shown by \Refs{Colaitis19a,Colaitis21}, reducing the influence of ray-discretization noise can result in large improvements in computational efficiency, even when accounting for the additional operations required by inverse ray-tracing, because less rays are needed to obtain the same field information. Inverse ray-tracing also opens the possibility for using complex rays that might allow MGO to model the evanescent fields that occur in caustic shadow regions. This is something that remains to be investigated.

% ==================== %
% ---- CONCLUSION ---- %
% ==================== %
%\vspace{3mm}
\section{Conclusions}

\label{sec:concl}

It is common practice to employ ray-tracing codes to optimize wave-plasma interactions in nuclear fusion research. Unfortunately, these codes are based on the GO approximation, which is not valid at caustics. This significantly limits their predictive capabilities and slows down design iteration by requiring the use of full-wave codes instead. Here we present a recently developed framework called metaplectic geometrical optics (MGO) for accurately modeling caustics within ray-tracing modules. Rather than evolving the wavefield amplitude in the traditional $\Vect{x}$ coordinates, which leads to caustic singularities, MGO evolves the wavefield amplitude in mixed coordinate-momentum variables that are optimally chosen to avoid caustics. These mixed-variable representations are obtained using metaplectic transforms (MT), for which MGO derives its name. (The trivial cases for mixed-variable representations, namely the pure $\Vect{x}$ or pure $\Vect{k}$ representation, are generated by the corresponding special cases of the MT: the identity and the Fourier transform.) MGO is also `caustic-agnostic' in the sense that it works on all types of caustics; hence, knowledge of the catastrophe-theoretical description of caustics (see \Sec{sec:paraxial}), which is often assumed to be a prerequisite for modeling caustics, is unnecessary for MGO. This makes MGO more robust and accessible as a theory.

In this work, we rederive the MGO theory using transformations that are both symplectic and orthogonal (orthosymplectic), rather than merely symplectic as was done in previous work. The formulas that result from this modified derivation are considerably simpler than those published previously in \Refs{Lopez20, Lopez21a}, which allows for MGO to be related to standard GO and to other published semiclassical caustic-removal schemes in a straightforward manner. Indeed, we present here the first explicit proof that MGO reduces to GO when evaluated away from caustics, which is an important result for instilling confidence in MGO but has thus far only been inferred from the results of numerical MGO calculations. We also present a new interpretation of MGO as a delta-windowed semiclassical integral that allows for arbitrary wavefunction profiles, rather than being restricted to bounded wavepackets as most semiclassical methods assume. We anticipate this observation will be the foundation for future dedicated comparison studies.

Besides outlining the basic theory, we also discuss several recently developed algorithms for MGO. These algorithms are: \textbf{(i)} an adaptive integration of the ray trajectories specifically tailored to MGO, \textbf{(ii)} an orthogonalization procedure to determine the optimal MT to prevent caustics at each point along a ray, \textbf{(iii)} a fast near-identity MT algorithm for evolving the optimal representation along a ray, and \textbf{(iv)} a specialized Gauss--Freud quadrature for performing the inverse MT that reverts the optimal representation along a ray back to the original $\Vect{x}$ variables. The foundations are now set for the development of an MGO-based ray-tracing code. The orthosymplectic transformations used here have consequences in this regard: the new MGO formalism is simpler to compute and more memory efficient, since the enhanced symmetry of $\Mat{S}$ means less elements need to be stored; however, errors may develop due to any erroneous non-orthogonality of $\Mat{S}$ introduced by the chosen numerical method for performing the orthogonalization. The tradeoff between these effects can be investigated when benchmarking a future MGO-based code. Further theoretical extensions of MGO are also planned, including coupling MGO with XGO~\cite{Ruiz15a,Ruiz15b,Ruiz17a} and quasioptics~\cite{Dodin19,Yanagihara19a,Yanagihara19b,Yanagihara21a,Yanagihara21b} to model mode-converting beams near caustics, and generalizing the MTs for complex-valued coordinate transformations~\cite{Wolf74} to model evanescent waves.

\section*{Acknowledgments}

This work was supported by the U.S.~DOE through Contract No.~DE-AC02-09CH11466

\section*{Disclosures}

The authors declare no conflicts of interest.

\section*{Data availability}

The data that support the findings of this study are available from the corresponding author upon reasonable request.

% ==================== %
% ----- APPENDIX ----- %
% ==================== %

\appendix

% ==================== %
% ---- WWT review ---- %
% ==================== %

\section{Overview of the Wigner--Weyl transform}
\label{sec:APPwwt}

Here we summarize the main identities for the Wigner--Weyl transform (WWT) that are necessary to derive the results presented in this work. (See \Refs{Tracy14, Littlejohn86a, Dodin19} for more detailed summaries.) The WWT (denoted $\Weyl$) maps a given operator $\oper{A}(\VectOp{z})$ to a corresponding phase-space function $\Symb{A}(\Vect{z})$ (called the Weyl symbol of $\oper{A}$) as
\begin{align}
    \hspace{-2mm}\Symb{A}(\Vect{z})
    &= \Weyl\left[ \oper{A}(\VectOp{z}) \right]
    \nonumber\\
    &\doteq
    \int \dd \Vect{\zeta} \,
    \frac{
        \exp\left(
            i\Vect{\zeta}^\intercal \JMat{2N} \Vect{z} 
        \right)
    }{(2\pi)^N} \, \Tr \left[
        \exp\left(
            -i\Vect{\zeta}^\intercal \JMat{2N} \VectOp{z}
        \right)
        \oper{A}
    \right] 
    ,
    \label{eq:wigner}
\end{align}

\noindent where $\Tr$ is the matrix trace and the integral is taken over phase space. The inverse WWT maps a phase-space function $\Symb{A}$ to an operator $\oper{A}$ as
\begin{align}
    \hspace{-4mm}\oper{A}(\VectOp{z})
    &=
    \WeylInv \left[ \Symb{A}(\Vect{z}) \right]
    \nonumber\\
    &\doteq
    \int \frac{\dd \Vect{z}' \, \dd \Vect{\zeta}}{(2\pi)^{2N}} \, 
    \Symb{A}(\Vect{z}') 
    \exp\left(
        -i\Vect{\zeta}^\intercal \JMat{2N} \Vect{z}'
        +i\Vect{\zeta}^\intercal \JMat{2N} \VectOp{z}
    \right) 
    ,
    \label{eq:weyl}
\end{align}

\noindent where both integrals are taken over phase space.

The WWT preserves hermiticity, \ie 
\begin{equation}
    \Weyl\left[\oper{A}^\dagger \right] = \Symb{A}^*
    ,
\end{equation}

\noindent and it preserves locality, in that two operators that are close approximations of each other map to two functions that are also close approximations of each other, and vice versa. The WWT of the product of two operators can be concisely represented as the so-called Moyal product $\star$ of their symbols:
\begin{equation}
    \Weyl[\oper{A}\oper{B}] = \Symb{A}(\Vect{z}) \star \Symb{B}(\Vect{z}) .
    \label{eq:weylMOYAL}
\end{equation}

\noindent This (non-commutative) product is given explicitly as
\begin{equation}
    \Symb{A}(\Vect{z}) \star \Symb{B}(\Vect{z}) 
    = \left. 
        \sum_{s = 0}^\infty 
        \frac{\left( \frac{i}{2} \pd{\Vect{z}}^\intercal \, \JMat{2N} \, \pd{\Vect{\zeta}} \right)^{s}}
        {s!}  
        \Symb{A}(\Vect{z}) \Symb{B}(\Vect{\zeta}) 
    \right|_{\Vect{\zeta} = \Vect{z}}
    .
    \label{eq:moyalDEF}
\end{equation}

\noindent These rules can be used to compute the following relevant WWT pairs:
\begin{gather}
    f(\Vect{q}) \Longleftrightarrow f(\VectOp{q}) 
    , \quad
    f(\Vect{p}) \Longleftrightarrow f(\VectOp{p})
    , \nonumber\\
    \Vect{p}^\intercal \Vect{v}(\Vect{q}) \Longleftrightarrow 
    \frac{\Vect{v}(\VectOp{q})^\intercal \VectOp{p} + \VectOp{p}^\intercal\Vect{v}(\VectOp{q})}{2}
    .
\end{gather}

% ==================== %
% - NonInvertible MGO - %
% ==================== %

\section{MGO for quasiuniform ray patterns}
\label{sec:APPsingular}

Here we present MGO formulas for the degenerate case when $\det \Mat{B} = 0$. We call such ray manifolds `quasiuniform ray patterns', since a sufficient condition of $\det \Mat{B} = 0$ is propagation in uniform medium. Suppose that $\Mat{B}$ has rank $\rho$ and corank $\varsigma = N - \rho$. We can perform a singular-value decomposition of $\Mat{B}$ as%
%%%
~\footnote{A less restrictive decomposition is used in \Ref{Littlejohn86a} to derive results analogous to those presented in \App{sec:APPsingular}. However, our use of the singular-value decomposition is more practical due to the plethora of efficient algorithms for its computation~\cite{Press07}.}%
%%%
\begin{equation}
    \Mat{B} = \Mat{L}_\textrm{s} \, \widetilde{\Mat{B}} \, \Mat{R}_\textrm{s}^\intercal ,
    \label{eq:bSVD}
\end{equation}

\noindent where $\widetilde{\Mat{B}}$ is a diagonal matrix given by
\begin{equation}
    \widetilde{\Mat{B}}
    =
    \begin{pmatrix}
        \Mat{\Lambda}_{\rho \rho} & \OMat{ \rho \varsigma} \\
        \OMat{\varsigma \rho} & \OMat{\varsigma \varsigma}
    \end{pmatrix}
    .
\end{equation}

\noindent The subscript $_{mn}$ is used to indicate a matrix sub-block is size $m \times n$. Correspondingly, $\Mat{\Lambda}_{\rho \rho}$ is a diagonal matrix containing all nonzero singular values of $\Mat{B}$ and hence has $\det \Mat{\Lambda}_{\rho \rho} \neq 0$ by definition. The matrices $\Mat{L}_\textrm{s}$ and $\Mat{R}_\textrm{s}$ are both orthogonal and can be written in terms of the left and right singular vectors $\{ \unit{\Vect{\ell}}_j \}$ and $\{ \unit{\Vect{r}}_j \}$ of $\Mat{B}$ as
\begin{equation}
    \Mat{L}_\textrm{s} 
    = 
    \begin{pmatrix}
        \uparrow &  & \uparrow \\[1mm]
        \unit{\Vect{\ell}}_1 & \ldots & \unit{\Vect{\ell}}_N \\[1mm]
        \downarrow &  & \downarrow
    \end{pmatrix}
    ,
    \quad
    \Mat{R}_\textrm{s} 
    = 
    \begin{pmatrix}
        \uparrow &  & \uparrow \\[1mm]
        \unit{\Vect{r}}_1 & \ldots & \unit{\Vect{r}}_N \\[1mm]
        \downarrow &  & \downarrow
    \end{pmatrix}
    .
\end{equation}

Similarly, let us introduce the matrix projection
\begin{equation}
    \widetilde{\Mat{A}} 
    \doteq \Mat{L}_\textrm{s}^\intercal \Mat{A} \Mat{R}_\textrm{s}
    \equiv
    \begin{pmatrix}
       \Mat{a}_{\rho \rho} & \OMat{\rho \varsigma} \\
        \OMat{\varsigma \rho} & \Mat{a}_{\varsigma \varsigma}
    \end{pmatrix}
    ,
    \label{eq:aSVD}
\end{equation}

\noindent and the vector projections
\begin{equation}
    \Mat{R}_\textrm{s}^\intercal\Vect{x}
    =
    \begin{pmatrix}
        \Vect{x}_\rho \\[1mm]
        \Vect{x}_\varsigma
    \end{pmatrix}
    , \quad
    \Mat{L}_\textrm{s}^\intercal \Vect{X}
    =
    \begin{pmatrix}
        \Vect{X}_\rho \\[1mm]
        \Vect{X}_\varsigma
    \end{pmatrix}
    .
\end{equation}

\noindent Importantly, $\Mat{a}_{\varsigma \varsigma}$ is orthogonal ($\Mat{a}_{\varsigma \varsigma}^{-1} = \Mat{a}_{\varsigma \varsigma}^{\intercal}$), which implies that $\det \Mat{a}_{\varsigma \varsigma} = \pm 1$. We should also note that the block-diagonal structure of $\widetilde{\Mat{A}}$ is only true when $\Mat{S}$ is orthosymplectic. Then, it can be shown that the inverse-MT matrix elements take the following form when $\det \Mat{B} = 0$~\cite{Lopez21a}:
\begin{equation}
    \braket{\Vect{x}(\Vect{y})}{\Vect{X}(\Vect{Y})}
    =
    \frac
    {
        \sigma \, 
        \exp
        \left[
            - i \,
            G^\rho(\Vect{y}_\rho, \Vect{Y}_\rho)
        \right]
        \, 
        \delta
        \left(
            \Vect{Y}_\varsigma - \Mat{a}_{\varsigma \varsigma} \Vect{y}_\varsigma
        \right)
    }
    {
        (-2\pi i)^{\rho/2}
        \sqrt
        {
            \det \Mat{\Lambda}_{\rho \rho}
            \, \det \Mat{a}_{\varsigma \varsigma}
        }
    }
    ,
\end{equation}

\noindent where we have defined 
\begin{align}
    G^\rho(\Vect{y}_\rho, \Vect{Y}_\rho)
    \doteq
    \frac{1}{2} \Vect{Y}_\rho^\intercal \, \Mat{a}_{\rho \rho} \Mat{\Lambda}_{\rho \rho}^{-1} \, \Vect{Y}_\rho
    &- \Vect{Y}_\rho^\intercal \, \Mat{\Lambda}_{\rho \rho}^{-1} \Vect{y}_\rho
    \nonumber\\
    &+ \frac{1}{2} \Vect{y}_\rho^\intercal \, \Mat{\Lambda}_{\rho \rho}^{-1} \Mat{a}_{\rho \rho} \, \Vect{y}_\rho
    ,
\end{align}

\noindent This representation of the MT ultimately yields the valid formulas for \Eqs{eq:MTpre} and \eq{eq:upsilon} when $\det \Mat{B} = 0$:
\begin{align}
    \MTnorm(\Vect{x})
    &=
    \frac
    {
        \sigma_\Vect{t} \, 
        \alpha_\Vect{t}
        \exp
        \left\{
            - i G^\rho_\Vect{t}[\Vect{x}_\rho, \Vect{X}_\Vect{t}^\rho(\Vect{t}) ]
        \right\}
    }
    {
        (- 2 \pi i)^{\rho/2}
        \sqrt
        {
            \det \Mat{\Lambda}_{\rho \rho}
            \, \det \Mat{a}_{\varsigma \varsigma}
        }
    }
    , \\
    \Upsilon_\Vect{t}(\Vect{x})
    &=
    \int_{\cont{0}} \dd \Vect{\epsilon}_\rho \,
    \fourier{\psi}
    \left[
        \Mat{L}_\textrm{s}
        \begin{pmatrix}
            \Vect{X}_\Vect{t}^\rho(\Vect{t}) + \Vect{\epsilon}_\rho \\
            \Mat{a}_{\varsigma \varsigma} \Vect{x}_\varsigma
        \end{pmatrix}
    \right]
    \nonumber\\
    &\hspace{30mm}\times
    \exp
    \left[
        - i \gamma_\Vect{t}^\rho(\Vect{\epsilon}_\rho, \Vect{x}_\rho)
    \right]
    ,
    \label{eq:mgoUPSILONb0}
\end{align}

\noindent where we have defined
\begin{align}
    \gamma_\Vect{t}^\rho(\Vect{\epsilon}_\rho, \Vect{x}_\rho)
    &\doteq
    \frac{1}{2} \Vect{\epsilon}_\rho^\intercal \, \Mat{a}_{\rho \rho} \Mat{\Lambda}_{\rho \rho}^{-1} \, \Vect{\epsilon}_\rho
    + \Vect{\epsilon}_\rho^\intercal \Mat{\Lambda}_{\rho \rho}^{-1}
    \left[
        \Mat{a}_{\rho \rho}^\intercal \Vect{X}_\Vect{t}^\rho(\Vect{t})
        - \Vect{x}_\rho
    \right] 
    .
\end{align}

\noindent Note that the $\Vect{t}$ dependence of the matrices has been suppressed for ease of notation.

% ==================== %
% --- Coherent MGO --- %
% ==================== %

\section{MGO with mixed-coherent-state representation of the MT}
\label{sec:APPcoherent}

Here we present an alternate formulation of MGO that is insensitive to the value of $\det \Mat{B}$. This is accomplished by using a mixture of configuration-space eigenstates and Gaussian coherent states~\cite{Scully12} to calculate the matrix elements of the inverse MT.

A Gaussian coherent state centered at phase-space location $\Stroke{\Vect{Z}}_0 = (\Vect{X}_0, \Vect{K}_0)$ is represented by the state vector $\ket{\Stroke{\Vect{Z}}_0}$, whose wavefunction has the explicit form
\begin{equation}
    \braket{\Vect{X}(\Vect{Y})}{\Stroke{\Vect{Z}}_0}
    =
    \frac
    {
        \exp
        \left[
            - \frac{|\Vect{Y} - \Vect{X}_0|^2}{2}
            + i \Vect{K}_0^\intercal 
            \left(
                \Vect{Y}
                - \frac{\Vect{X}_0}{2}
            \right)
        \right] 
    }
    {
        \pi^{N/4}
    }
    .
    \label{eq:coherentX}
\end{equation}

\noindent These states satisfy a completeness relation analogous to \Eq{eq:norm}:
\begin{equation}
    \IdentOp = \int \frac{\dd \Vect{X}_0 \dd \Vect{K}_0}{(2\pi)^N} \, \ket{\Stroke{\Vect{Z}}_0}
    \bra{\Stroke{\Vect{Z}}_0}
    ,
\end{equation}

\noindent which allows the inverse-MT matrix elements to be expressed as
\begin{equation}
    \braket{\Vect{x}(\Vect{y})}{\Vect{X}(\Vect{Y})}
    =
    \int \frac{\dd \Vect{X}_0 \dd \Vect{K}_0}{(2\pi)^N} \, \braket{\Vect{x}(\Vect{y})}{\Stroke{\Vect{Z}}_0}
    \braket{\Stroke{\Vect{Z}}_0}{\Vect{X}(\Vect{Y})}
    .
    \label{eq:coherMT}
\end{equation}

\noindent After using \Eq{eq:coherentX} along with the known result~\cite{Littlejohn86a}
\begin{align}
    &\braket{\Vect{x}(\Vect{y})}{\Stroke{\Vect{Z}}_0}
    =
    \frac
    {
        \sigma
        \,
        \exp
        \left(
            -\frac{1}{2} \Vect{y}^\intercal \Vect{y}
        \right)
    }
    {
        \pi^{N/4} 
        \sqrt{ \det(\Mat{A}_\Vect{t} - i \Mat{B}_\Vect{t} ) }
    }
    \nonumber\\
    &\times
    \exp
    \left[
        \left(
            \Vect{y}
            - \frac{i}{2} \Mat{B}_\Vect{t}^\intercal \Vect{\zeta}
        \right)^\intercal
        \left(
            \Mat{A}_\Vect{t}
            - i \Mat{B}_\Vect{t}
        \right)^{-1}
        \Vect{\zeta}
        - \frac{1}{2} \Vect{X}_0^\intercal \Vect{\zeta}
    \right]
    ,
\end{align}

\noindent then integrating over $\Vect{X}_0$, \Eq{eq:coherMT} takes the form
\begin{align}
    \hspace{-1mm}\braket{\Vect{x}(\Vect{y})}{\Vect{X}(\Vect{Y})}
    = 
    \int \dd \Vect{K}_0 \,
    \frac
    {
        \sigma \,
        \exp
        \left[
            \fourier{G}(\Vect{y}, \Vect{\xi})
            - |\Vect{K}_0|^2
        \right]
    }
    {
        (\sqrt{2} \pi)^N 
        \sqrt{ \det(2 \Mat{A}_\Vect{t} - i \Mat{B}_\Vect{t} ) }
    }
    ,
    \label{eq:gaussMT}
\end{align}

\noindent where we have introduced the complex vectors $\Vect{\zeta} \doteq \Vect{X}_0 + i \Vect{K}_0$ and $\Vect{\xi} \doteq \Vect{Y} + 2 i \Vect{K}_0$, and we have defined
\begin{align}
    \hspace{-1mm}\fourier{G}(\Vect{y}, \Vect{\xi})
    & \doteq
    - \frac{1}{2} \Vect{y}^\intercal 
    \left(
        2\Mat{A}_\Vect{t} 
        - i \Mat{B}_\Vect{t}
    \right)^{-1}
    \left(
        \Mat{A}_\Vect{t}
        - 2 i \Mat{B}_\Vect{t}
    \right)
    \Vect{y}
    \nonumber\\
    &\hspace{4mm}
    +
    \left(
        \Vect{y}
        - \frac{1}{2} \Mat{A}_\Vect{t}^\intercal \Vect{\xi}
    \right)^\intercal
    \left(
        2 \Mat{A}_\Vect{t}
        - i \Mat{B}_\Vect{t}
    \right)^{-1} \Vect{\xi}
    .
\end{align}

\noindent The complex matrix $2 \Mat{A} - i \Mat{B}$ is always invertible~\cite{Littlejohn87}. 

Ultimately, this approach yields the following alternate formulas for \Eqs{eq:MTpre} and \eq{eq:upsilon}:
\begin{align}
    \hspace{-1mm}\MTnorm(\Vect{x})
    &=
    \frac
    {
        \sigma_\Vect{t} \,
        \alpha_\Vect{t}
        \exp
        \left\{
            \fourier{G}_\Vect{t}[ \Vect{x}, \Vect{\xi}_\Vect{t}(\Vect{t})]
            - |\Vect{K}_\Vect{t}(\Vect{t})|^2 
        \right\}
    }
    {
        (2\sqrt{2} \, \pi)^N 
        \sqrt{ \det(2 \Mat{A}_\Vect{t} - i \Mat{B}_\Vect{t} ) }
    }
    , \\
    \hspace{-1mm}\Upsilon_\Vect{t}(\Vect{x})
    &=
    \int_{\cont{0}} \dd \Vect{\epsilon}_r \, \dd \Vect{\epsilon}_i \,
    \fourier{\psi}[\Vect{\epsilon}_r + \Vect{X}_\Vect{t}(\Vect{t})]
    \exp
    \left[
        - \fourier{\gamma}_\Vect{t}(\Vect{\epsilon}, \Vect{x})
    \right]
    ,
    \label{eq:MGOpsiGAUSS}
\end{align}

\noindent where we have defined $\Vect{\epsilon} \doteq \Vect{\epsilon}_r + i \Vect{\epsilon}_i$,
\begin{subequations}
    \begin{equation}
        %%%
        \Vect{\xi}_\Vect{t}(\Vect{t}) \doteq \Vect{X}_\Vect{t}(\Vect{t}) + 2 i \Vect{K}_\Vect{t}(\Vect{t})
        ,
    \end{equation}
    \noindent and also,
    \begin{align}
        \fourier{\gamma}_\Vect{t}(\Vect{\epsilon}, \Vect{x})
        &\doteq
        \frac{1}{2} \Vect{\epsilon}^\intercal \Mat{A}_\Vect{t}
        \left(
            2 \Mat{A}_\Vect{t}
            - i \Mat{B}_\Vect{t}
        \right)^{-1} \Vect{\epsilon}
        + \frac
        {
            |\Vect{\epsilon}_i|^2
        }{4}
        + \Vect{\epsilon}_i^\intercal \Vect{K}_\Vect{t}(\Vect{t})
        %%%
        \nonumber\\
        &\hspace{4mm}
        - \Vect{\epsilon}^\intercal 
        \left(
            2 \Mat{A}_\Vect{t}
            - i \Mat{B}_\Vect{t}
        \right)^{-\intercal}
        \left[
            \Vect{x}
            - \Mat{A}_\Vect{t}^\intercal \Vect{\xi}_\Vect{t}(\Vect{t})
        \right]
        .
    \end{align}
\end{subequations}

\noindent As mentioned, this formulation of MGO does not require any special treatment for quasiuniform ray patterns with $\det \Mat{B} = 0$ and is thus simpler to implement in a code. However, it involves computing a $2N$-D integral \eq{eq:MGOpsiGAUSS} rather than an integral \eq{eq:mgoUPSILONb0} over the non-singular subspace of $\Mat{B}$ ($N$-D at most). The tradeoff between these two factors will ultimately determine which formulation of MGO will be optimal for a specific application.

\section{Derivation of the MGO continuity factor}
\label{app:deriv}

In \Ref{Lopez20} it was shown that \Eq{eq:alphaEQ} can be formally solved along a ray in the limit that the discrete step size $h \to 0$. The solution takes the integral form
\begin{equation}
	\alpha_\Vect{t} 
	= 
	\alpha_{\left(0,\Vect{t}_\perp \right)} 
	\exp
	\left[
		- \frac{i}{2}
		\int_0^{t_1} \dd \xi \, \fourier{\eta}_{(\xi,\Vect{t}_\perp)}
	\right]
	,
	\label{eq:alphaINTapp}
\end{equation}

\noindent where $\fourier{\eta}_\Vect{t}$ is given as~\cite{Lopez20,Lopez21a}
\begin{align}
	\fourier{\eta}_\Vect{t}
	&\doteq
	\Vect{K}^\intercal _\Vect{t}(\Vect{t}) \Mat{W}_\Vect{t} \Vect{K}_\Vect{t}(\Vect{t})
	+ \Vect{X}^\intercal _\Vect{t}(\Vect{t}) \Mat{U}_\Vect{t} \Vect{X}_\Vect{t}(\Vect{t})
	+ 2 \Vect{K}^\intercal_\Vect{t}(\Vect{t})\Mat{V}_\Vect{t}^\intercal \Vect{X}_\Vect{t}(\Vect{t})
	%%%%%
	\nonumber\\
	&\hspace{4mm}
	%%%%%
	+ 2 i \pd{\Vect{X}} \Phi_\Vect{t}\left[
			\Vect{X}_\Vect{t}(\Vect{t})
		\right]^\intercal
	\left[
		\dot{\Vect{X}}_\Vect{t}(\Vect{t}) 
		- \Mat{V}_\Vect{t}^\intercal \Vect{X}_\Vect{t}(\Vect{t}) 
		- \Mat{W}_\Vect{t}^\intercal \Vect{K}_\Vect{t}(\Vect{t})
		\nullFrac
	\right]
	\nonumber\\
	&\hspace{4mm}
	+ i \Tr\left(\Mat{V}_\Vect{t} \right) 
	- 2 \Vect{K}^\intercal_\Vect{t}(\Vect{t})\dot{\Vect{X}}_\Vect{t}(\Vect{t}) 
	.
	\label{eq:etaDEFapp}
\end{align}

\noindent Here, we have introduced the $N \times N$ matrices $\Mat{U}_\Vect{t}$, $\Mat{V}_\Vect{t}$, and $\Mat{W}_\Vect{t}$ as the block elements of the Hamiltonian matrix
\begin{equation}
    \dot{\Mat{S}}_\Vect{t} \Mat{S}_\Vect{t}^{-1}
    \equiv \Mat{H}_\Vect{t}
    \doteq
    \begin{pmatrix}
		\Mat{V}_\Vect{t}^\intercal & \Mat{W}_\Vect{t} \\
		- \Mat{U}_\Vect{t} & - \Mat{V}_\Vect{t}
	\end{pmatrix}
	,
\end{equation}

\noindent with $\Mat{U}_\Vect{t}$ and $\Mat{W}_\Vect{t}$ also being symmetric. Recall that a matrix $\Mat{H}$ is Hamiltonian if it satisfies 
\begin{equation}
    \JMat{2N} \Mat{H} = - \Mat{H}^\intercal \JMat{2N}
    .
\end{equation}

\noindent One can readily show that $\dot{\Mat{S}}_\Vect{t} \Mat{S}_\Vect{t}^{-1}$ is Hamiltonian by differentiating the symplectic condition \eq{eq:symplecDEF}; moreover, the orthogonal property \eq{eq:orthoDEF} implies that $\dot{\Mat{S}}_\Vect{t} \Mat{S}_\Vect{t}^{-1}$ is also antisymmetric (as seen by differentiating the relation $\Mat{S}^\intercal \Mat{S} = \IMat{2N}$). Hence, when $\Mat{S}$ is orthosymplectic, one has that $\Mat{U}_\Vect{t} = \Mat{W}_\Vect{t}$ and also that $\Mat{V}_\Vect{t}$ is antisymmetric and thereby traceless as well.

Using these observations, the top line of \Eq{eq:etaDEFapp} can be simplified as
\begin{align}
	- \Stroke{\Vect{Z}}^\intercal_\Vect{t}(\Vect{t}) \,
	\Mat{J}_{2N}
	\dot{\Mat{S}}_\Vect{t} \Mat{S}_\Vect{t}^{-1} \,
	\Stroke{\Vect{Z}}_\Vect{t}(\Vect{t}) 
	=
	- \Vect{z}^\intercal(\Vect{t}) \,
	\Mat{S}_\Vect{t}^\intercal
	\Mat{J}_{2N}
	\dot{\Mat{S}}_\Vect{t} \,
	\Vect{z}(\Vect{t})
	.
\end{align}

\noindent Furthermore, note that integration by parts yields the simplification
\begin{align}
	&2\int_0^{t_1} \dd \xi \, \Vect{K}^\intercal_{(\xi,\Vect{t}_\perp)}(\xi,\Vect{t}_\perp)\dot{\Vect{X}}_{(\xi,\Vect{t}_\perp)}(\xi,\Vect{t}_\perp) 
	%%%
	\nonumber\\
	&=
	\Vect{K}^\intercal_{\Vect{t}}(\Vect{t}) \Vect{X}_{\Vect{t}}(\Vect{t})
	- \Vect{K}^\intercal_{(0,\Vect{t}_\perp)}(0,\Vect{t}_\perp) \Vect{X}_{(0,\Vect{t}_\perp)}(0,\Vect{t}_\perp)
	%%%
	\nonumber\\
	&\hspace{4mm}
	%%%
	- \int_0^{t_1} \dd \xi \,
	\Stroke{\Vect{Z}}^\intercal_{(\xi, \Vect{t}_\perp)}(\xi, \Vect{t}_\perp)
    \,
    \Mat{J}_{2N} 
    \,
    \dot{\Stroke{\Vect{Z}}}_{(\xi, \Vect{t}_\perp)}(\xi, \Vect{t}_\perp)
	\nonumber\\
	&=
	\Vect{K}^\intercal_{\Vect{t}}(\Vect{t}) \Vect{X}_{\Vect{t}}(\Vect{t})
	- \Vect{K}^\intercal_{(0,\Vect{t}_\perp)}(0,\Vect{t}_\perp) \Vect{X}_{(0,\Vect{t}_\perp)}(0,\Vect{t}_\perp)
	%%%
	\nonumber\\
	&\hspace{4mm}
	%%%
	- \int_0^{t_1} \dd \xi \,
	\left[
	    \Vect{z}^\intercal(\xi, \Vect{t}_\perp)
        \,
	    \Mat{S}^\intercal_{\Vect{t}}
    	\Mat{J}_{2N} 
	    \dot{\Mat{S}}_\Vect{t}
    	\, 
	    \Vect{z}(\xi, \Vect{t}_\perp)
	    %%%
    	\right.\nonumber\\
    	&\left.\hspace{22mm}
    	%%%
	    +
    	\Vect{z}^\intercal(\xi, \Vect{t}_\perp)
	    \,
    	\Mat{J}_{2N} 
	    \,
    	\dot{\Vect{z}}(\xi, \Vect{t}_\perp)
    \right]
	.
\end{align}

\noindent Also, let us note that the identity
\begin{equation}
    \dot{\Stroke{\Vect{Z}}}_\Vect{t}(\Vect{t})
	- \dot{\Mat{S}}_\Vect{t} \Mat{S}^{-1} 
	\Stroke{\Vect{Z}}_\Vect{t}(\Vect{t})
	=
	\Mat{S}_\Vect{t}
	\dot{\Vect{z}}(\Vect{t})
\end{equation}

\noindent implies that
\begin{align}
    \dot{\Vect{X}}_\Vect{t}(\Vect{t}) 
	+ \Mat{V}_\Vect{t} \Vect{X}_\Vect{t}(\Vect{t}) 
	- \Mat{W}_\Vect{t} \Vect{K}_\Vect{t}(\Vect{t})
	= \Mat{A}_\Vect{t} \dot{\Vect{x}}(\Vect{t})
	+ \Mat{B}_\Vect{t} \dot{\Vect{k}}(\Vect{t})
	.
\end{align}

\noindent Hence, after defining the overall constant function \begin{equation}
    f(\Vect{t}_\perp)\doteq \alpha_{\left(0,\Vect{t}_\perp \right)} \exp \left[ - \frac{i}{2} \Vect{K}^\intercal_{(0,\Vect{t}_\perp)}(0,\Vect{t}_\perp) \Vect{X}_{(0,\Vect{t}_\perp)}(0,\Vect{t}_\perp) \right]
    ,
\end{equation}

\noindent one obtains \Eqs{eq:alphaINT} and \eq{eq:etaDEF}.

\section{Proof of \Eq{eq:tangent}}
\label{app:tangent}

By \Eq{eq:pdX}, the matrix $\pd{\Vect{\tau}} \Vect{X}_\Vect{t}(\Vect{t})$ is invertible, so one can express $\pd{\Vect{X}} \Vect{K}_\Vect{t}\left[ \Vect{X}_\Vect{t}(\Vect{t}) \right]$ as
\begin{align}
    \pd{\Vect{X}} \Vect{K}_\Vect{t}\left[ \Vect{X}_\Vect{t}(\Vect{t}) \right]
    &=
    \pd{\Vect{\tau}} \Vect{K}_\Vect{t}\left( \Vect{t} \right)
    \left[
        \pd{\Vect{\tau}} \Vect{X}_\Vect{t}\left(\Vect{t} \right)
    \right]^{-1}
    .
    \label{eq:pdKx}
\end{align}

\noindent Analogous to \Eq{eq:pdX}, the matrix $\pd{\Vect{\tau}} \Vect{K}_\Vect{t}(\Vect{\tau})$ is given as
\begin{equation}
    \pd{\Vect{\tau}} \Vect{K}_\Vect{t}(\Vect{\tau})
    = 
    -\Mat{B}_\Vect{t} \pd{\Vect{\tau}} \Vect{x}(\Vect{\tau}) + \Mat{A}_\Vect{t} \pd{\Vect{\tau}} \Vect{k}(\Vect{\tau})
    =
    \Mat{Q}_\Vect{t}^\intercal \JMat{2N} \Mat{Q}_\Vect{\tau} \Mat{R}_\Vect{\tau}
    ,
    \label{eq:pdKtau}
\end{equation}

\noindent where we have used also \Eqs{eq:zQR} and \eq{eq:QRab}. Since the columns of $\Mat{Q}_\Vect{t}$ are vectors tangent to the ray manifold, they satisfy $\Mat{Q}_\Vect{t}^\intercal \JMat{2N} \Mat{Q}_\Vect{t} = \OMat{N}$ [cf.~\Eq{eq:tangentLAGRANG}]. Then, \Eq{eq:pdKtau} gives 
\begin{equation}
    \pd{\Vect{\tau}} \Vect{K}_\Vect{t}(\Vect{t}) = \OMat{N},
\end{equation}

\noindent so \Eq{eq:pdKx} leads to the sought result \eq{eq:tangent}.

There are two ways to intuitively understand this result. First, one expects that the graph of the ray manifold, \ie that of $\Vect{K}_\Vect{t}(\Vect{X})$, should be locally flat when viewed in a frame that is rotated to be parallel with the local tangent plane, as we have done. Second, the assumption that $\Vect{\tau}$ comprise an independent set of intrinsic coordinates on the ray manifold implies that the matrix $\pd{\Vect{\tau}} \Stroke{\Vect{Z}}_\Vect{t}(\Vect{t})$ has full column rank. Since the top-half block of $\pd{\Vect{\tau}} \Stroke{\Vect{Z}}_\Vect{t}(\Vect{t})$, \ie $\pd{\Vect{\tau}} \Vect{X}_\Vect{t}(\Vect{t})$, also has full rank, one expects the lower-half block of $\pd{\Vect{\tau}} \Stroke{\Vect{Z}}_\Vect{t}(\Vect{t})$, \ie $\pd{\Vect{\tau}} \Vect{K}_\Vect{t}(\Vect{t})$, to have rank of zero. This is only satisfied by the null matrix.

% Bibliography
\bibliography{Biblio.bib}
\bibliographystyle{apsrev4-1}
\end{document}